\documentclass[12pt,preprint]{aastex}
\begin{document}

\title{Leptonic or Hadronic Emission: X-Ray Radiation Mechanism of Large-scale Jet Knots in 3C 273}
\author{ Zhen-Jie Wang\altaffilmark{1}, Jin Zhang\altaffilmark{2}, Xiao-Na Sun\altaffilmark{3}, En-Wei,Liang\altaffilmark{1}}
\altaffiltext{1}{Guangxi Key Laboratory for Relativistic Astrophysics, School of Physical Science and Technology, Guangxi University, Nanning 530004, People's Republic of
China}
\altaffiltext{2}{Key Laboratory of Space Astronomy and Technology, National Astronomical Observatories, Chinese Academy of Sciences, Beijing 100012, People's Republic of
China; jinzhang@bao.ac.cn}
\altaffiltext{3}{School of Astronomy and Space Science, Nanjing University, Nanjing 210093, People's Republic of China}

\begin{abstract}

A comprehensively theoretical analysis on the broadband spectral energy distributions (SEDs) of large-scale jet knots in 3C 273 is presented for revealing their X-ray radiation mechanism. We show that these SEDs cannot be explained with a single electron population model when the Doppler boosting effect is either considered or not. By adding a more energetic electron (the leptonic model) or proton (the hadronic model) population, the SEDs of all knots are well represented. In the leptonic model, the electron population that contributes the X-ray emission is more energetic than the one responsible for the radio--optical emission by almost two orders of magnitude; the derived equipartition magnetic field strengths ($B_{\rm eq}$) are $\sim0.1$ mG. In the hadronic model, the protons with energy of $\sim 20$ PeV are required to interpret the observed X-rays; the $B_{\rm eq}$ values are several mG, larger than that in the leptonic model. Based on the fact that no resolved substructures are observed in these knots and the fast cooling-time of the high-energy electrons is difficult to explain the observed X-ray morphologies, we argue that two distinct electron populations accelerated in these knots are unreasonable and their X-ray emission would be attributed to the proton synchrotron radiation accelerated in these knots. In case of these knots have relativistic motion towards the observer, the super-Eddington issue of the hadronic model could be avoided. Multiwavelength polarimetry and the $\gamma$-ray observations with high resolution may be helpful to discriminate these models.

\end{abstract}


\section{Introduction}           
\label{sect:intro}

The jets of some active galactic nuclei (AGNs) can extend to kpc--Mpc scale. The substructures of these large-scale jets have been resolved in the radio, optical and X-ray bands and are defined as knots, hot-spots, and lobes (Harris \& Krawczynski 2006, and references therein). The radio--optical radiation of the substructures in large-scale jets is believed to be produced by the synchrotron process of relativistic electrons on account of the polarimetry, however, the X-ray emission mechanism is still a debated issue since the detection of X-ray radiation in these substructures (Harris \& Krawczynski 2006, and references therein). The consistency between the X-ray spectrum and the extrapolation of the radio--optical synchrotron emission indicates the same synchrotron radiation origin of X-ray emission (Sambruna et al. 2007; Zhang et al. 2010, 2018b). But for most substructures, the hard spectra in the X-ray band require a new radiation component different from the low energy band, and then the inverse Compton (IC) scattering process is suggested to explain the X-ray emission (e.g., Kataoka \& Stawarz 2005; Zhang et al. 2010, 2018b), i.e., the synchrotron-self-Compton (SSC, Stawarz et al. 2007) model and the IC scattering of the cosmic microwave background (IC/CMB, Georganopoulos \& Kazanas 2003; Abdo et al. 2010; McKeough et al. 2016; Wu et al. 2017; Zhang et al. 2018a; Guo et al. 2018) model. The X-ray emission may be produced by the synchrotron radiation of the second electron population different from the radio--optical emission (e.g., Zhang et al. 2009; Zargaryan et al. 2017; Sun et al. 2018). It can also be from the synchrotron radiation of protons in the extended regions of large-scale jets (Aharonian 2002; Kundu \& Gupta 2014).

3C 273 is a typical $\gamma$-ray emitting flat spectrum radio quasar (FSRQ) located at $z=0.158$ (Schmidt 1963). Its one-side knotty jet in the radio band extends to $\sim21\arcsec$ from the nucleus, but its optical emission is observed only from $12\arcsec$ outward (Jester et al. 2005; Uchiyama et al. 2006). The observed morphology at wavelengths from 3.6 cm to 300 nm is similar (Jester et al. 2005). Using ground-based imaging in the radio (Conway et al. 1993), near-infrared (Neumann et al. 1997), and optical bands (Meisenheimer et al. 1996; R\"{o}ser et al. 2000), the radio--IR--optical continuums are obtained for the hot-spot and the brightest knots, and they can be explained by the synchrotron radiation of a single power-law electron population (Jester et al. 2005). On the basis of the deep very large array (VLA) and HST observations of the large-scale jet, the excess near-ultraviolet over the radio--optical synchrotron radiation is revealed (Jester et al. 2005), and thus a two-component model is necessary to describe the broadband spectral energy distributions (SEDs) of these substructures. This is further confirmed by the far-ultraviolet observations at $\sim$ 150 nm with the HST; the far-ultraviolet fluxes of these substructures are compatible with the extrapolation of the X-ray power-law down to the ultraviolet (UV) band (Jester et al. 2007). The optical polarization is consistent with the radio polarization in degree and orientation, indicating that the optical emission is likely of synchrotron origin as the radio emission (Uchiyama et al. 2006). Recently, Zhang et al. (2018b) reported that the broadband SEDs from radio to X-ray bands of the large-scale jet knots in 3C 273 cannot well be represented by the radiations of a single electron population, including the synchrotron emission, SSC and IC/CMB processes.

In this paper, we use three different models to reproduce the SEDs in the radio--IR--optical--UV--X-ray bands of nine large-scale jet knots in 3C 273 for revealing their X-ray radiation mechanism. The SED data of the nine knots are taken from Jester et al. (2007). The SED modeling is given in Section 2. A discussion on the fitting results and a summary are presented in Section 3 and Section 4, respectively. Throughout, $H_0=$71 km s$^{-1}$ Mpc$^{-1}$, $\Omega_{m}=0.27$, and $\Omega_{\Lambda}=0.73$
are adopted.

\section{SED Modeling}

Doppler booting effect is critical in modeling the SEDs of jets. The Doppler factor ($\delta$) of the relativistic jet at the pc-scale in 3C 273 has been estimated using the flux-density variations\footnote{Comparing with the variability of the core radiation, the emission of substructures in large-scale jets almost does not show any variation, which is also used to estimate the origin of the $\gamma$-ray emission (e.g., Zhang et al. 2018a; Guo et al. 2018; Meyer et al. 2019).} at the radio band (Hovatta et al. 2009; Liodakis et al. 2017) or by the broadband SED modeling (e.g., Zhang et al. 2014). It was also proposed that the jets would be seriously decelerated at the kpc-scale (e.g., Arshakian \& Longair 2004; Uchiyama et al. 2006; Mullin \& Hardcastle 2009; Meyer et al. 2016). Recently, using the multi-epoch images observed by the \emph{Hubble Space Telescope} (HST) over the past twenty years, Meyer et al. (2016) reported that the kpc-scale knots in 3C 273 are compatible with being stationary, with a mean speed of $-0.2\pm0.5c$ over the whole jet. Due to the uncertainty of this issue, we model the SEDs in two cases of without considering the beaming effect ($\delta=1$) and considering the beaming effect ($\delta>1$). Note that $\delta=1/[\Gamma-(\Gamma^2-1)^{1/2}\cos\theta]$, where $\theta$ is the viewing angle and $\Gamma$ is the bulk Lorenz factor of the emission region. We take $\theta=3.3^{\circ}$ (Hovatta et al. 2009) for all the knots .

The synchrotron radiation, SSC, and IC/CMB processes of relativistic electrons (leptonic model) and protons (hadronic model) are the candidates of radiation mechanisms in our analysis. The radiation region is assumed to be a sphere with radius $R$, which is derived from the angular radius and taken from Jester et al. (2005), as listed in Table 1. The CMB peak frequency at $z=0$ is $\nu_{\rm CMB}=1.6\times10^{11}$ Hz and the CMB energy density in the comoving frame is $U^{'}_{\rm CMB}=\frac{4}{3}\Gamma^2U_{\rm CMB}(1+z)^4$ (Dermer \& Schlickeiser 1994; Georganopoulos et al. 2006), where $U_{\rm CMB}=4.2\times10^{-13}$ erg cm$^{-3}$. The number distributions of the electrons and/or protons are taken as an exponential cutoff power-law or a broken power-law.

\begin{itemize}
\item Broken power-law:
\begin{equation}
N_1(E)=A_{1}\left\{ \begin{array}{ll}
(\frac{E}{E_{0}})^{-p_1}  &  \mbox{ $E\leq E_{\rm b}$} \\
(\frac{E_{\rm b}}{E_{0}})^{p_2-p_1}(\frac{E}{E_{0}})^{-p_2}  &  \mbox{$E> E_{\rm b}$,}
\end{array}
\right.
\end{equation}

\item Exponential cutoff power-law:
\begin{equation}
N_0(E)=A_{0}~(\frac{E}{E_{0}})^{-p_{0}}~{\rm exp}\left[-\left(\frac{E}{E_{\rm c}}\right)^{\beta }\right],
\end{equation}
where $E_{0}=1$ TeV. $\beta$ is fixed to 2, in which the electrons are accelerated up to 100 TeV and beyond and the maximum energy of electron is resulted from the competition between the acceleration and energy loss rates (Zirakashvili \& Aharonian 2007; Aharonian et al. 2017).
\end{itemize}

We consider three scenarios in our SED fits, i.e., a single electron population, two electron populations, and an electron population plus a proton population. The numerical package of Naima (Zabalza 2015), which includes a set of non-thermal radiation models and the spectral fitting procedure, is used in this paper. The best-fit and uncertainty of the model parameters are derived via the Markov Chain Monte Carlo (MCMC; Foreman-Mackey et al. 2013) method.

\subsection{Scenario I: A Single Electron Population}

\begin{figure*}
\includegraphics[angle=0,scale=0.26]{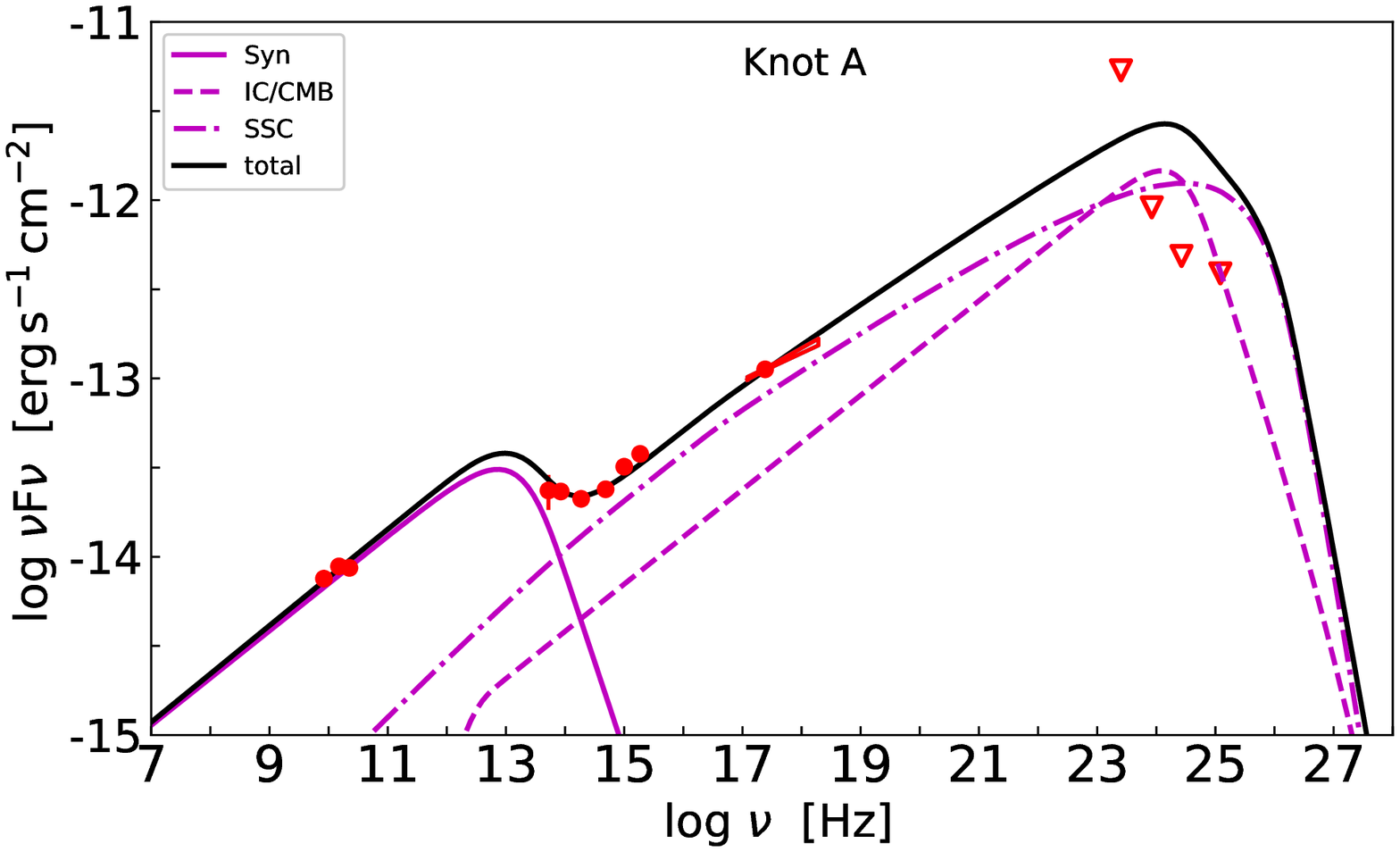}
\includegraphics[angle=0,scale=0.26]{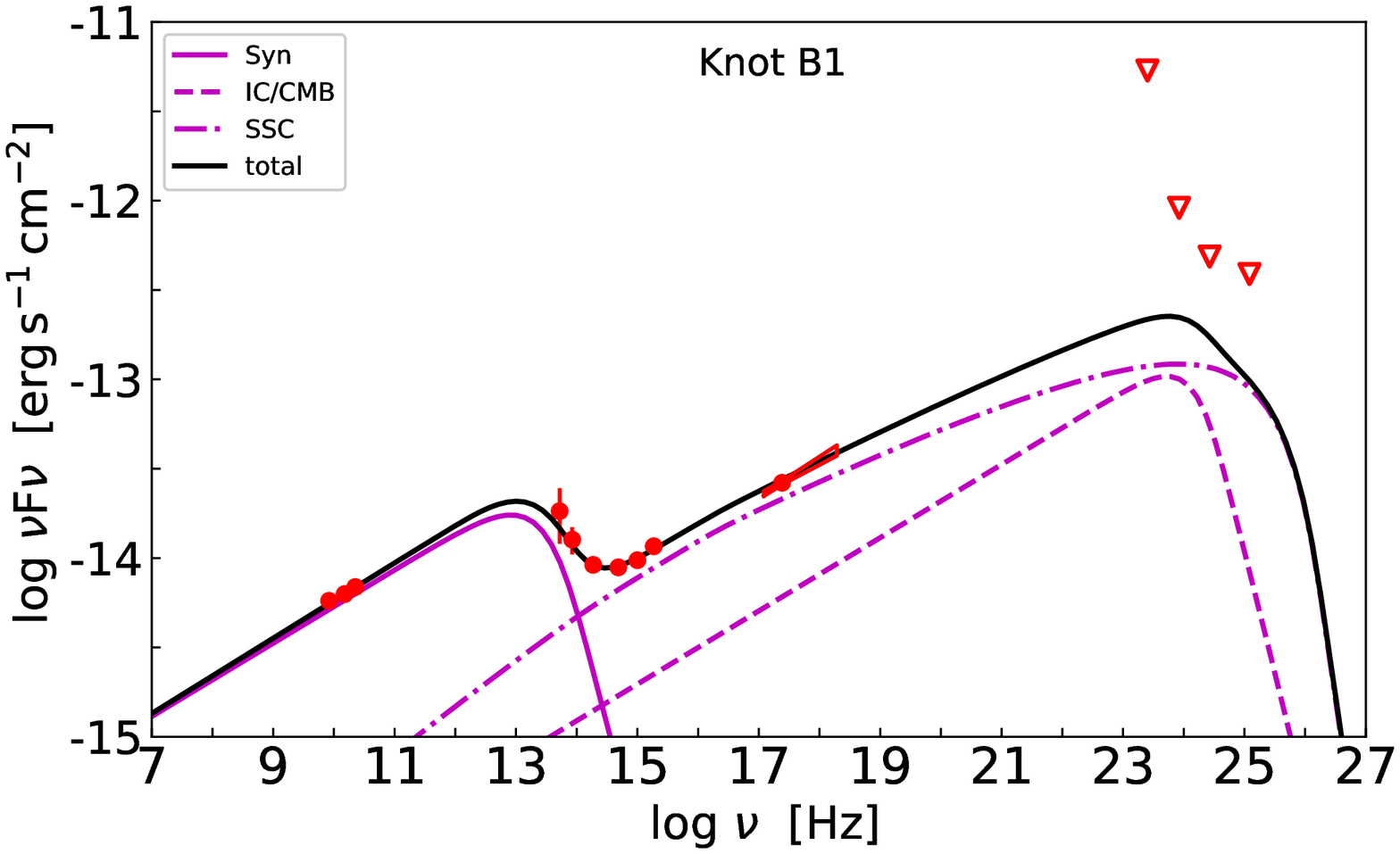}
\includegraphics[angle=0,scale=0.26]{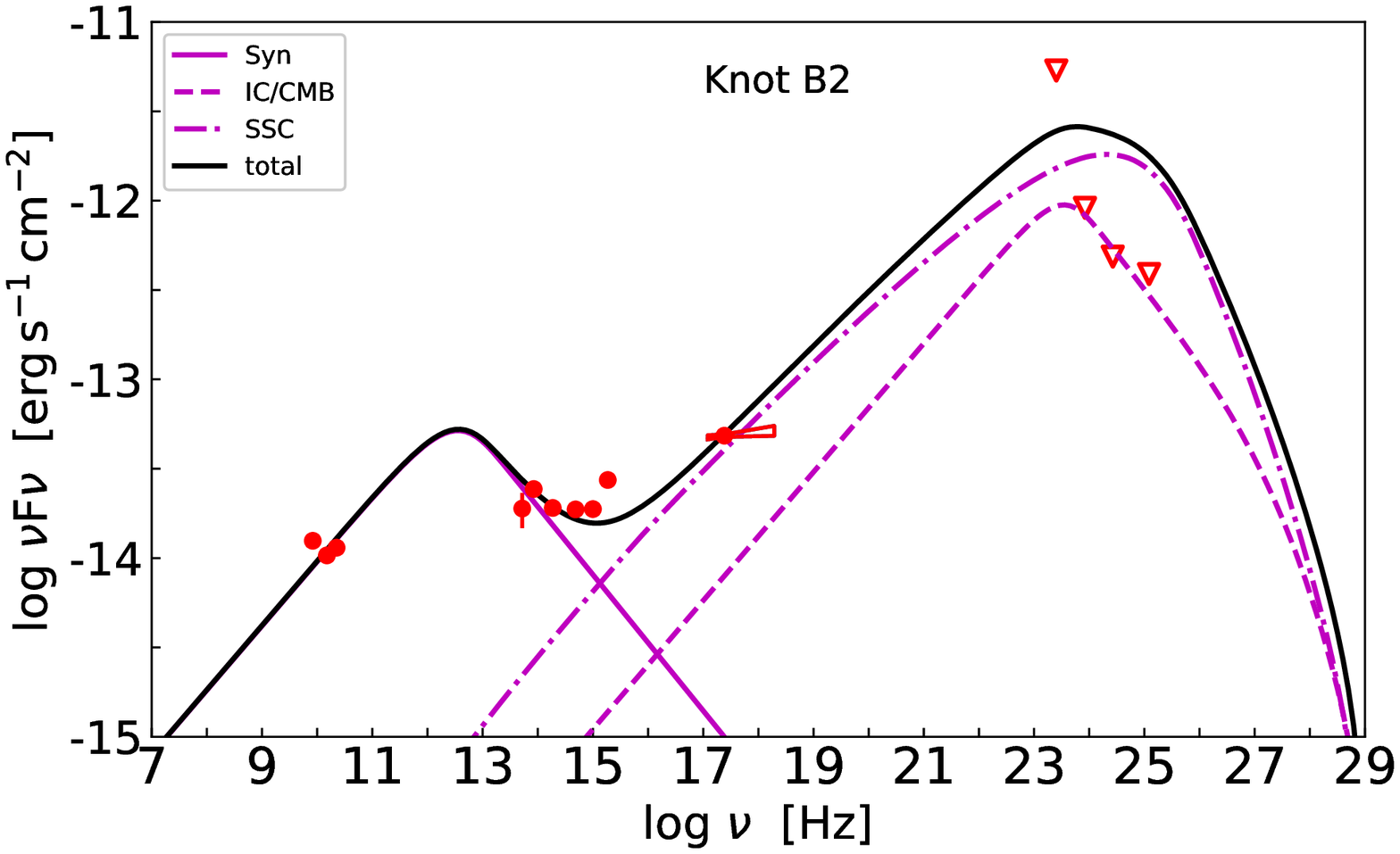}
\includegraphics[angle=0,scale=0.26]{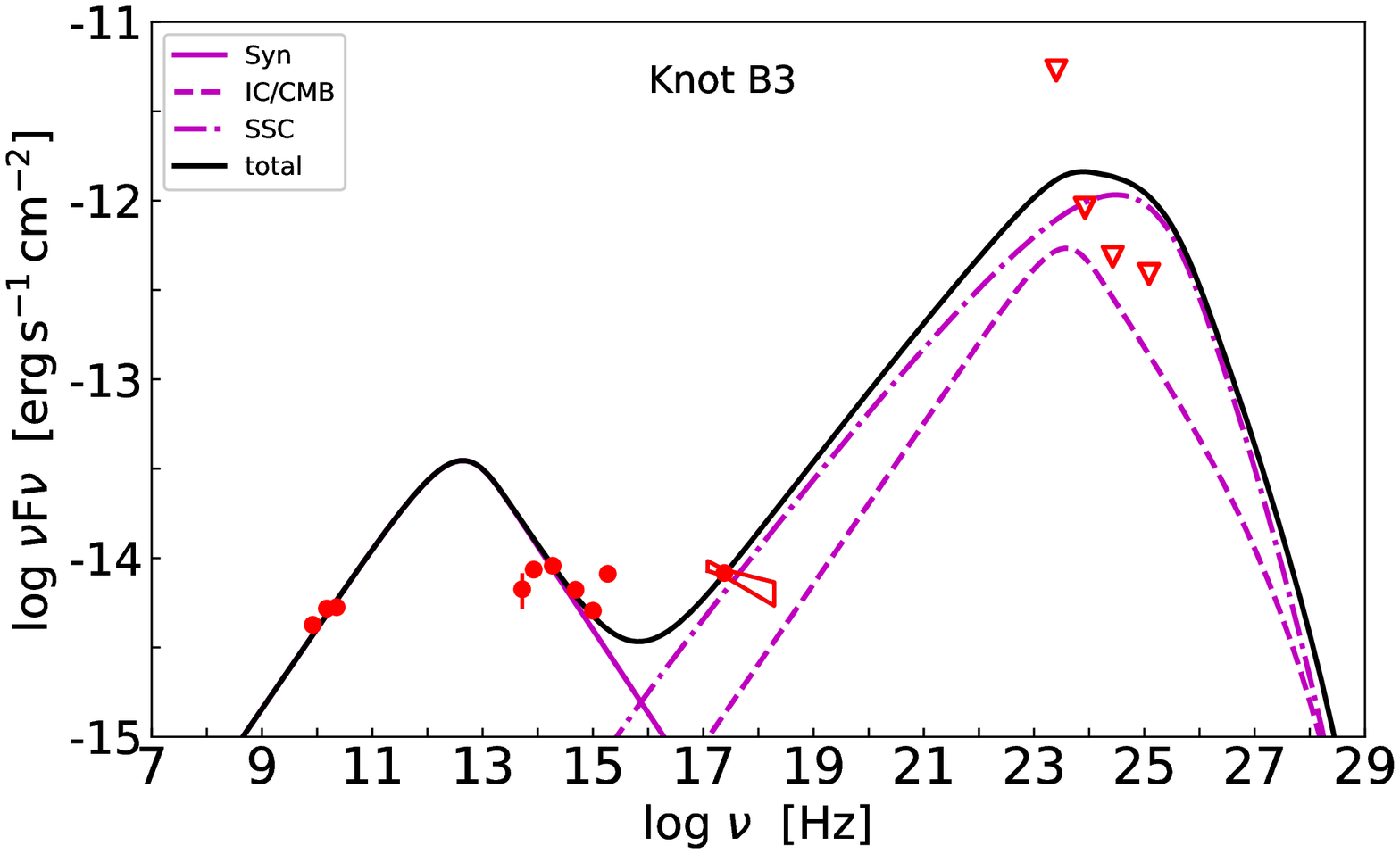}
\includegraphics[angle=0,scale=0.26]{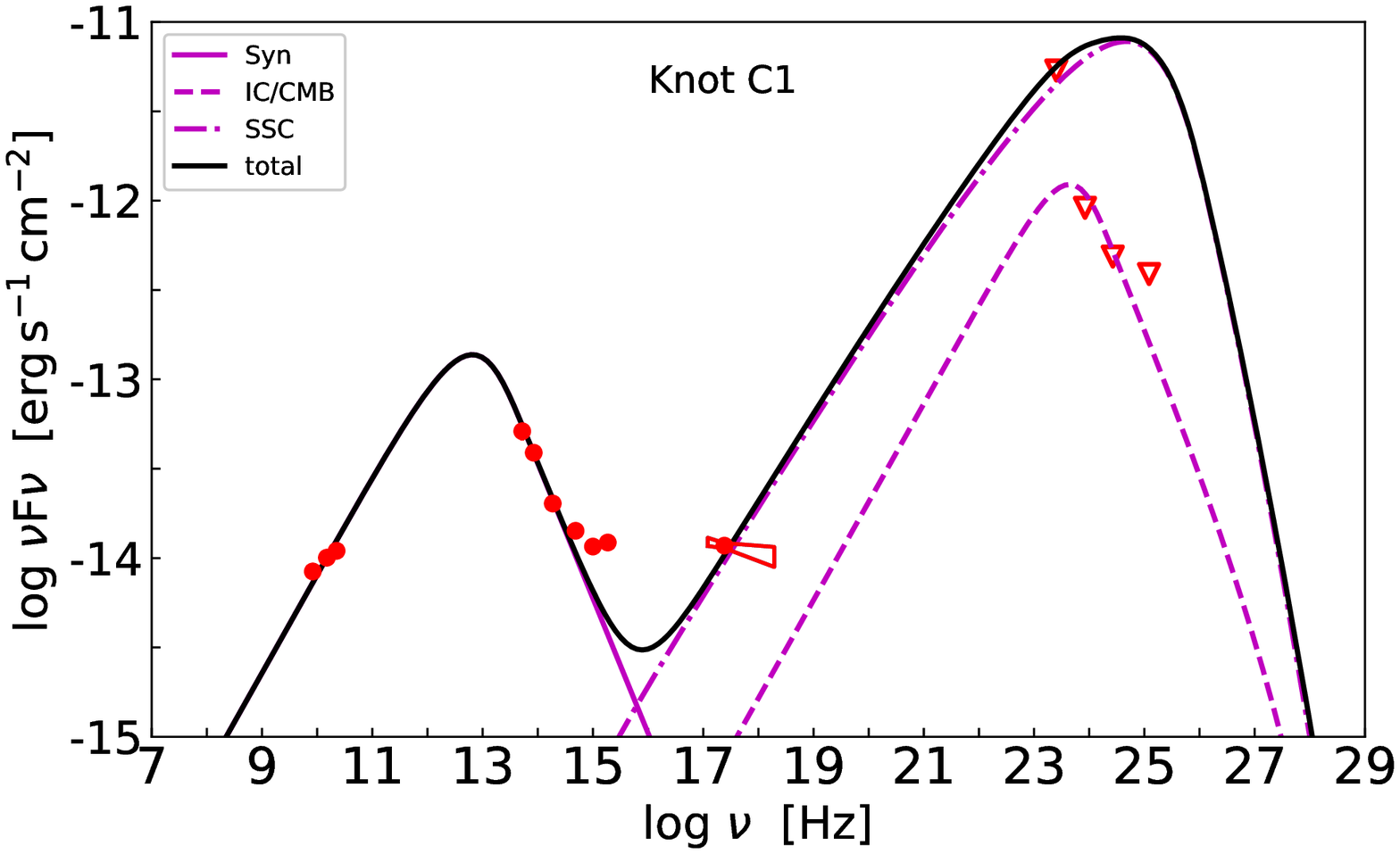}
\includegraphics[angle=0,scale=0.26]{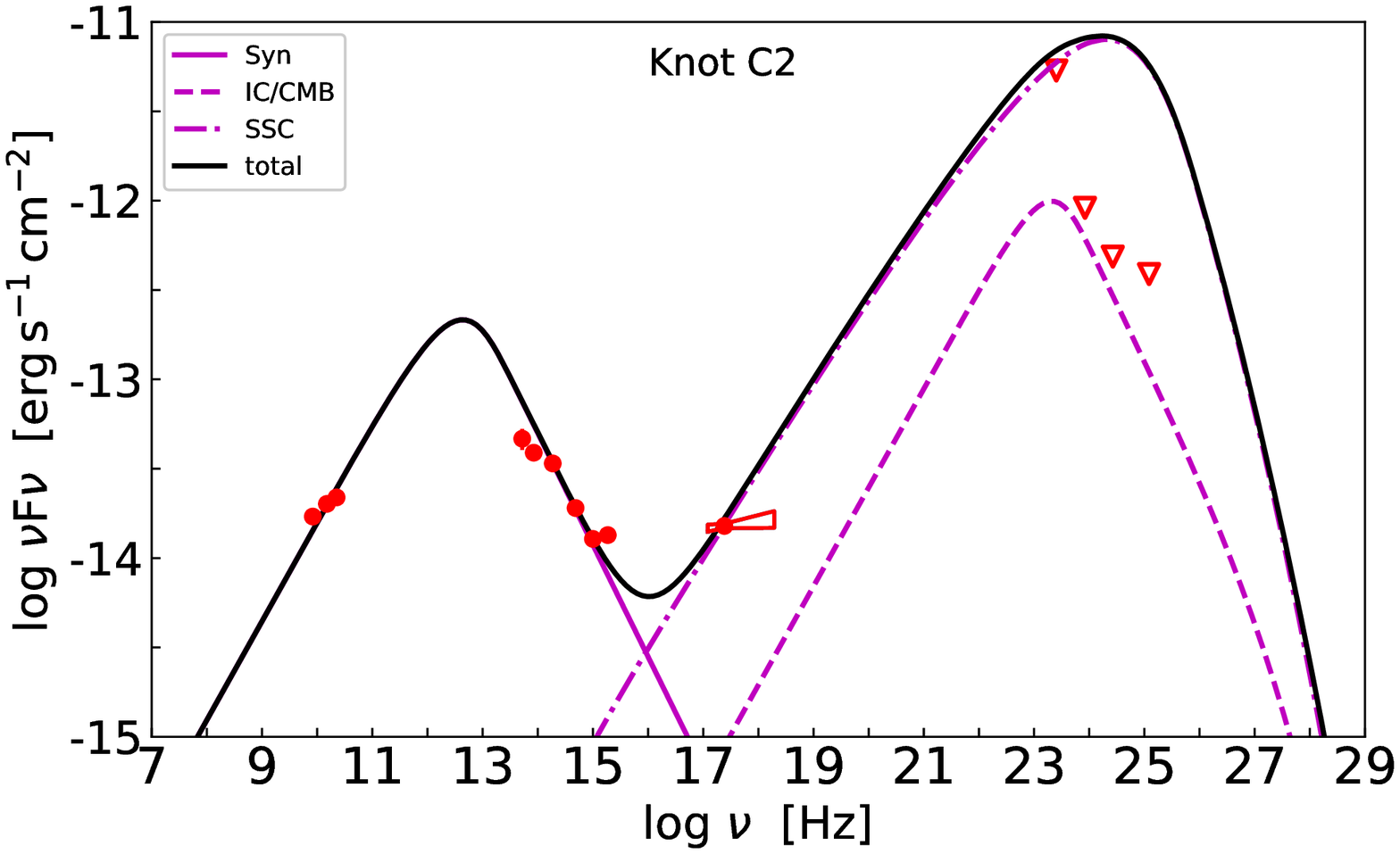}
\includegraphics[angle=0,scale=0.26]{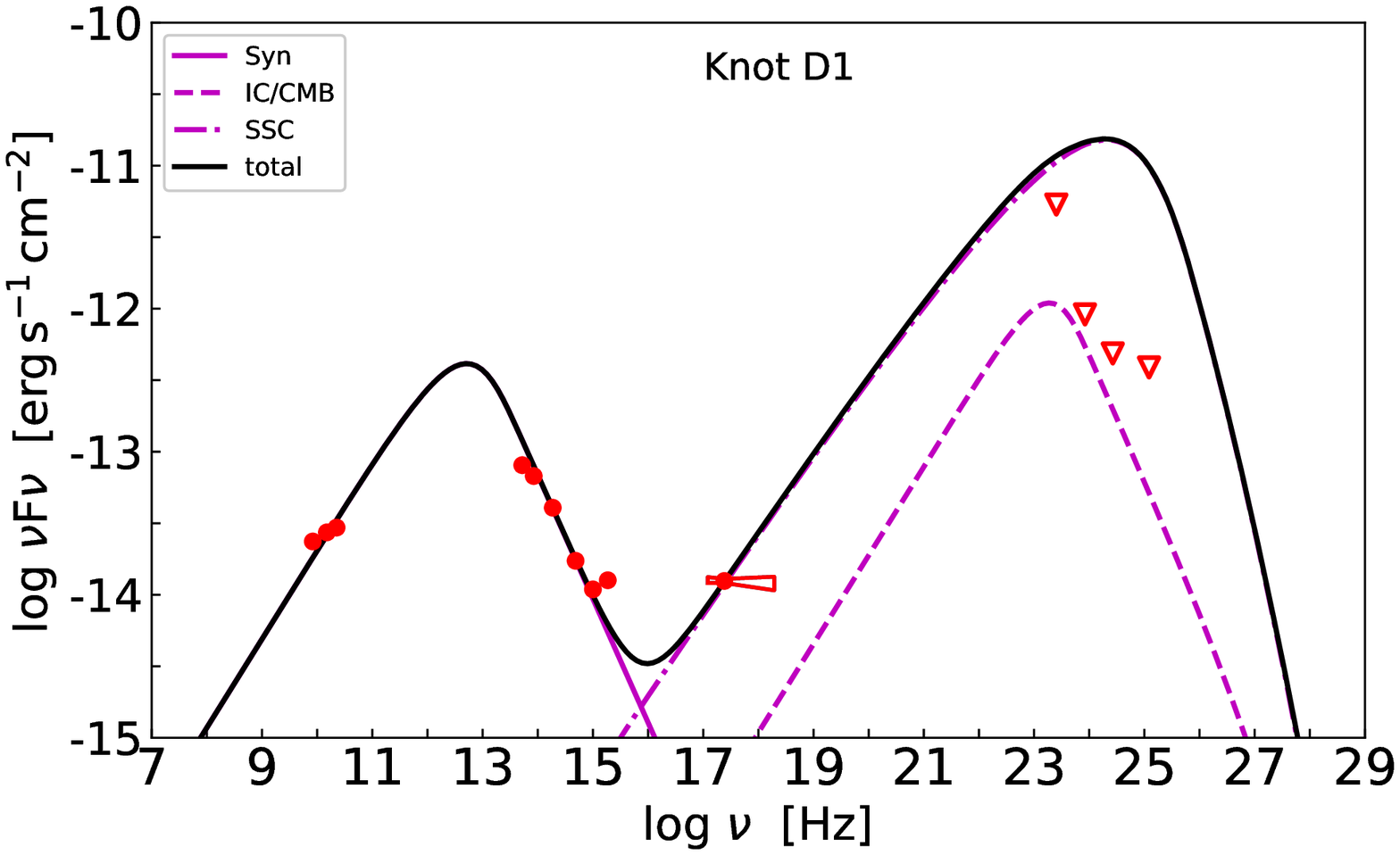}
\includegraphics[angle=0,scale=0.26]{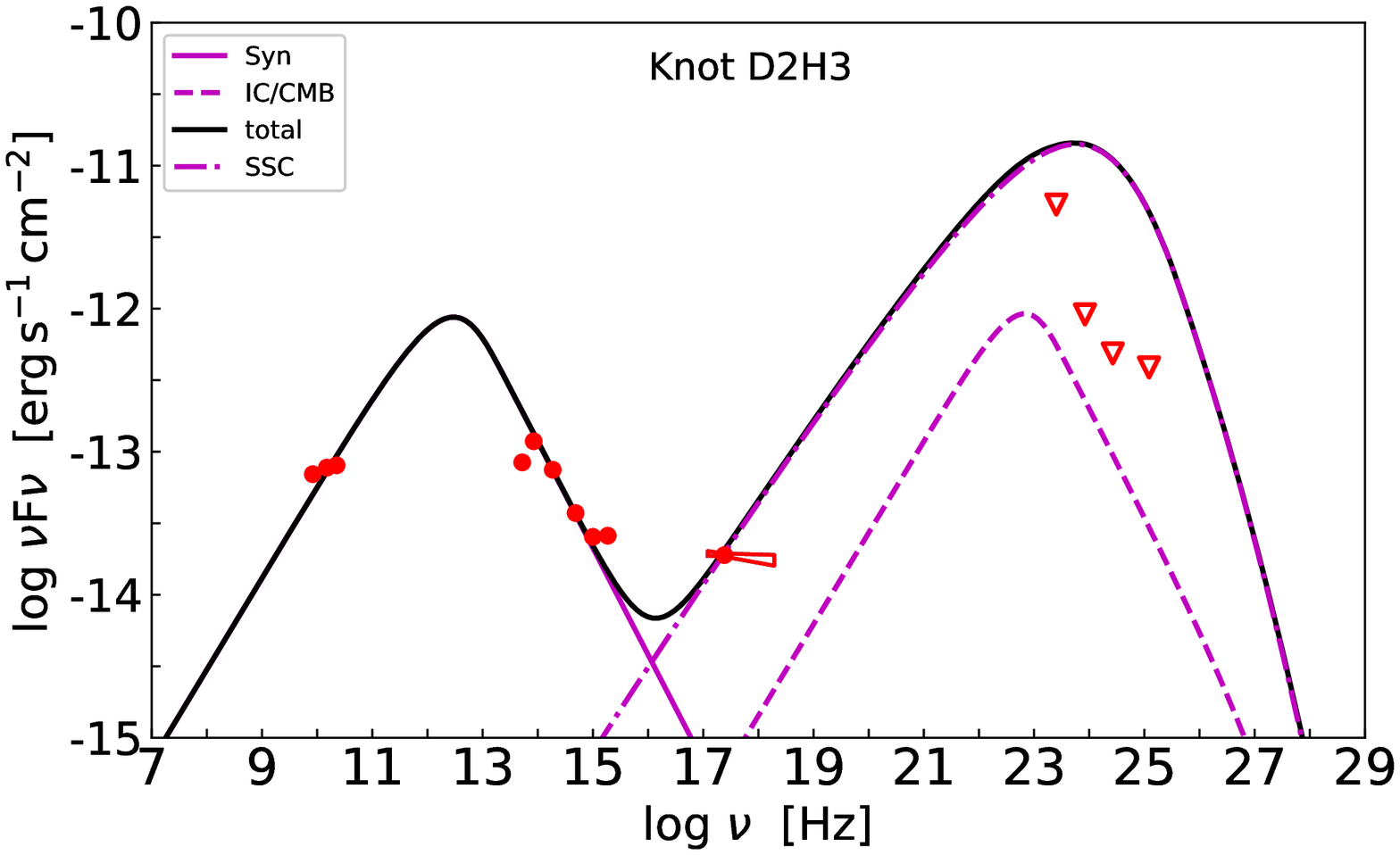}
\hfill
\includegraphics[angle=0,scale=0.26]{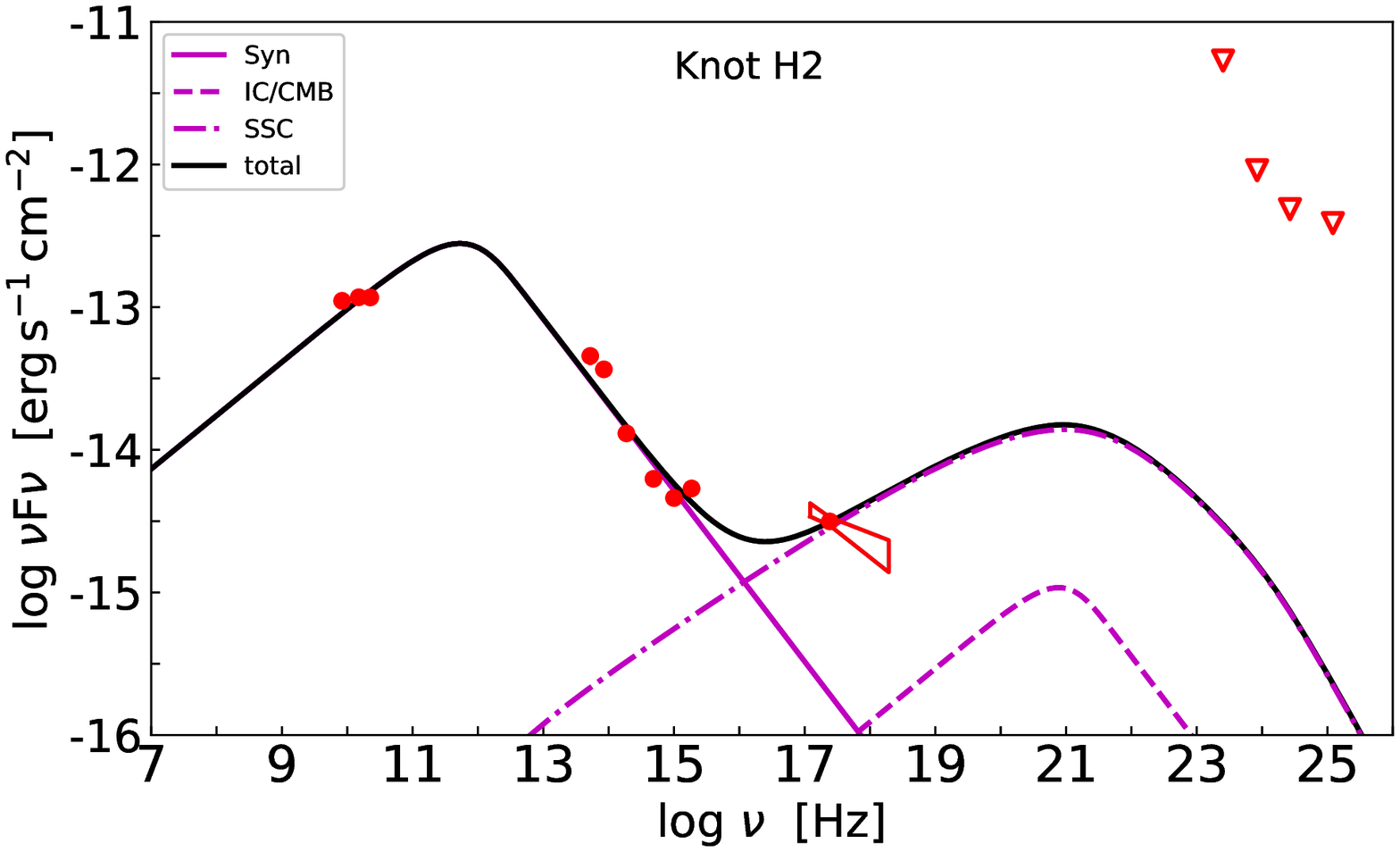}
\caption{Observed broadband SEDs (circles, bowties, and triangles) of the jet knots together with the model fits (black solid lines) of a single electron population (scenario I) in case of $\delta=1$. The purple solid lines, dash-dotted lines, and dashed lines display the synchrotron radiation, SSC, and IC/CMB components, respectively. The upper-limit data (red opened triangles) observed by the \emph{Fermi}/LAT are taken from Meyer et al. (2015).}\label{single1}
\end{figure*}

\begin{figure*}
\includegraphics[angle=0,scale=0.26]{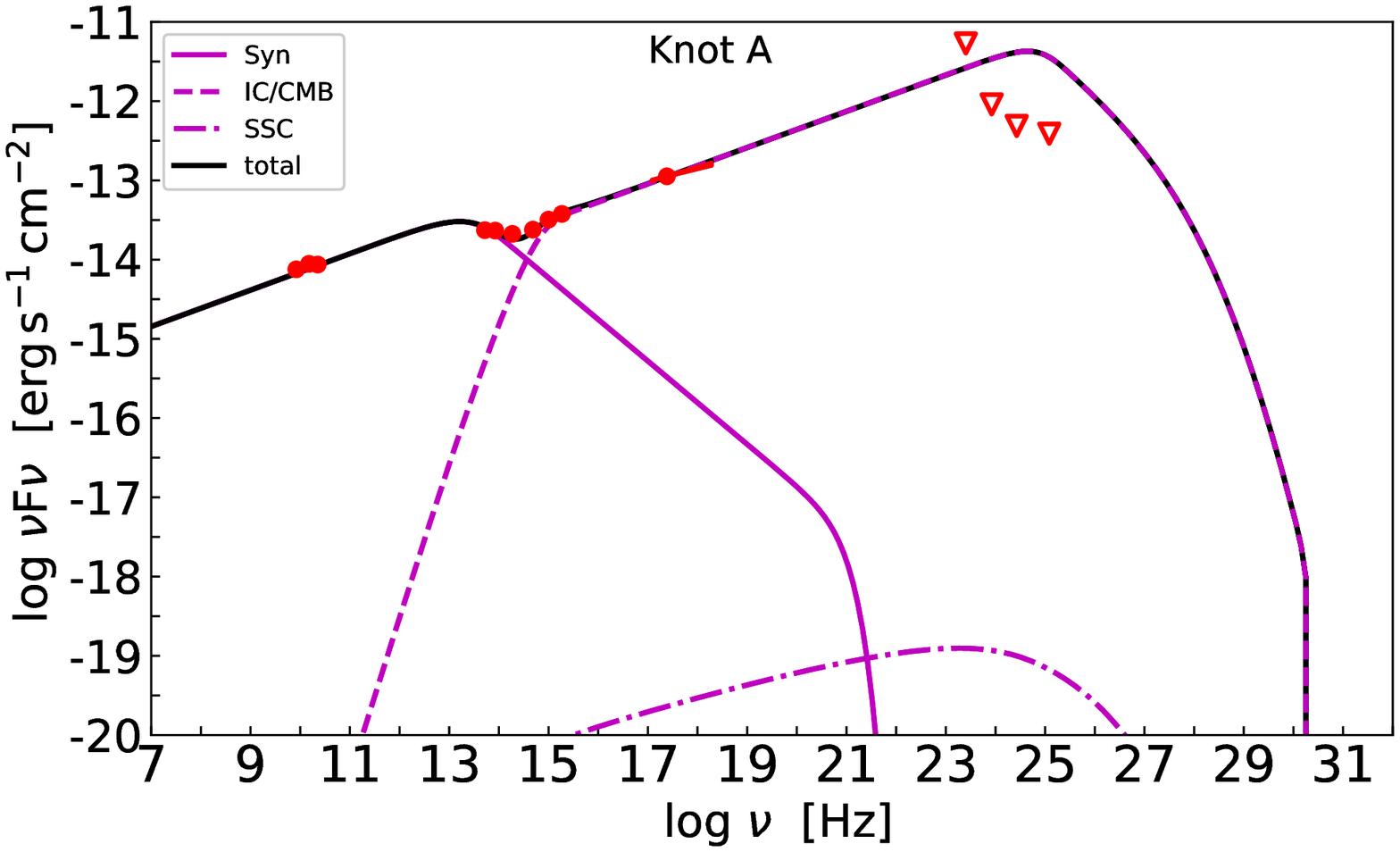}
\includegraphics[angle=0,scale=0.26]{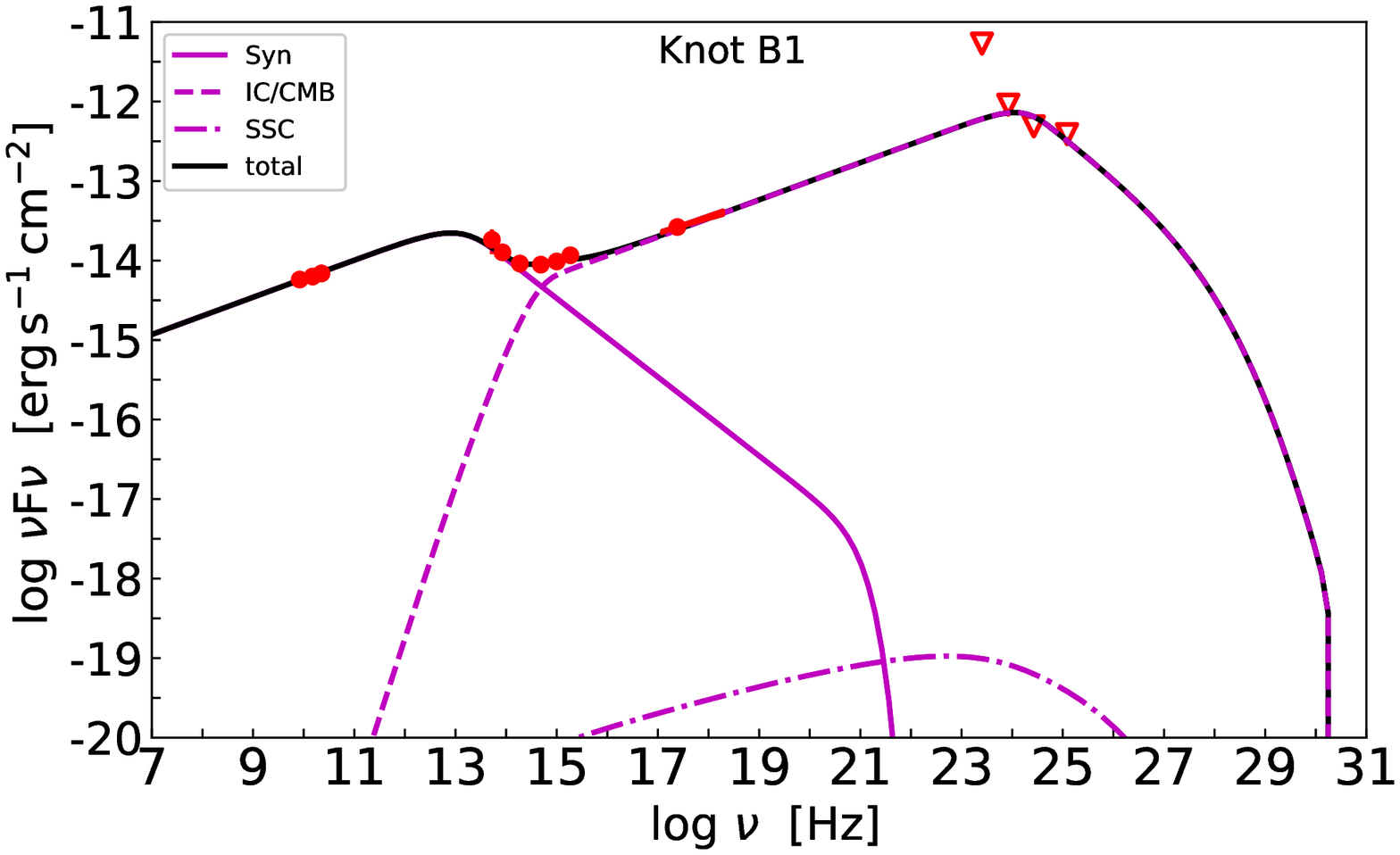}
\includegraphics[angle=0,scale=0.26]{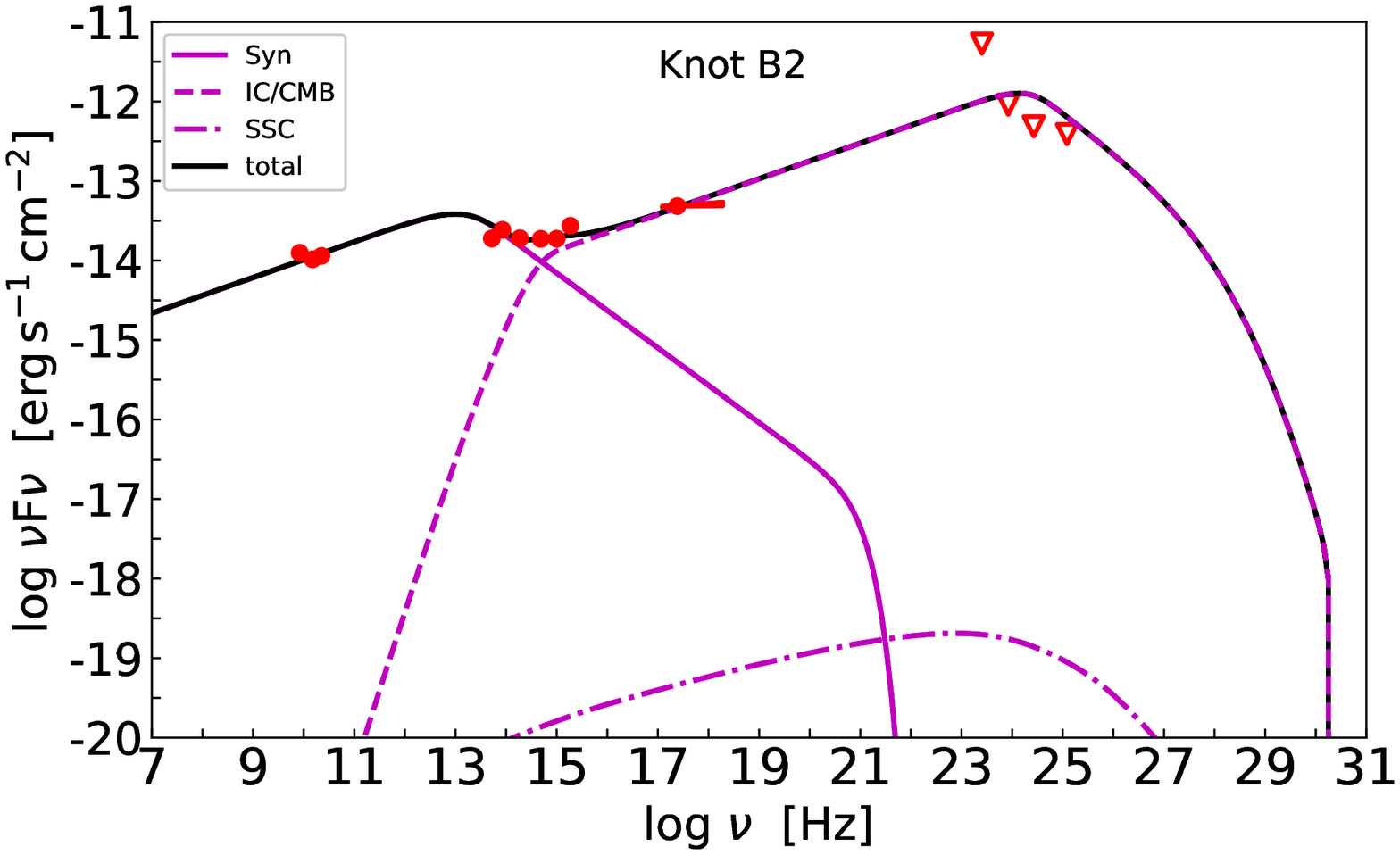}
\includegraphics[angle=0,scale=0.26]{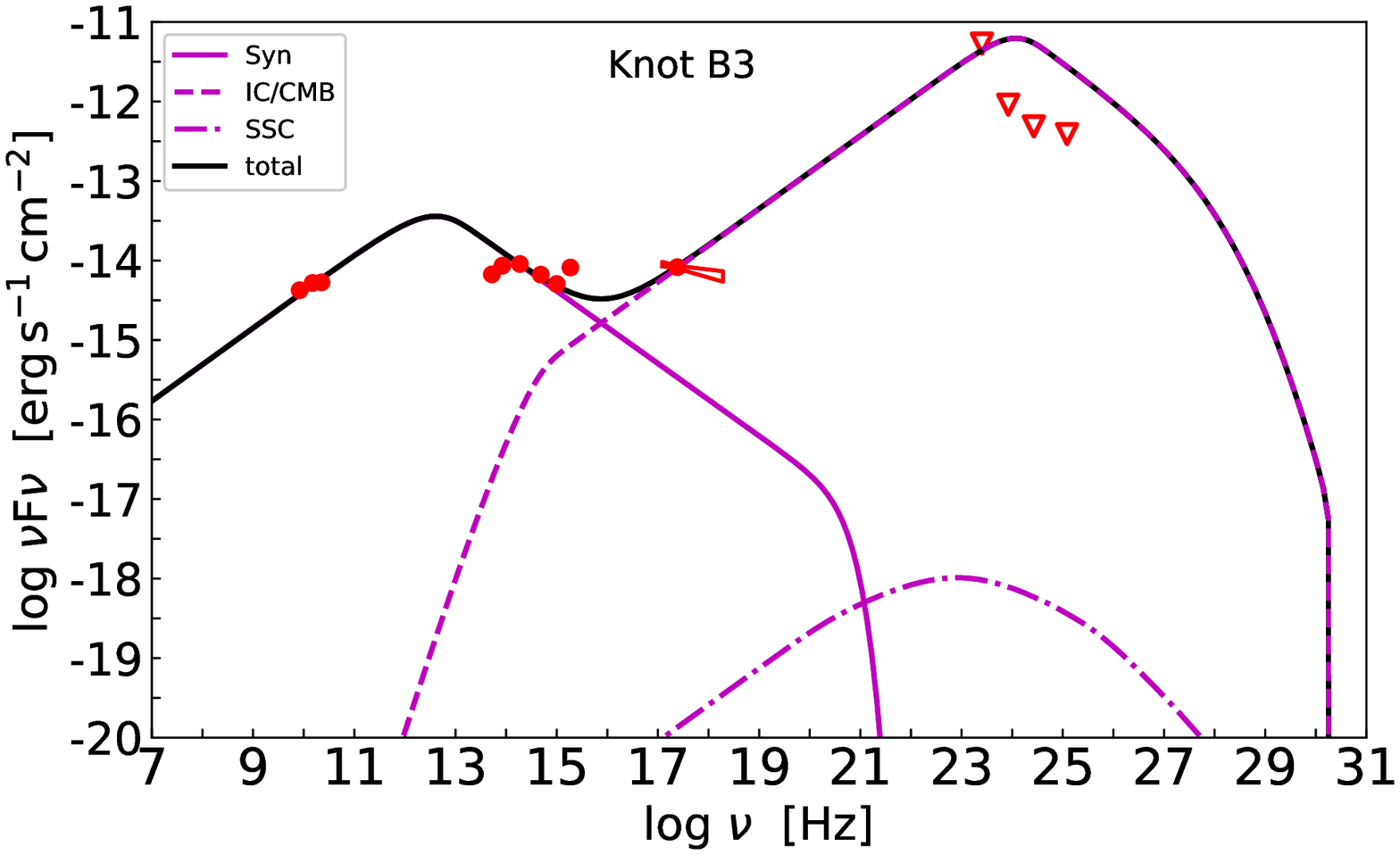}
\includegraphics[angle=0,scale=0.26]{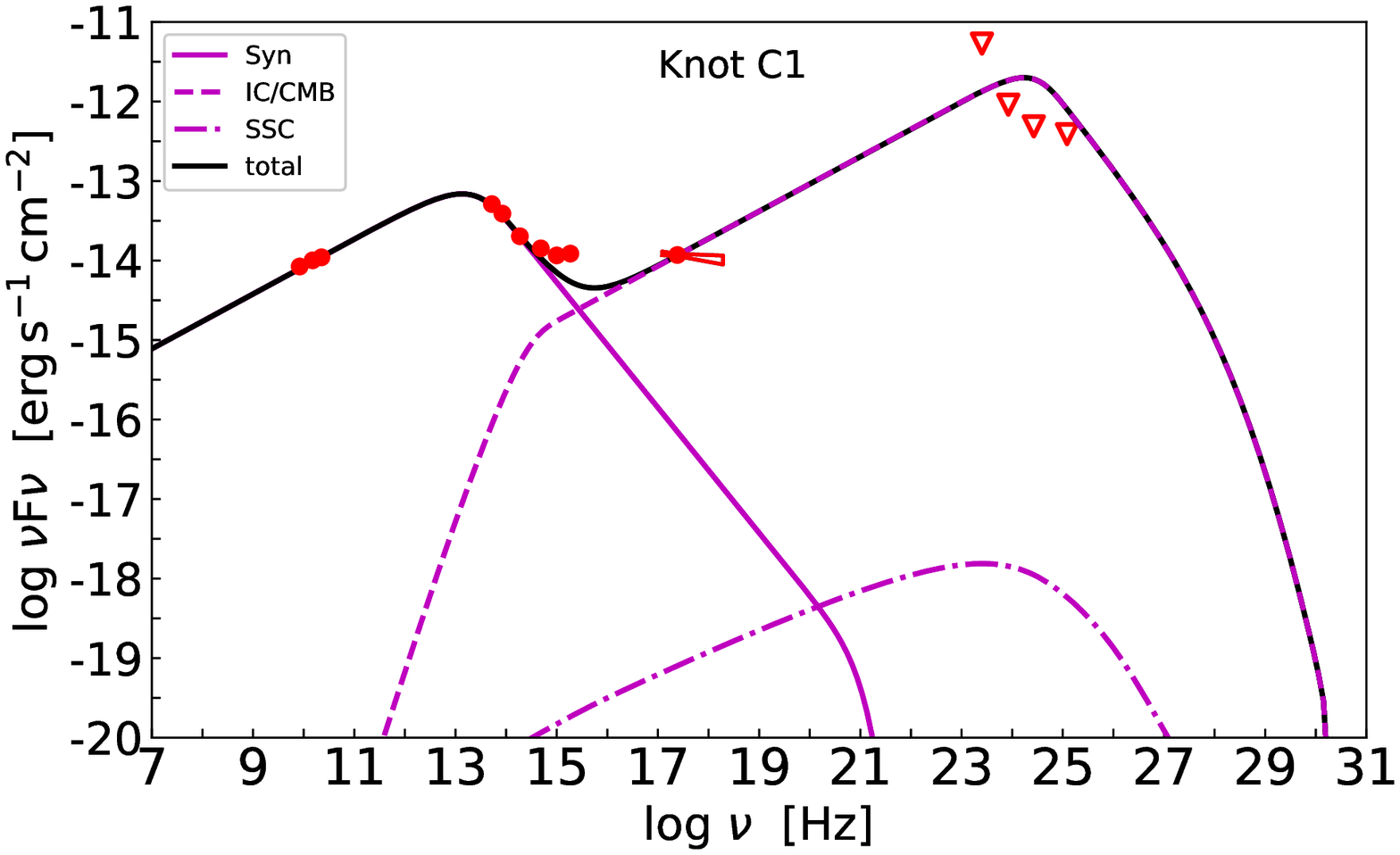}
\includegraphics[angle=0,scale=0.26]{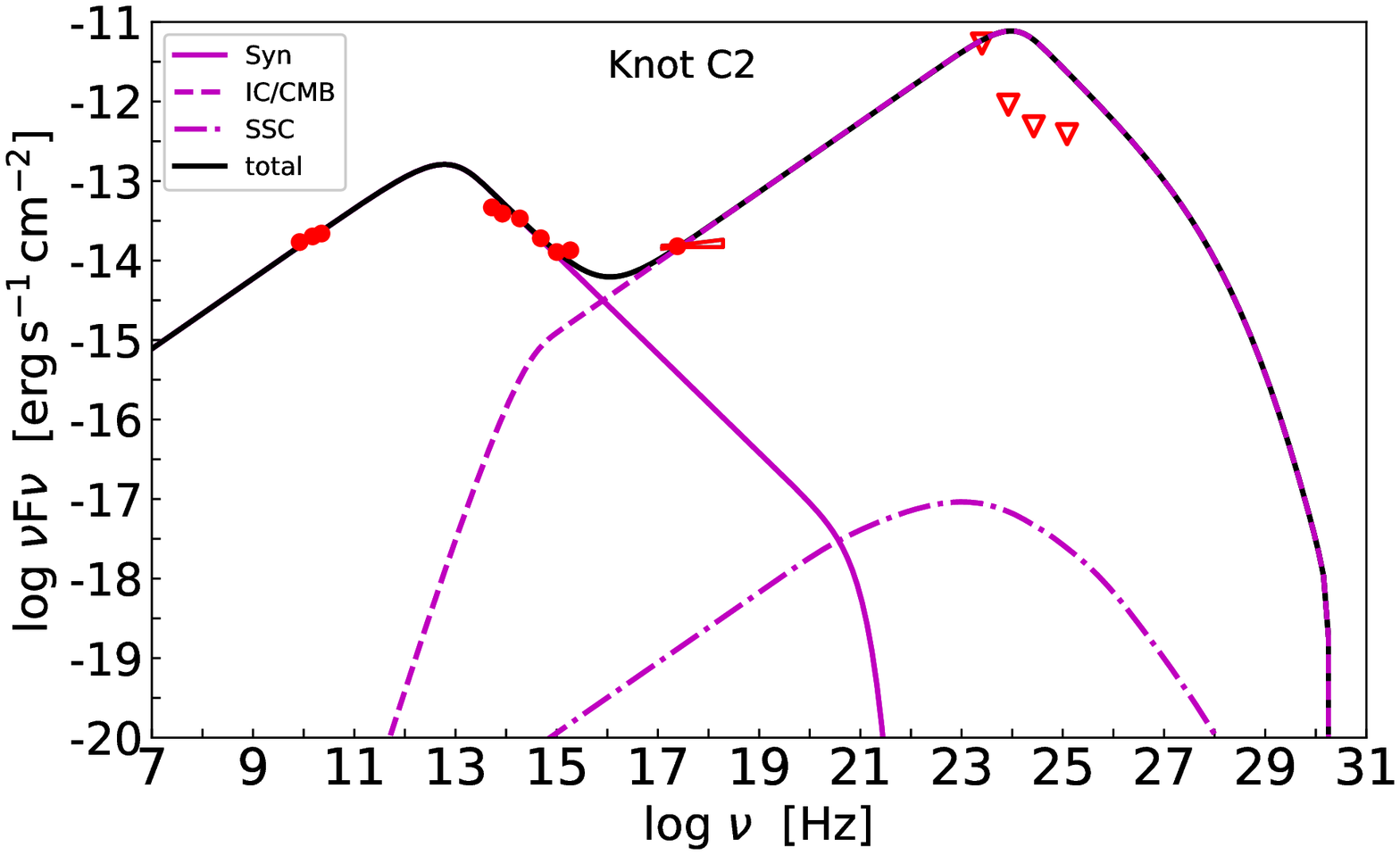}
\includegraphics[angle=0,scale=0.26]{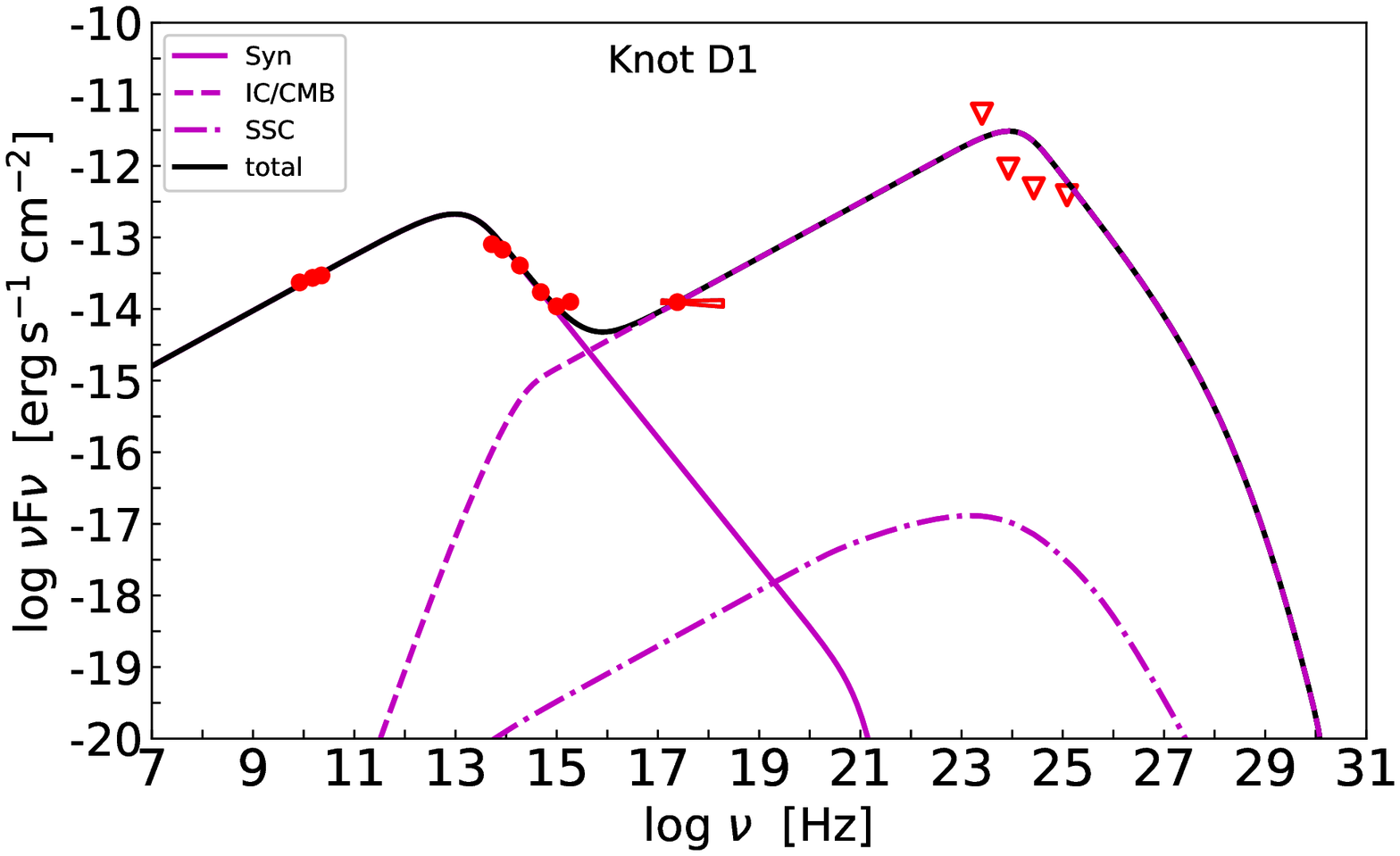}
\includegraphics[angle=0,scale=0.26]{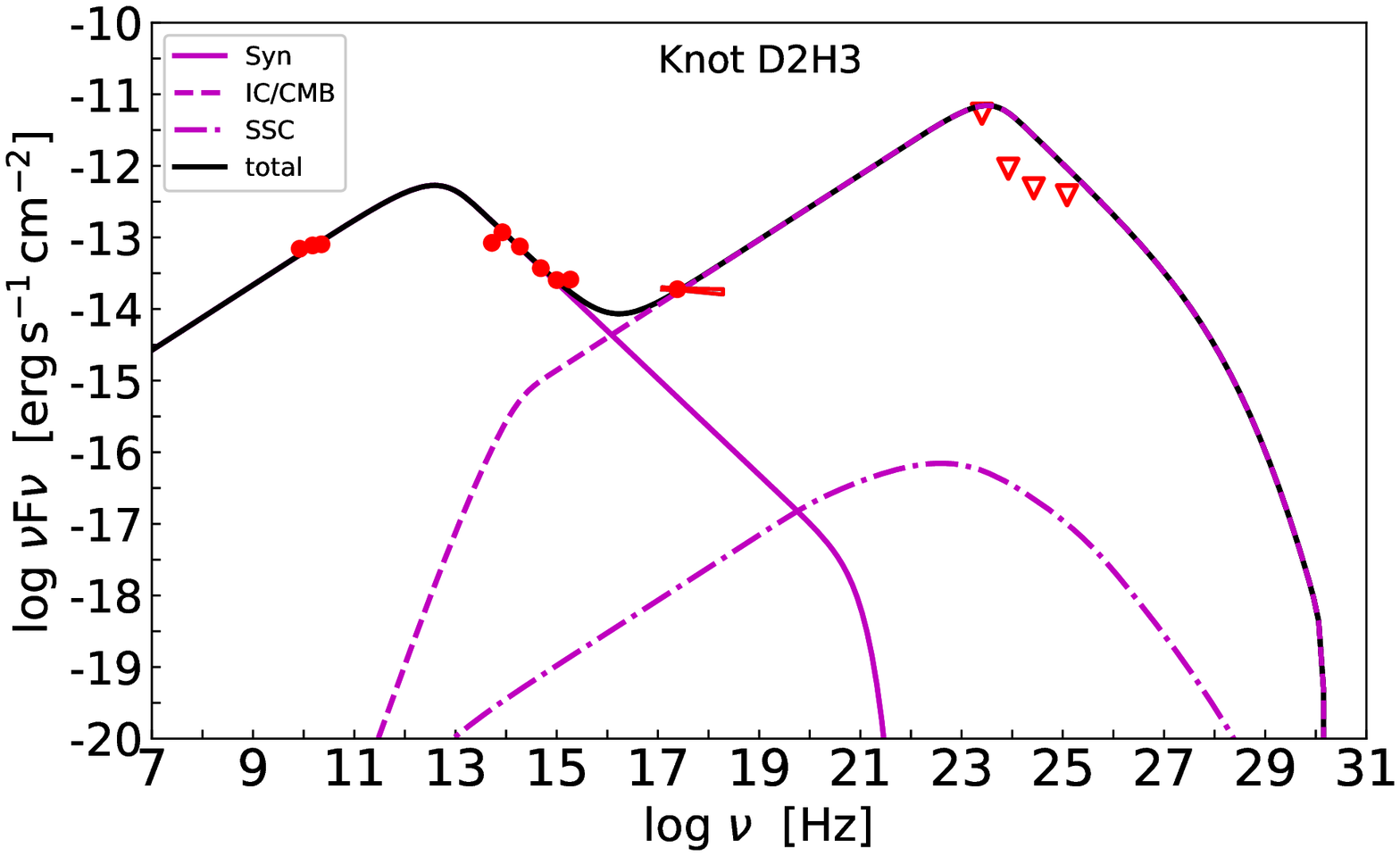}
\hfill
\includegraphics[angle=0,scale=0.26]{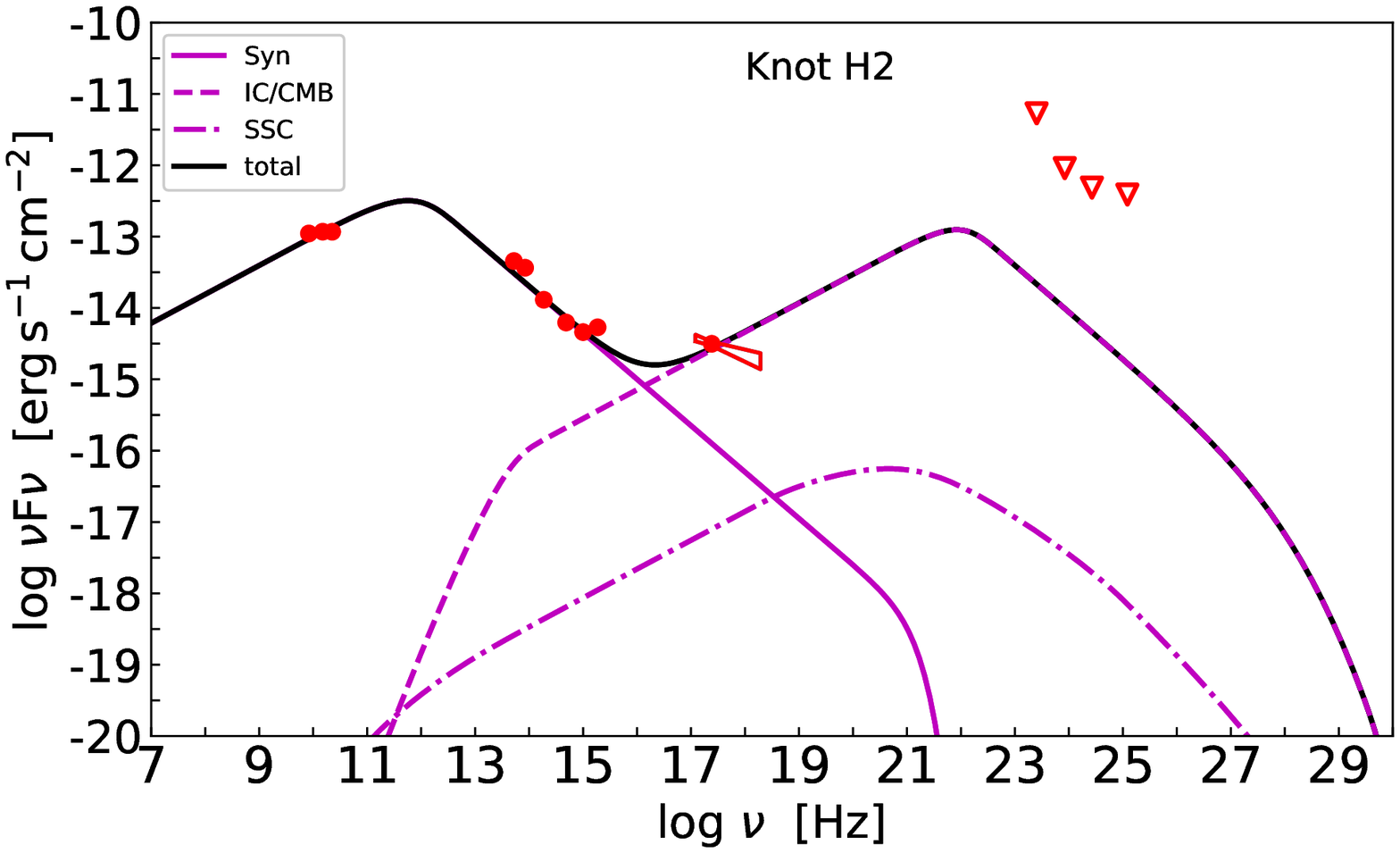}
\caption{The same as Figure \ref{single1}, but for the model fits of a single electron population (scenario I) in case of $\delta>1$. }\label{single2}
\end{figure*}

In this scenario, the synchrotron, SSC, and IC/CMB radiations of a single electron population are used to reproduce the broadband SEDs of knots. The radiating electrons are assumed to have the number distribution as Equation (1). The minimum and maximum energies of electrons are taken as $E_{\rm e,min}$=1 MeV and $E_{\rm e,max}$=510 TeV, which are respectively corresponding to $\gamma_{\rm e}\sim2$ and $\gamma_{\rm e}\sim10^9$, where $\gamma_{\rm e}$ is the Lorenz factor of electrons. In case of the knots do not have the relativistic motions, i.e., $\delta=\Gamma=1$, the IC component would be dominated by the SSC process since the energy density of the synchrotron radiation photon field ($U_{\rm syn}$) is higher than $U^{'}_{\rm CMB}$. A magnetic field strength ($B$) lower than the equipartition value ($B_{\rm eq}$) is also needed (e.g., Kataoka \& Stawarz 2005; Harris \& Krawczynski 2006; Zhang et al. 2010, 2018), and thus we do not take the equipartition condition into account in this scenario. The free parameters of the SED modeling are $B$, $A_1$, $p_1$, $p_2$, $E_{\rm b}$. The SED fitting results are shown in Figure \ref{single1} and the derived parameters are listed in Table 1. In case of these knots have relativistic motions towards us, i.e., $\delta>1$, the IC/CMB process may dominate the high energy emission of the SEDs. The equipartition condition is usually taken into account to constrain the model parameters, i.e., the energy density of relativistic electrons ($U_{\rm e}$) equal to the energy density of magnetic fields ($U_{B}$). Thus the free parameters in the SED modeling are $\delta$ (or $\Gamma$), $A_1$, $p_1$, $p_2$, $E_{\rm b}$. The SED fitting results are displayed in Figure \ref{single2} and the derived parameters are listed in Table 1.

As illustrated in Figures \ref{single1} and \ref{single2}, although the X-rays together with the UV emission of knot-A and knot-B1 can be explained with a single electron population via the IC processes either in case of $\delta=1$ or $\delta>1$, the SEDs of the other knots cannot be represented in this scenario.

\subsection{Scenario II: Two Electron Populations}

\begin{figure*}
\includegraphics[angle=0,scale=0.26]{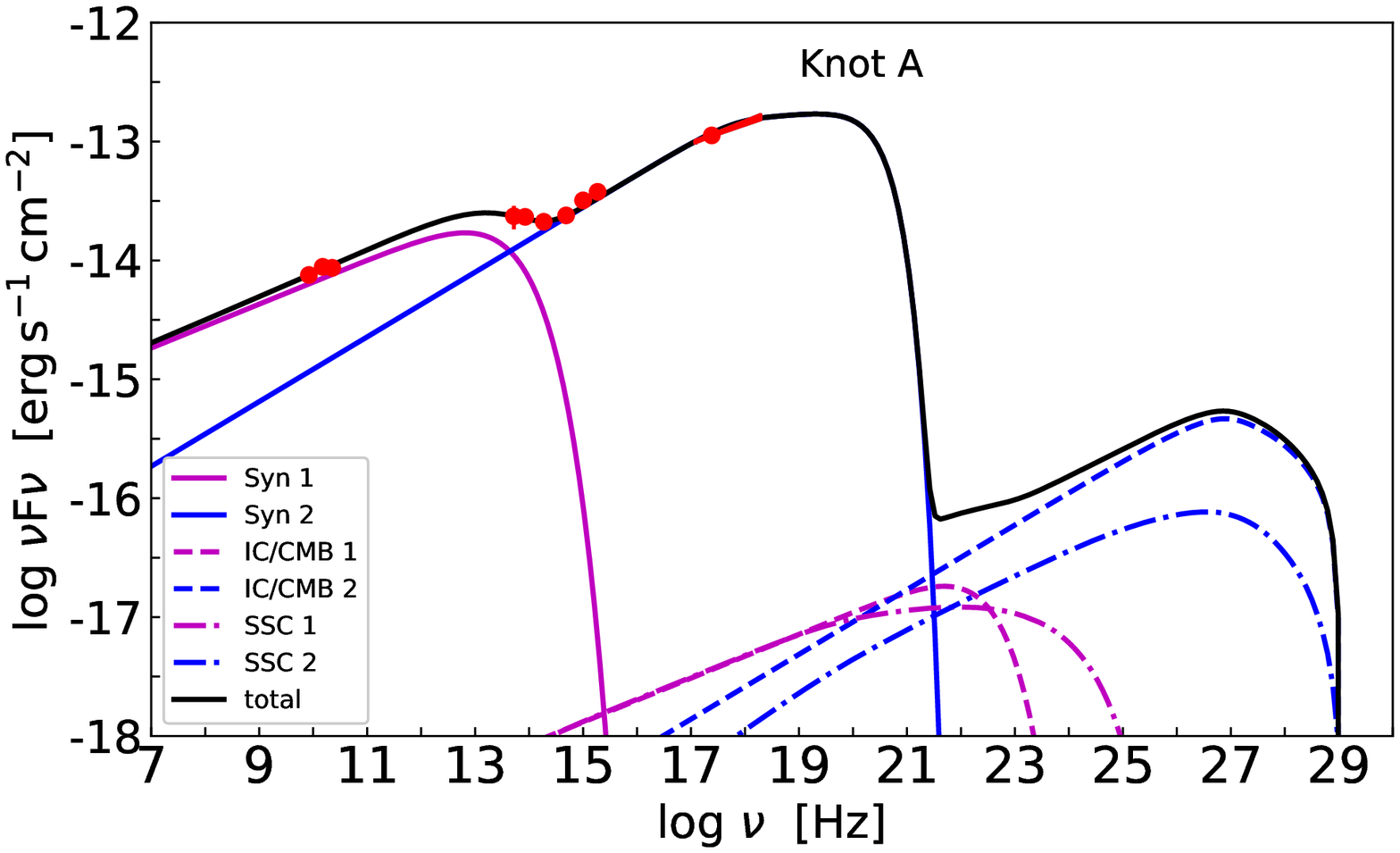}
\includegraphics[angle=0,scale=0.26]{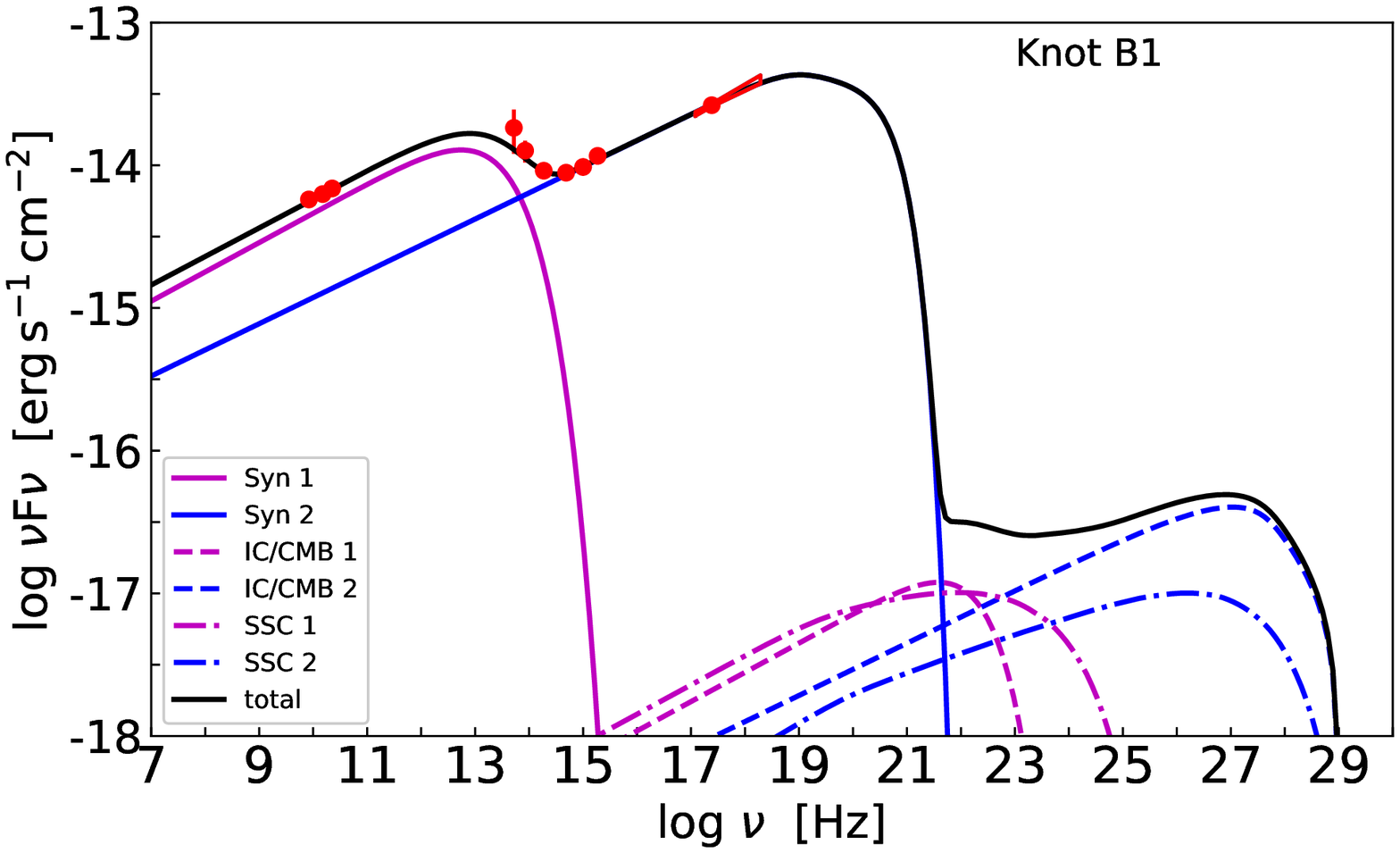}
\includegraphics[angle=0,scale=0.26]{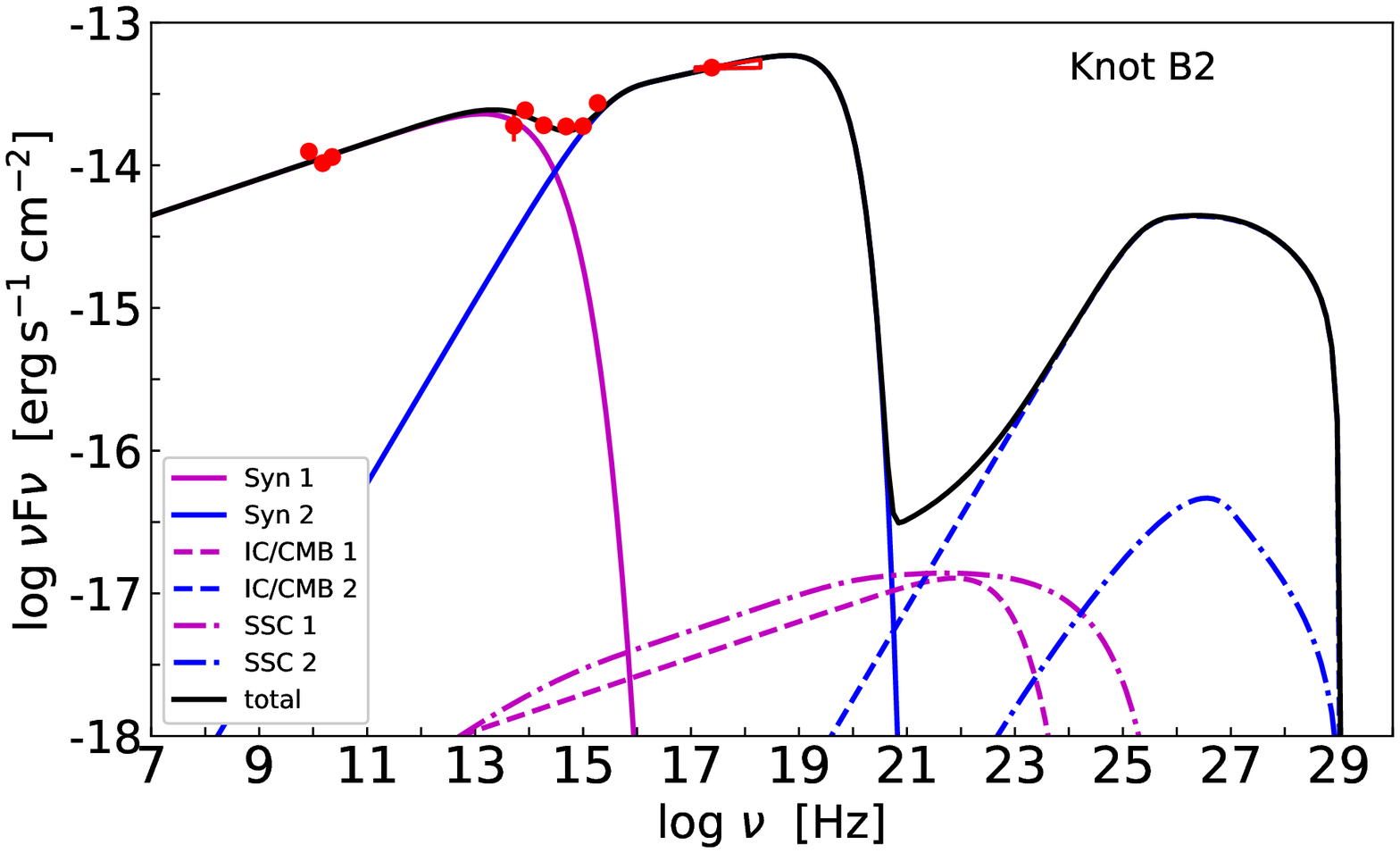}
\includegraphics[angle=0,scale=0.26]{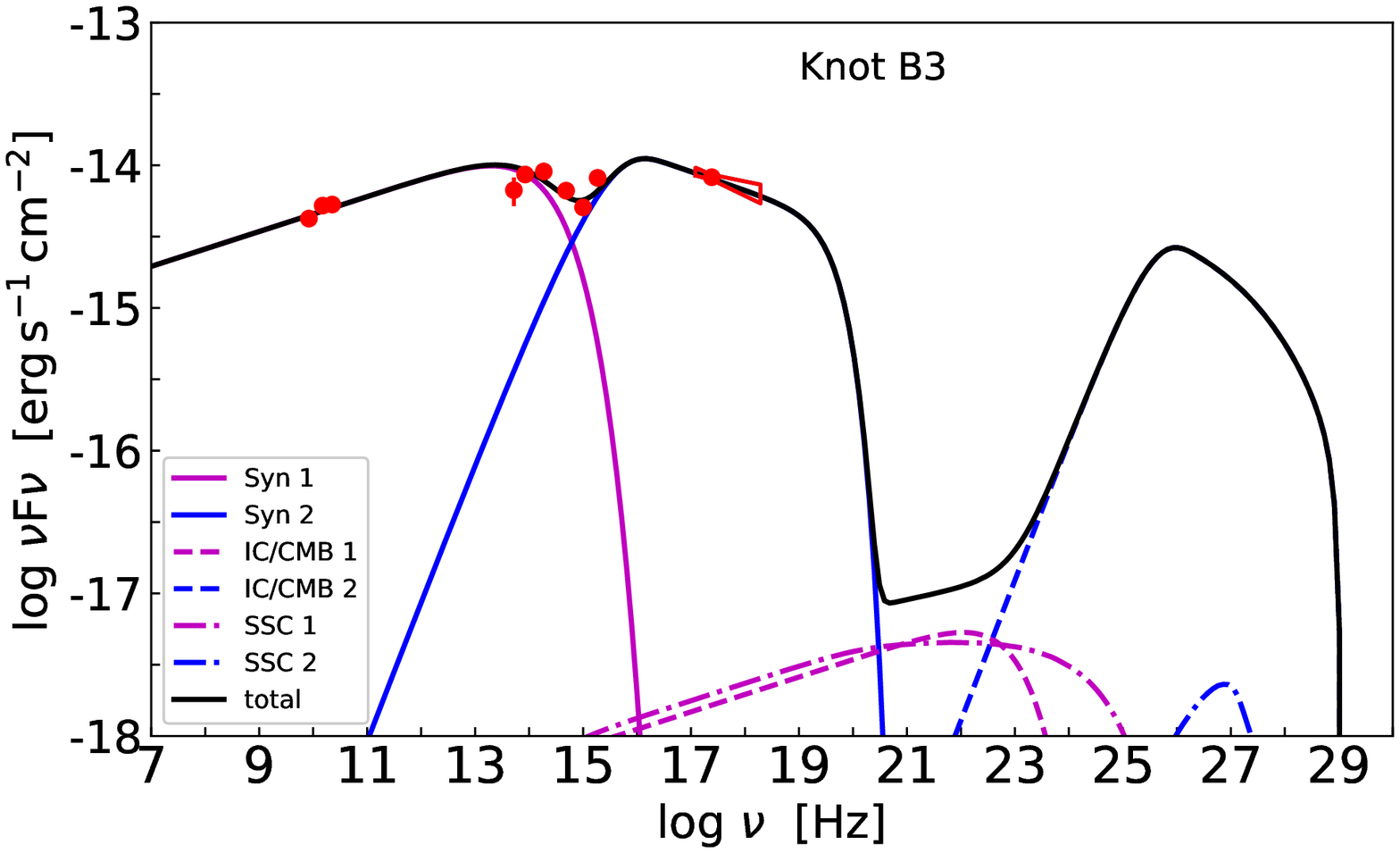}
\includegraphics[angle=0,scale=0.26]{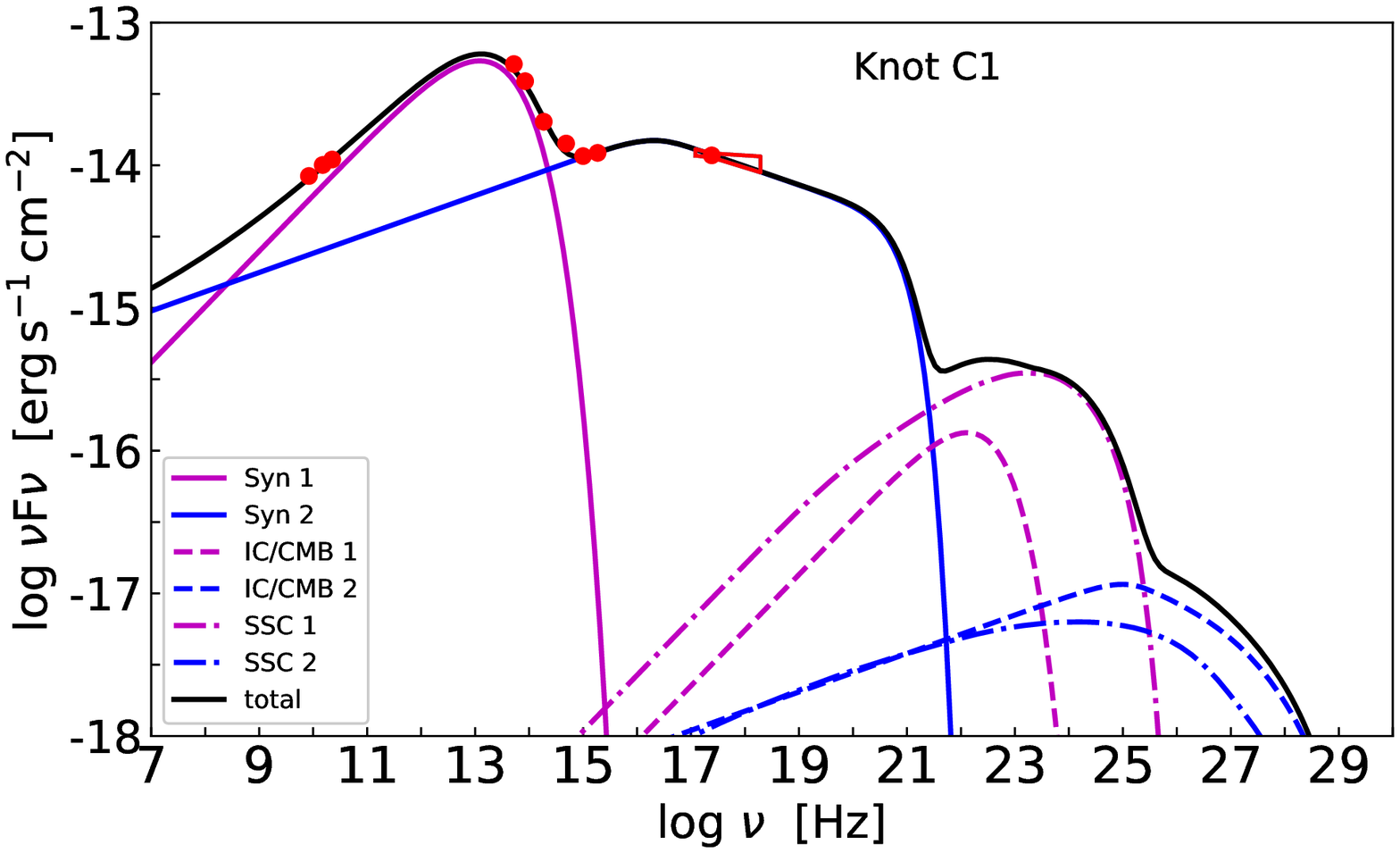}
\includegraphics[angle=0,scale=0.26]{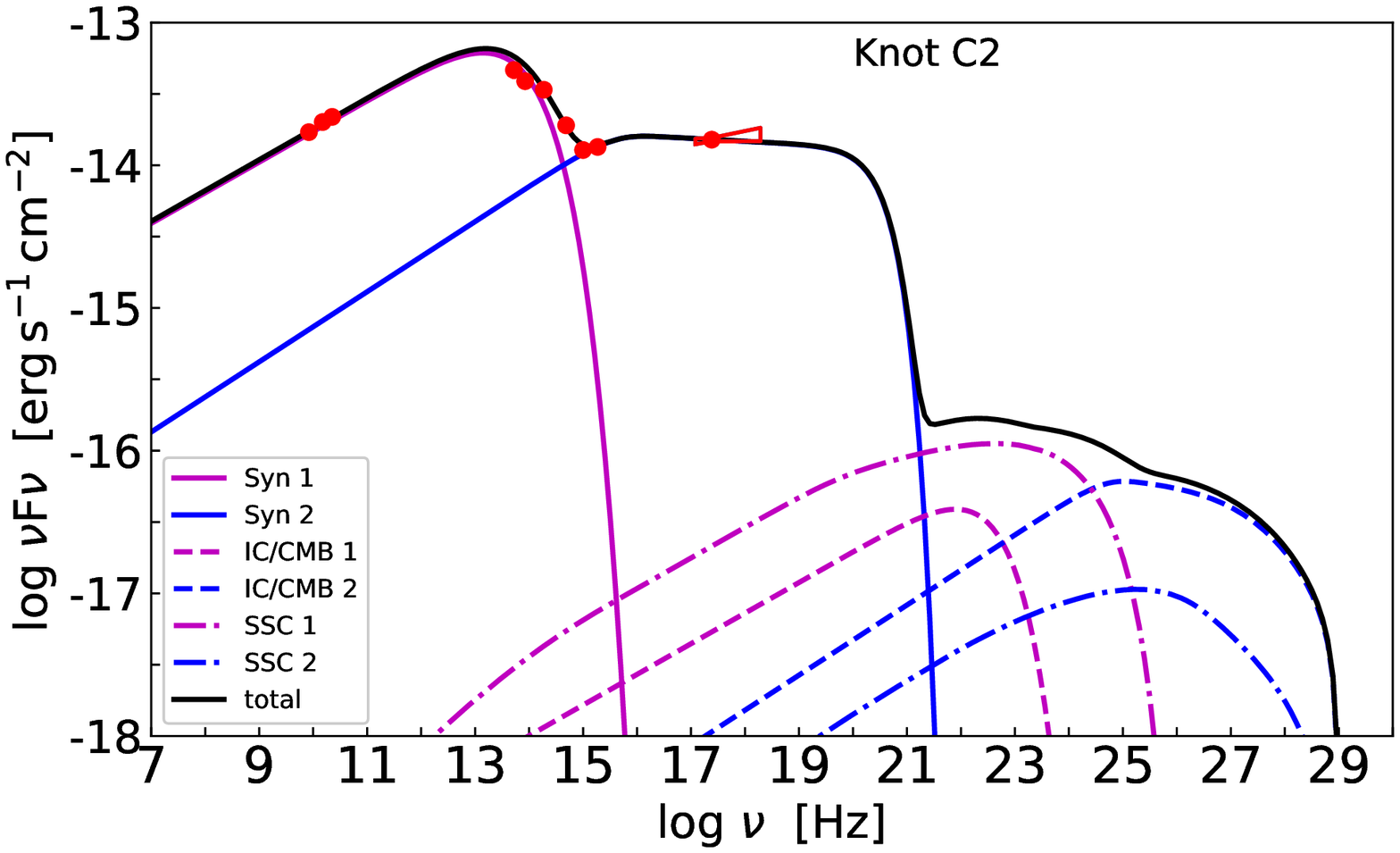}
\includegraphics[angle=0,scale=0.26]{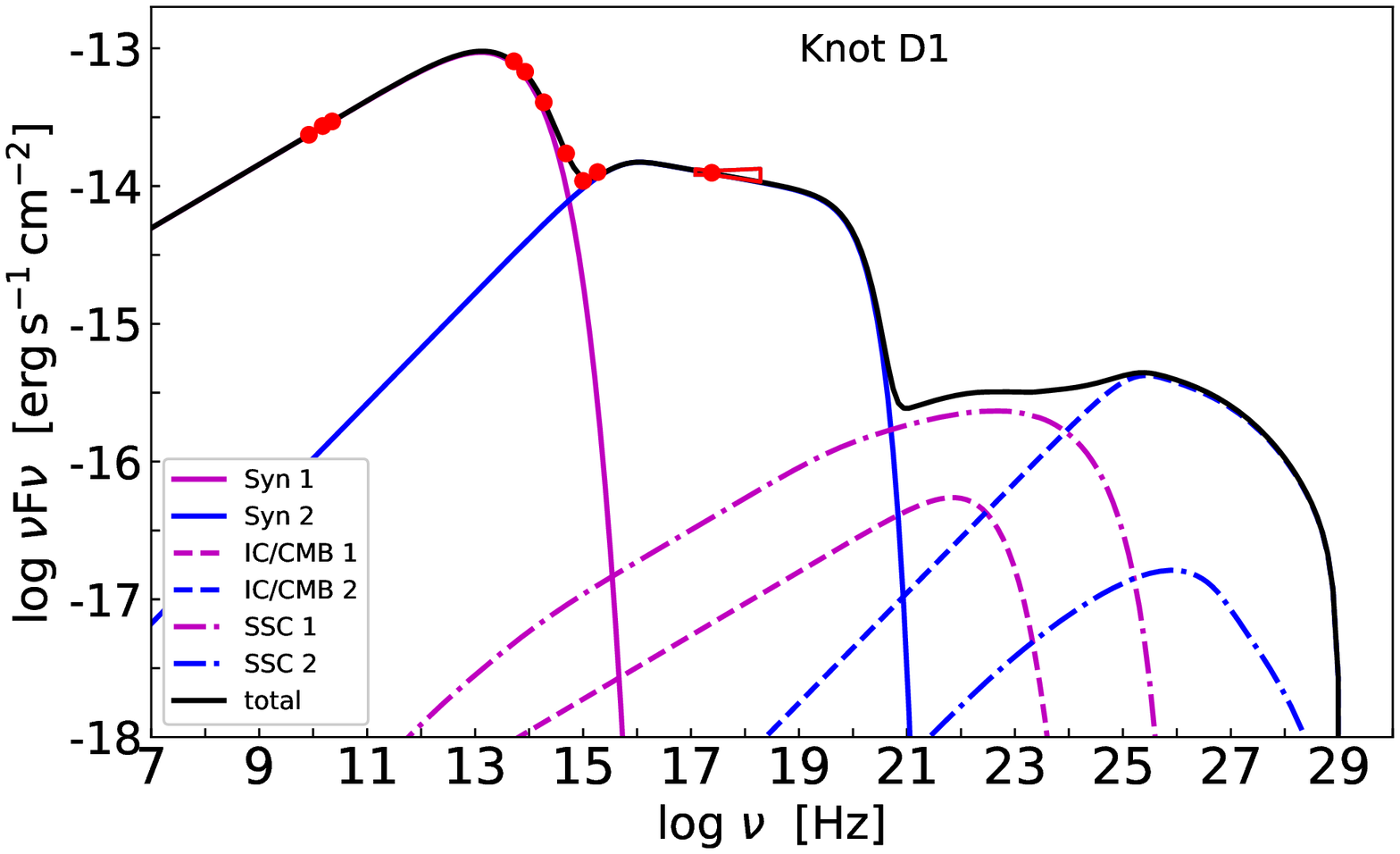}
\includegraphics[angle=0,scale=0.26]{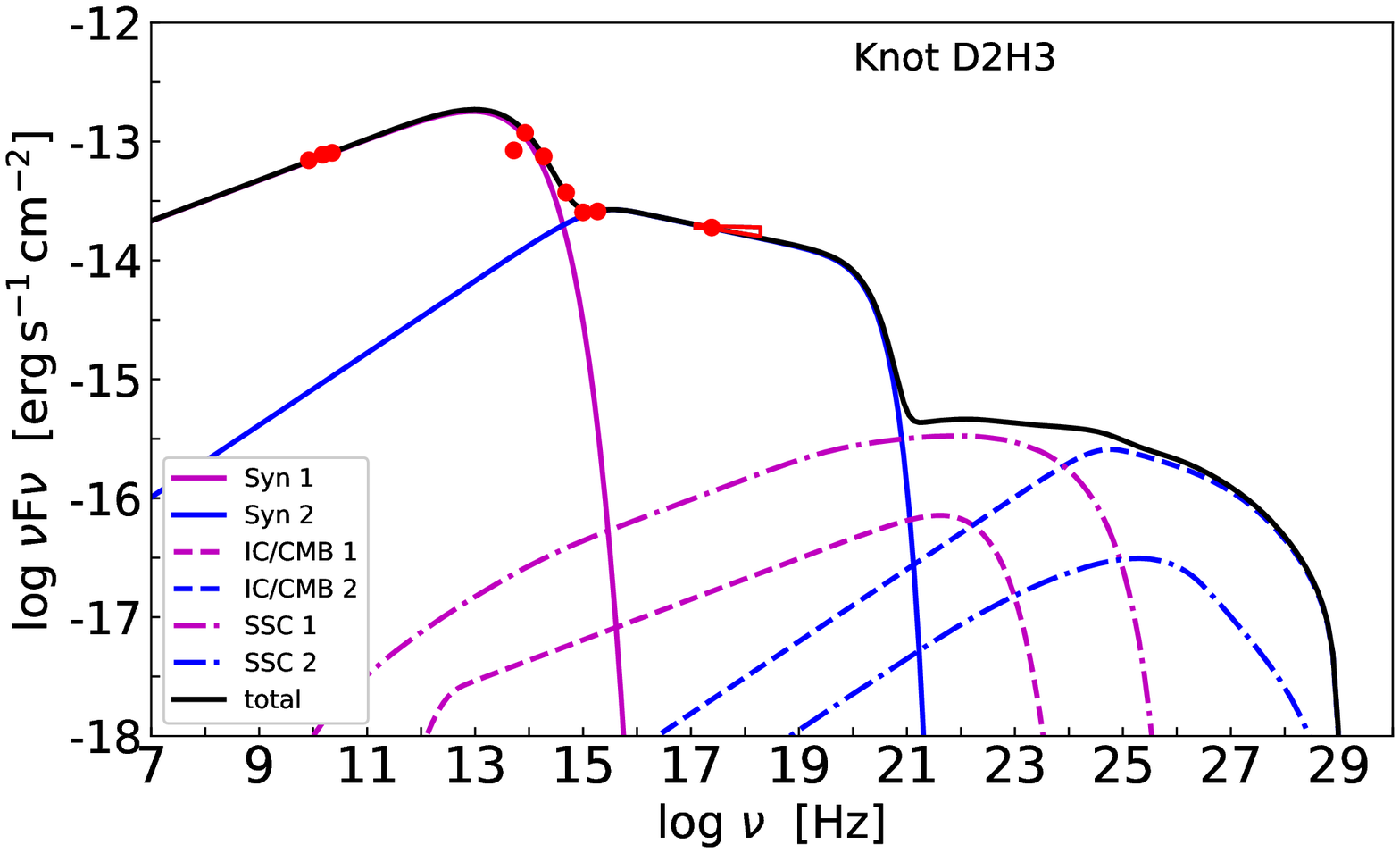}
\hfill
\includegraphics[angle=0,scale=0.26]{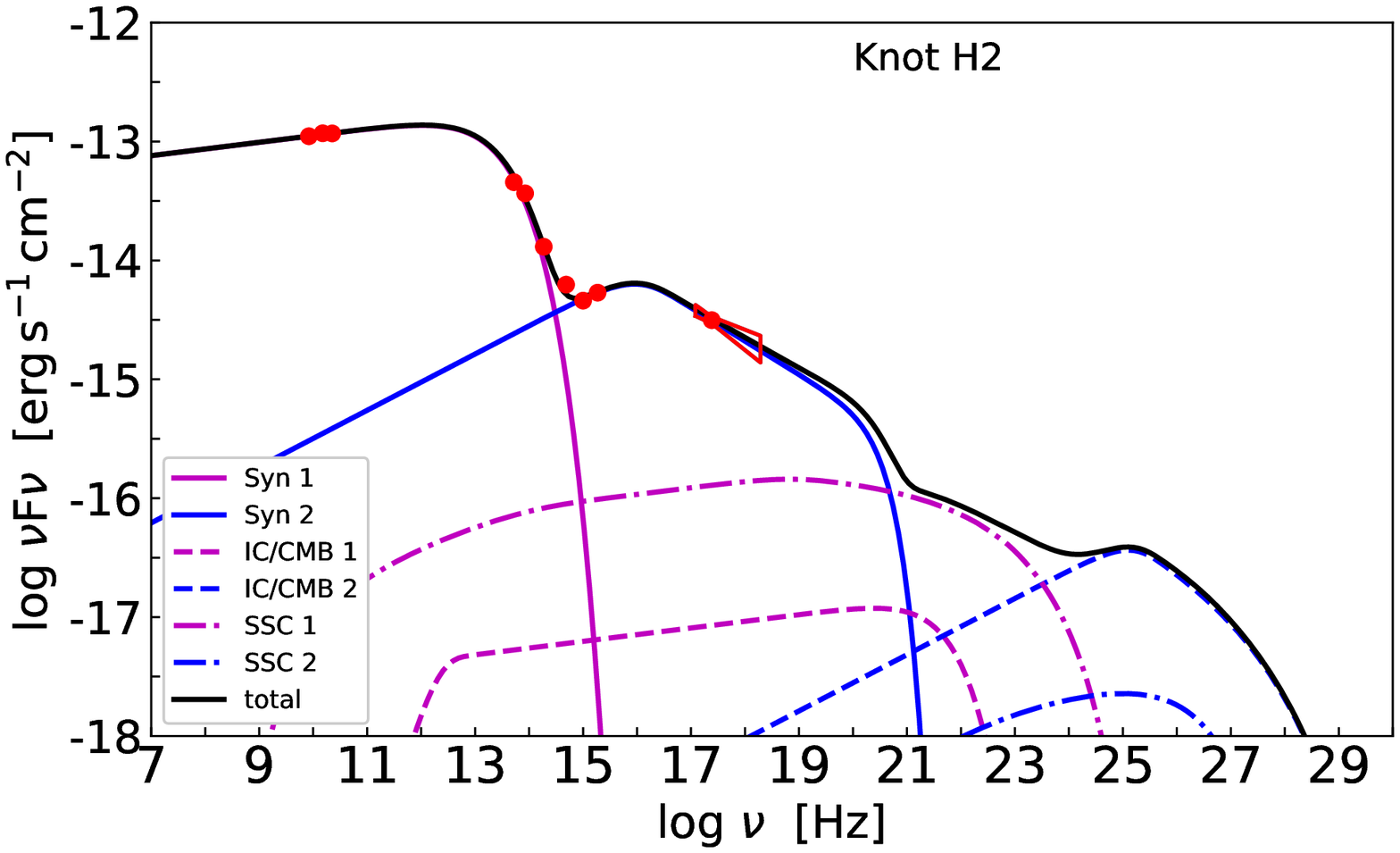}
\caption{The same as Figure \ref{single1}, but for the model fits of two electron populations (scenario II) in case of $\delta=1$. The purple and blue lines respectively display the radiation components from the two electron populations.}\label{Two_E1}
\end{figure*}

In this scenario, we add another electron population and try to explain the broadband SEDs with the synchrotron radiations of two electron populations, i.e., an exponential cutoff power-law electron population (Equation (2)) plus a broken power-law electron population. The minimum and maximum energies of electrons in the two electron populations are taken as $E_{\rm e,min}$=1 MeV and $E_{\rm e,max}$=510 TeV. We assume that the two electron populations are totaly independent (see also Zargaryan et al. 2017; Sun et al. 2018) and riskily assume that the two radiation regions of an individual knot have the same size with $R_1=R_2=R$ to calculate the equipartition magnetic field strength, i.e., $U_{B}=U_{\rm e}$. The two synchrotron components are calculated under the equipartition condition. In case of $\delta=1$, the free parameters of the SED modeling are $A_{0}$, $p_0$, $E_{\rm c}$, $A_1$, $p_1$, $p_2$, $E_{\rm b}$. The SED fitting results are shown in Figure \ref{Two_E1} and the corresponding SSC and IC/CMB contributions of two electron populations are also presented in Figure \ref{Two_E1}. The fitting parameters are reported in Table 2.

\begin{figure*}
\includegraphics[angle=0,scale=0.26]{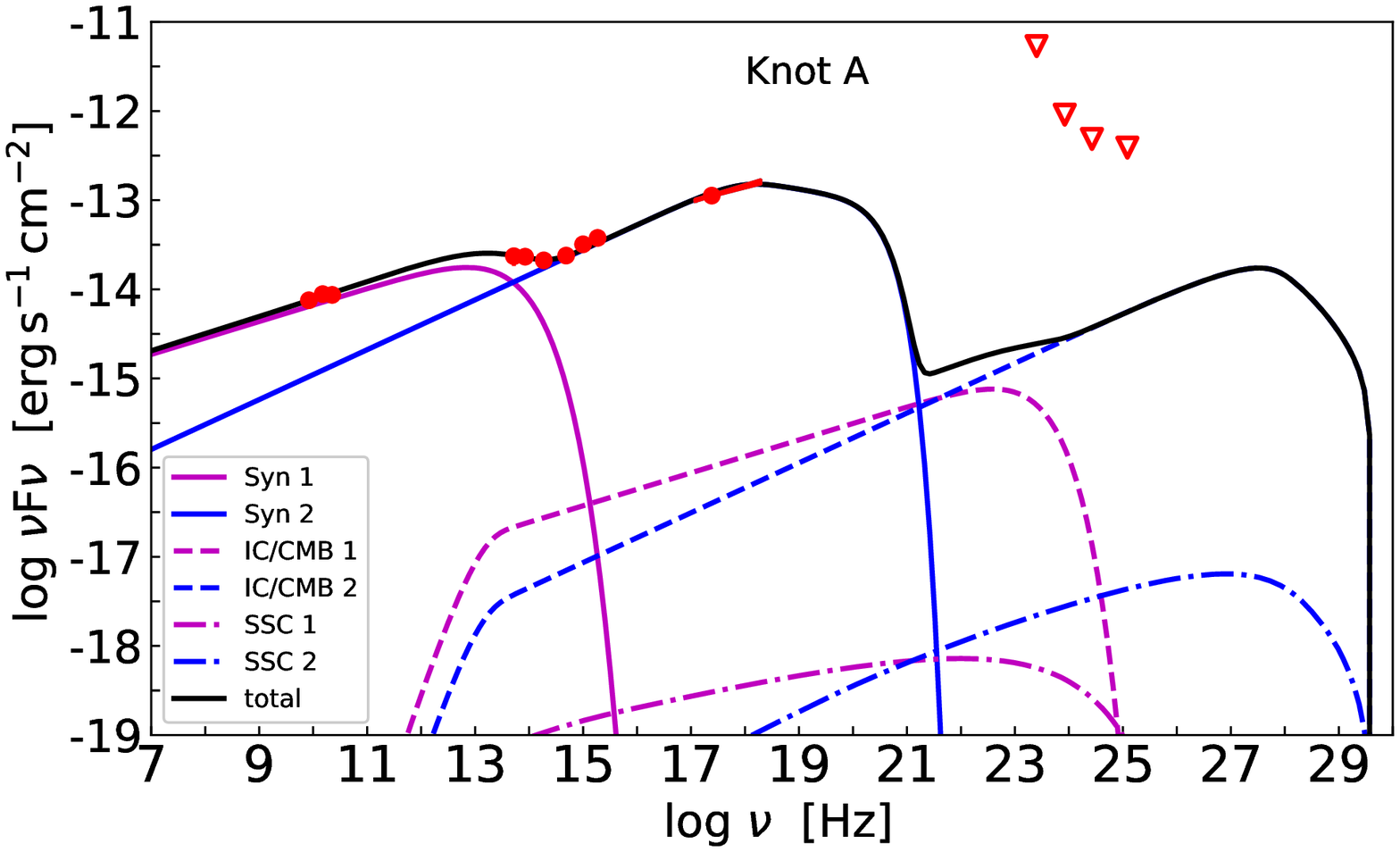}
\includegraphics[angle=0,scale=0.26]{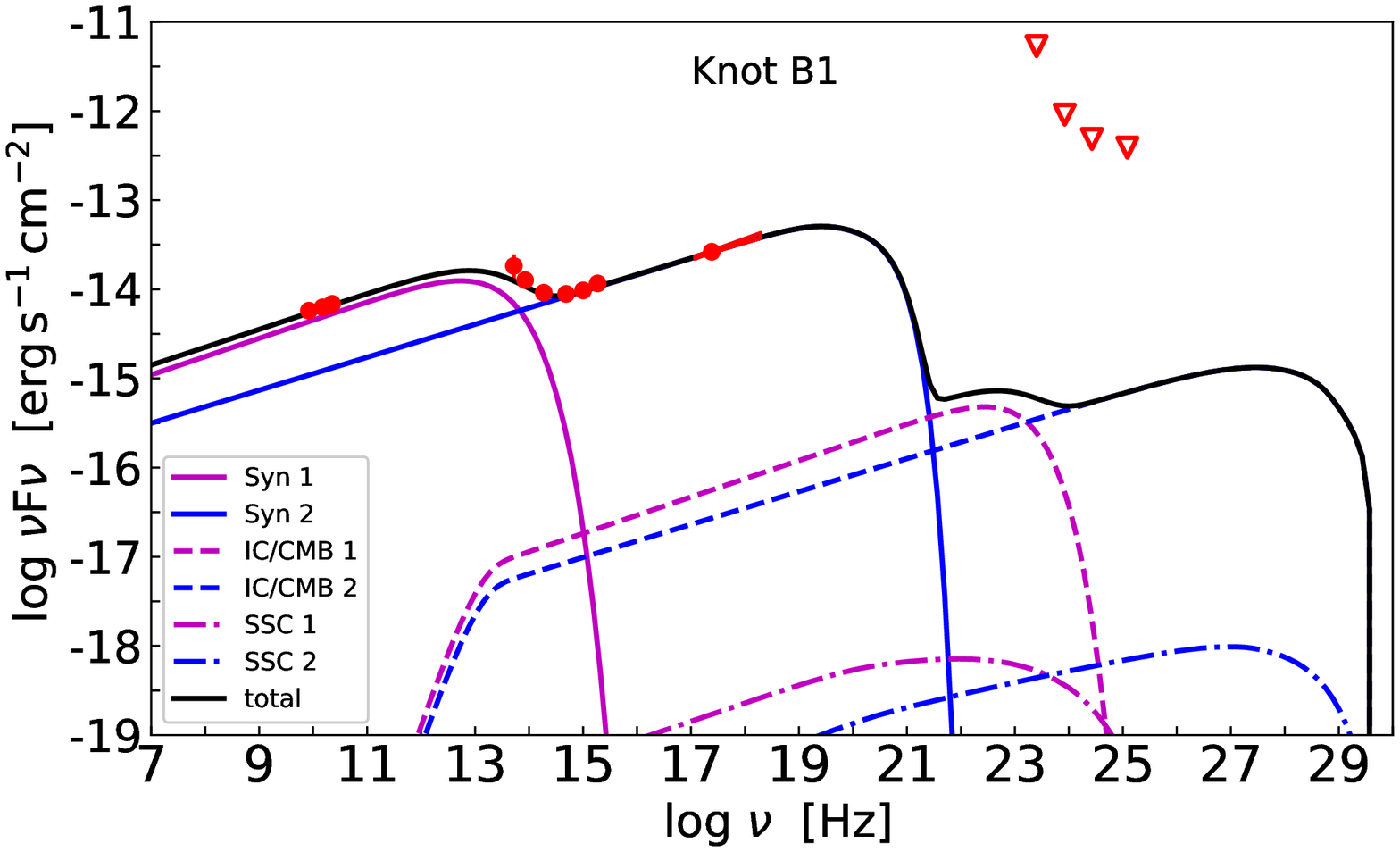}
\includegraphics[angle=0,scale=0.26]{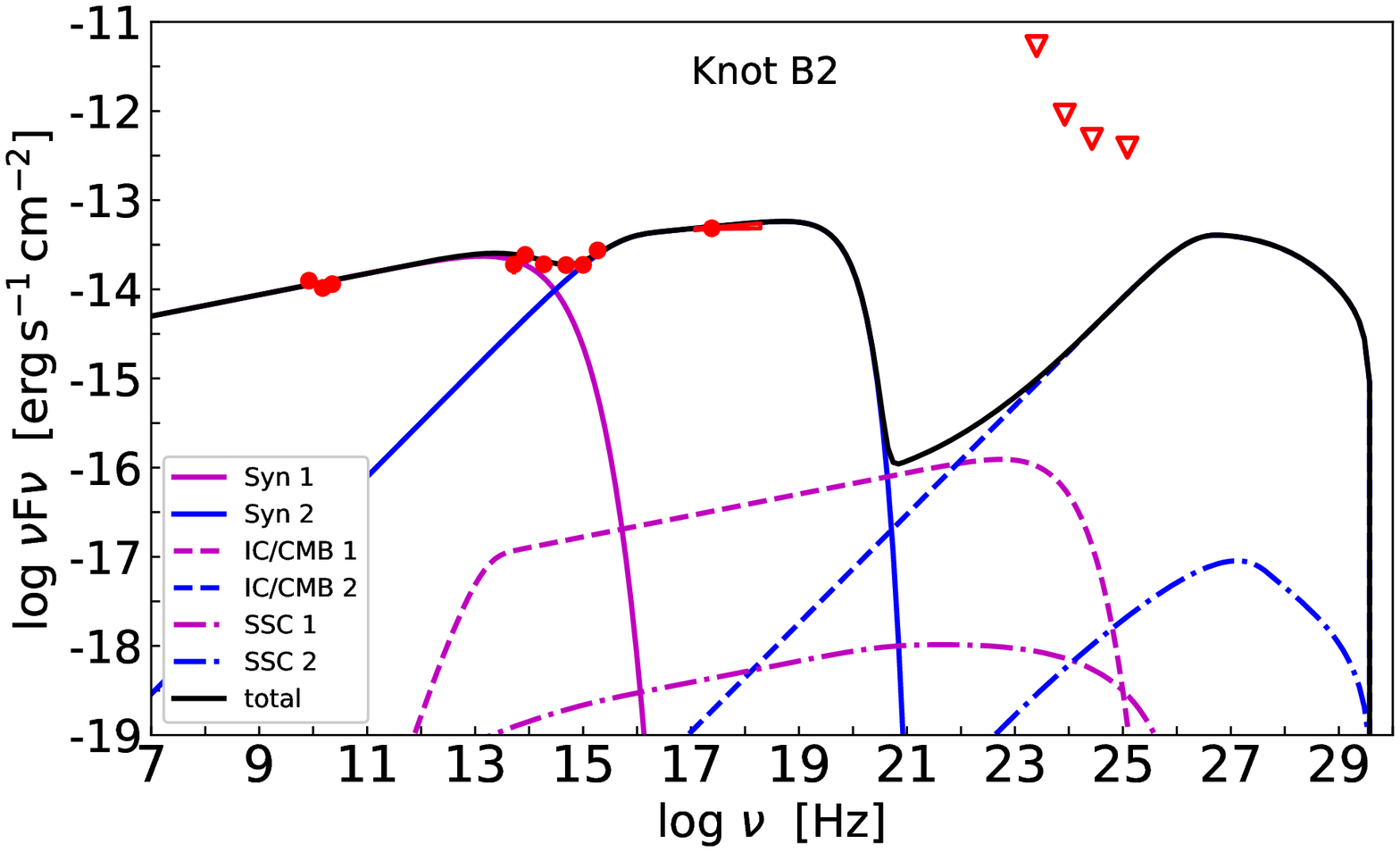}
\includegraphics[angle=0,scale=0.26]{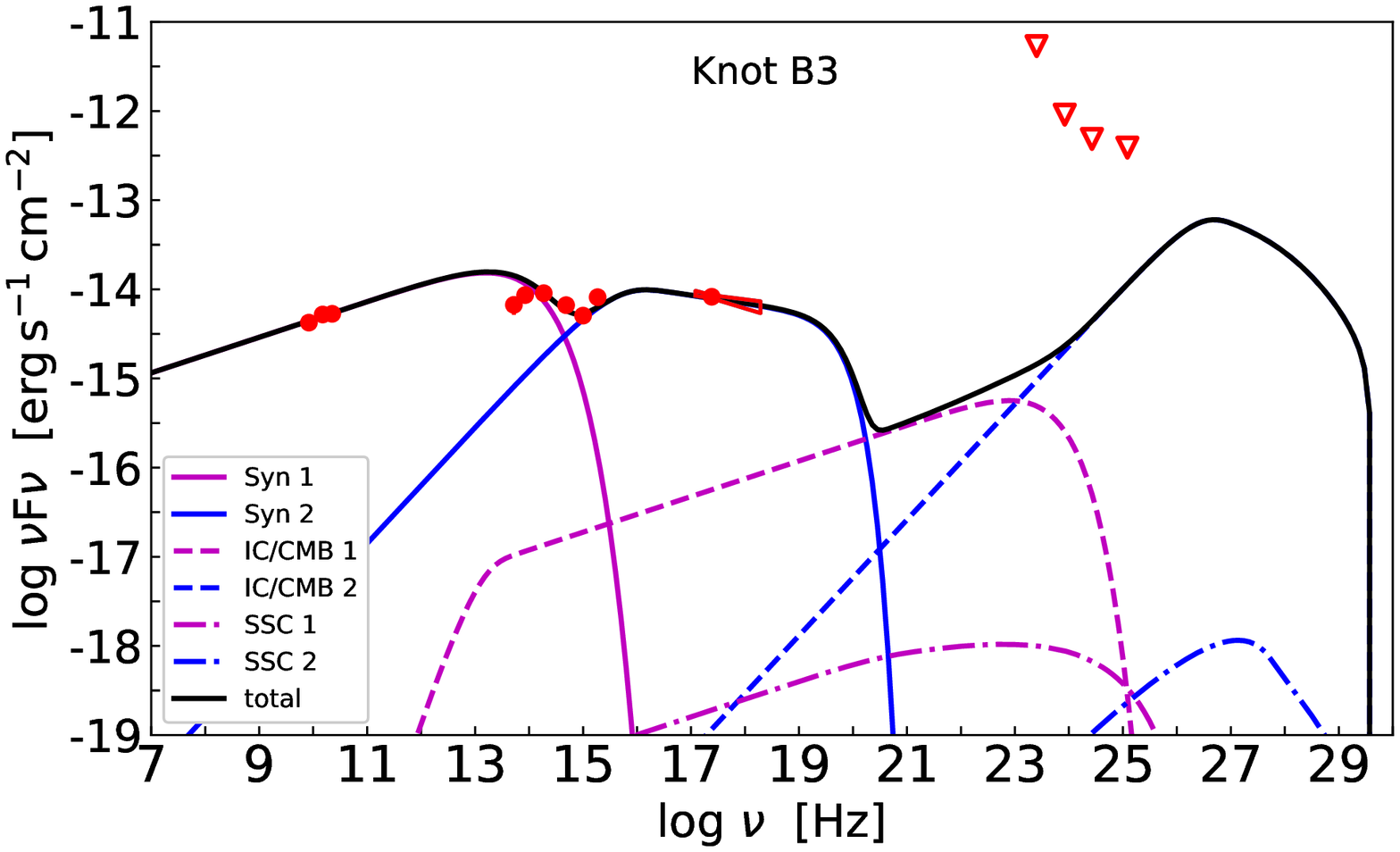}
\includegraphics[angle=0,scale=0.26]{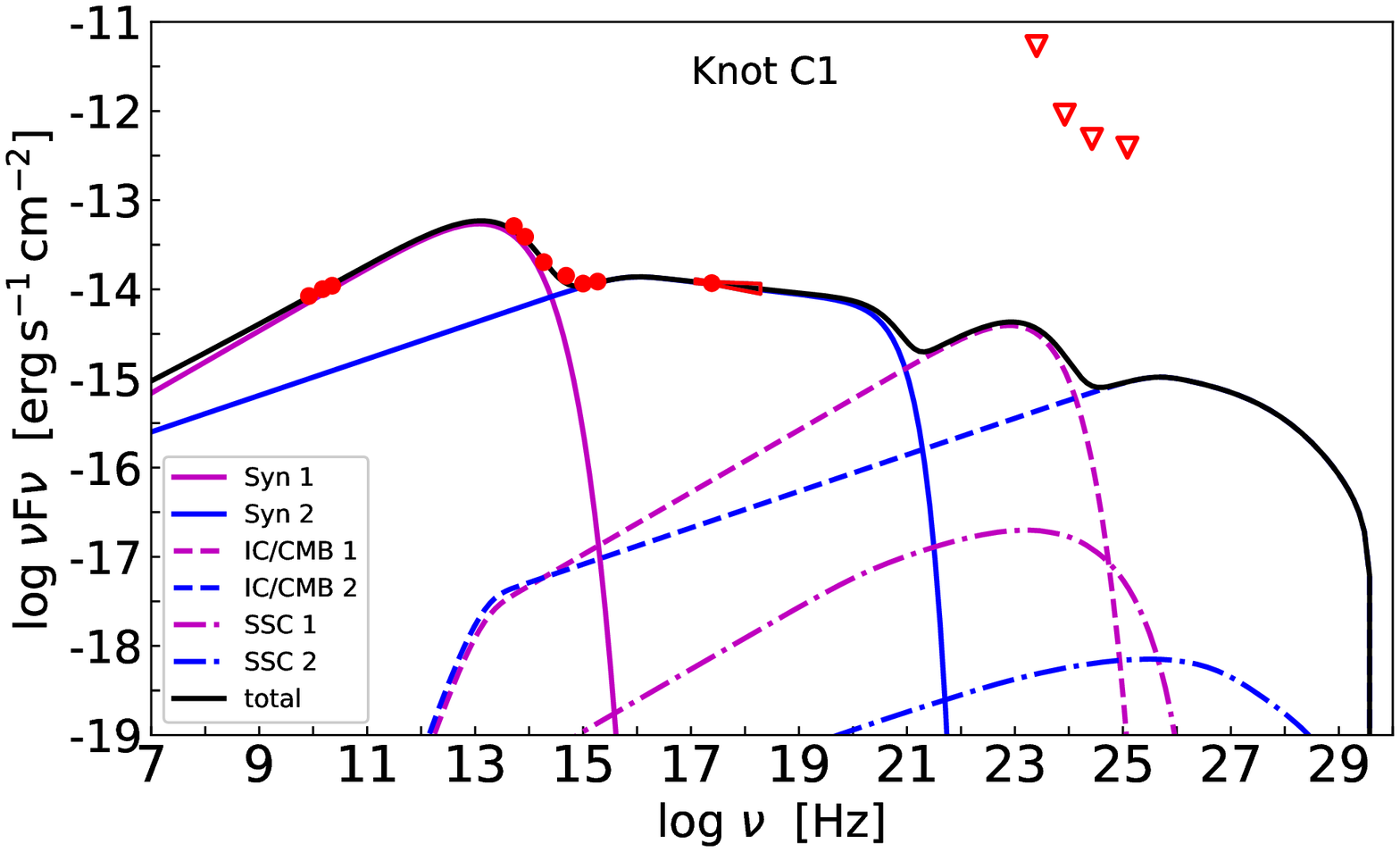}
\includegraphics[angle=0,scale=0.26]{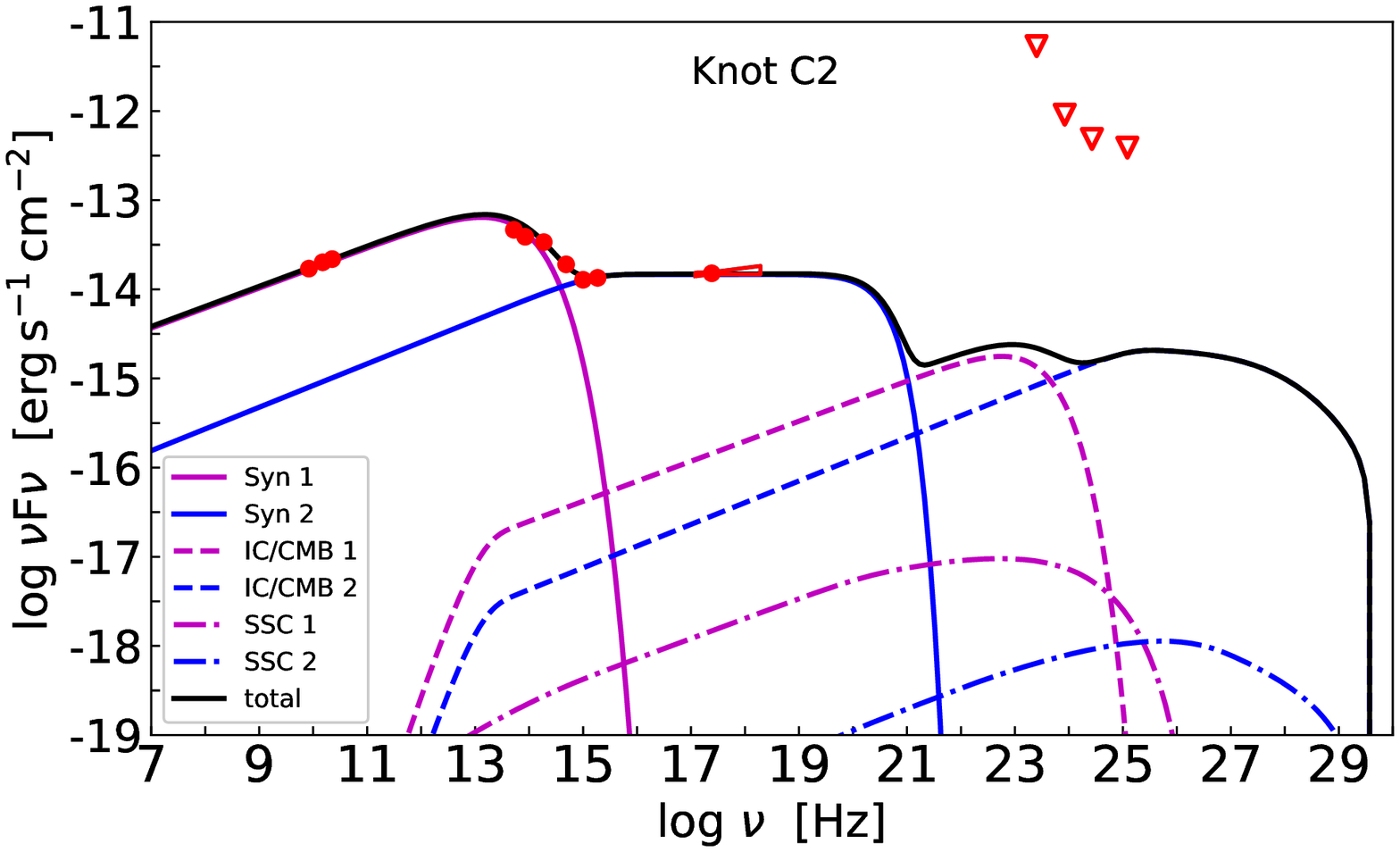}
\includegraphics[angle=0,scale=0.26]{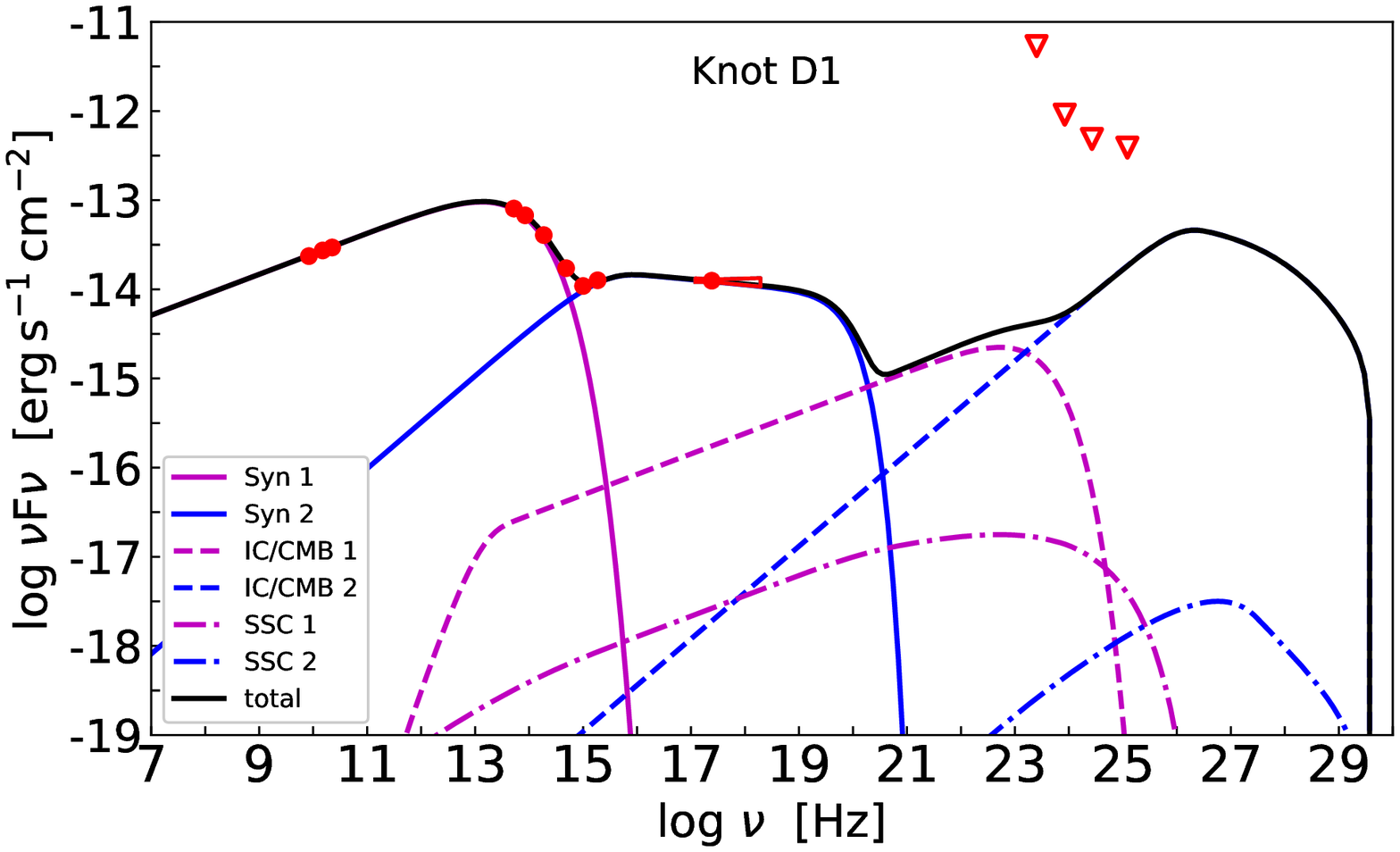}
\includegraphics[angle=0,scale=0.26]{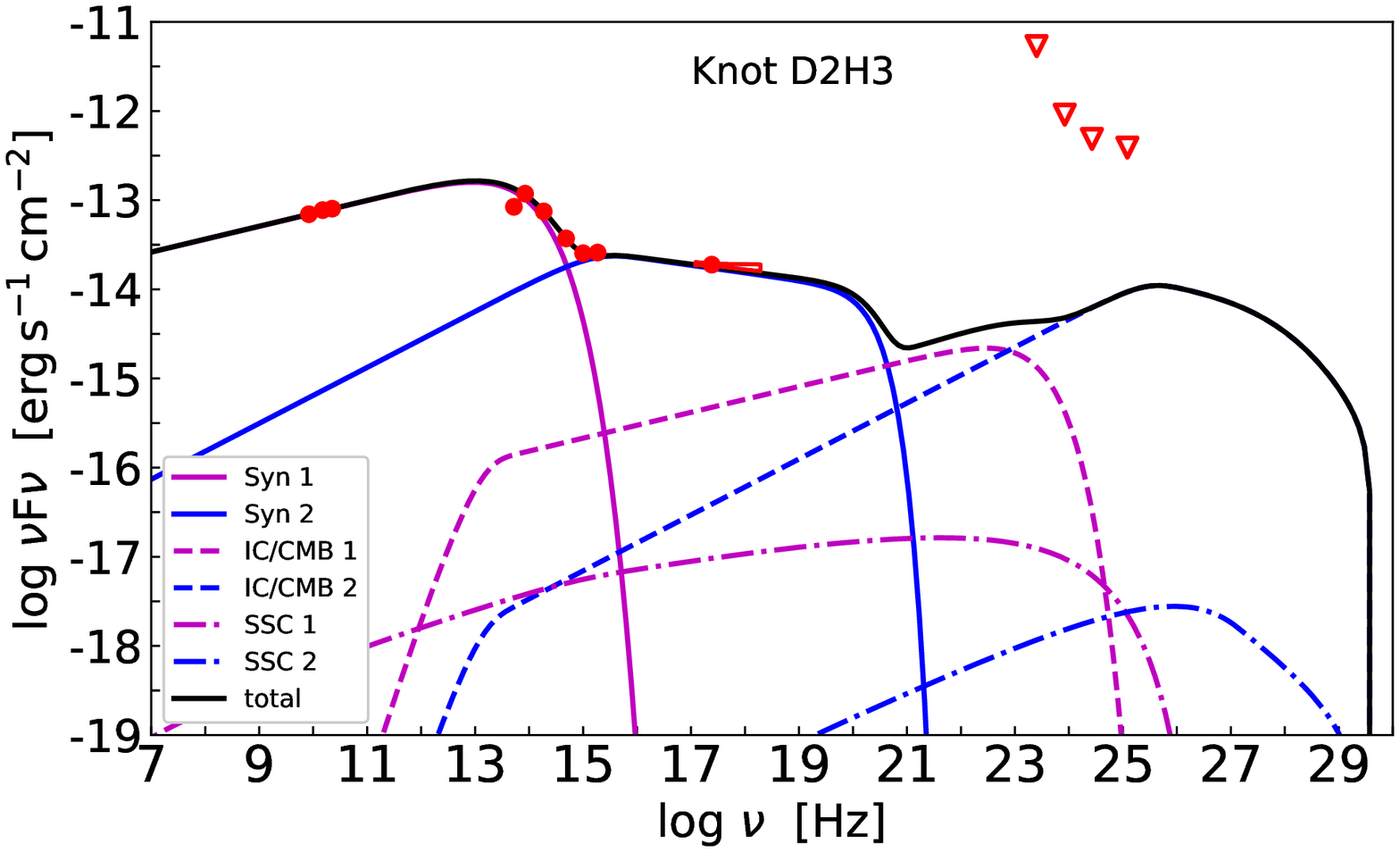}
\hfill
\includegraphics[angle=0,scale=0.26]{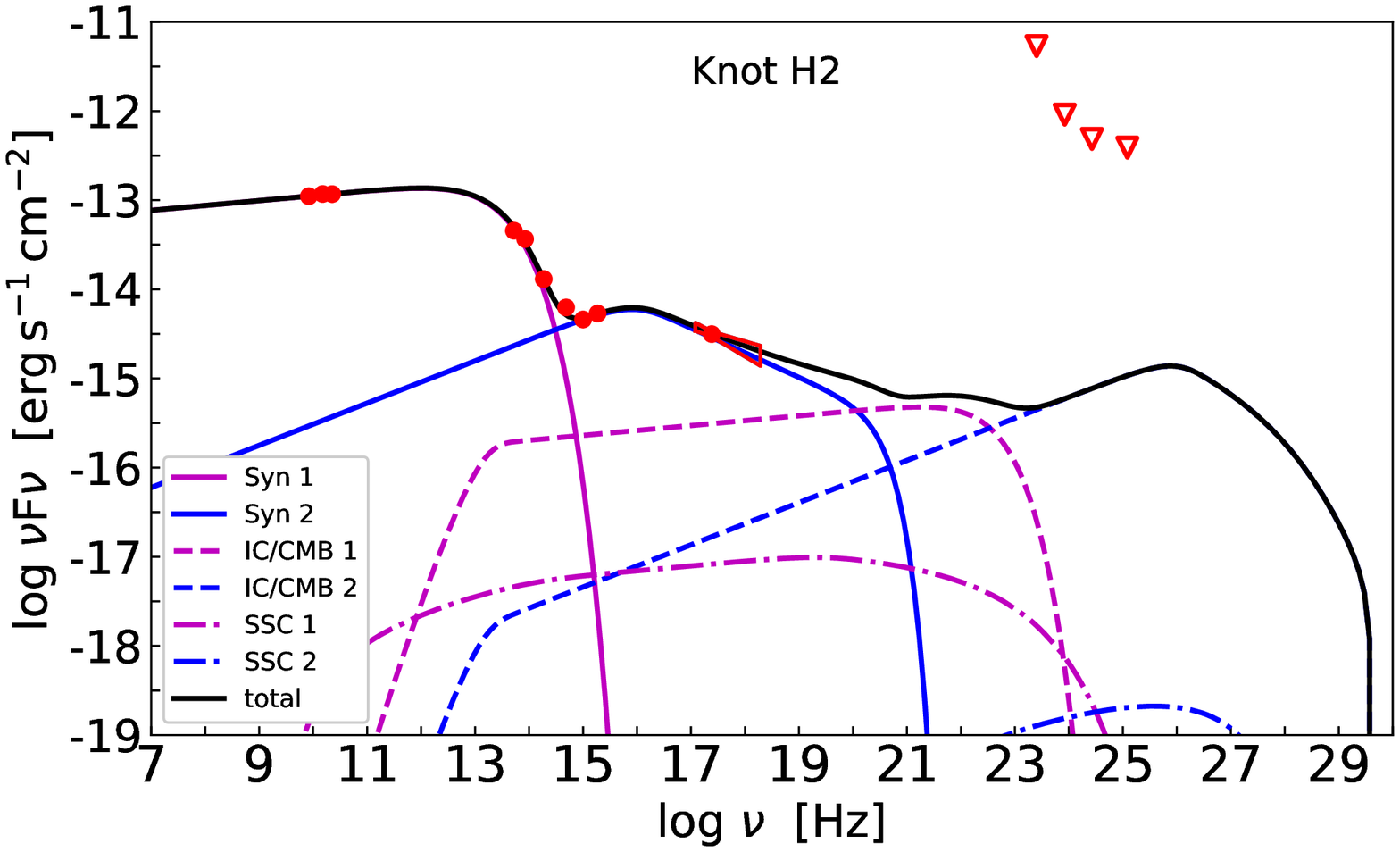}
\caption{The same as Figure \ref{single1}, but for the model fits of two electron populations (scenario II) in case of $\delta>1$. The purple and blue lines respectively display the radiation components from the two electron populations. }\label{Two_E2}
\end{figure*}

In case of $\delta>1$, the model loses the constraint on the $\delta$ value since no observation of the IC component in these knots. We adopt $\delta=3.7$ (the variability Doppler factor in Liodakis et al. 2017) and $\theta=3.3^{\circ}$ (Hovatta et al. 2009), and thus obtain $\Gamma=2$, which is consistent with $\Gamma<2.9$ by assuming that the large-scale jet knots are packets of moving plasma with an upper-limit of $1c$ (Meyer et al. 2016). In this case, the free parameters of the SED modeling are $A_{0}$, $p_0$, $E_{\rm c}$, $A_1$, $p_1$, $p_2$, $E_{\rm b}$. The SED fitting results are shown in Figure \ref{Two_E2} and the corresponding SSC and IC/CMB contributions of two electron populations are also presented in Figure \ref{Two_E2}. The fitting parameters are listed in Table 2. One can observe that the model can well reproduce the broadband SEDs of knots in both cases of $\delta=1$ and $\delta>1$.

\subsection{Scenario III: an Electron Population plus a Proton Population}

\begin{figure*}
\includegraphics[angle=0,scale=0.26]{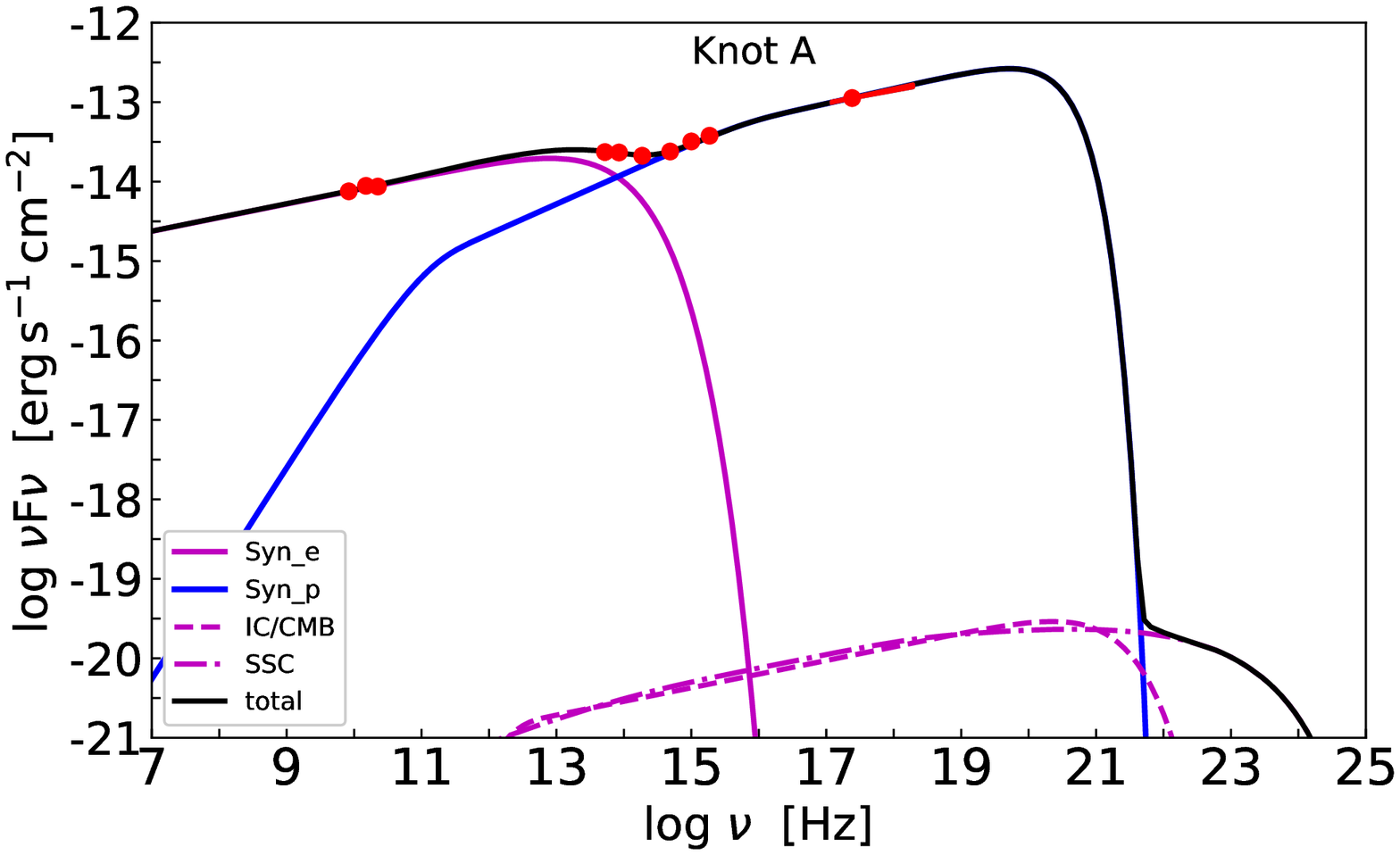}
\includegraphics[angle=0,scale=0.26]{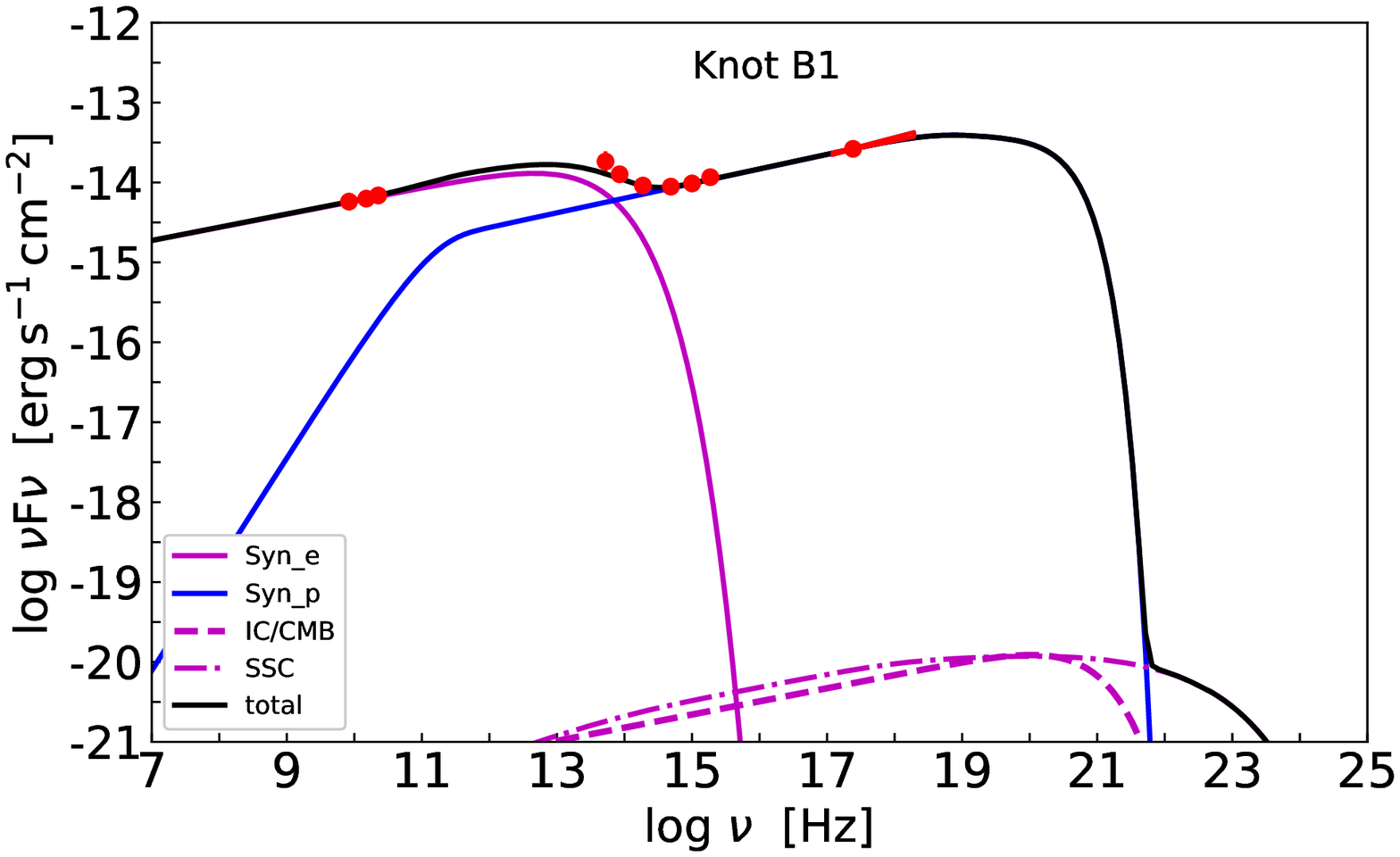}
\includegraphics[angle=0,scale=0.26]{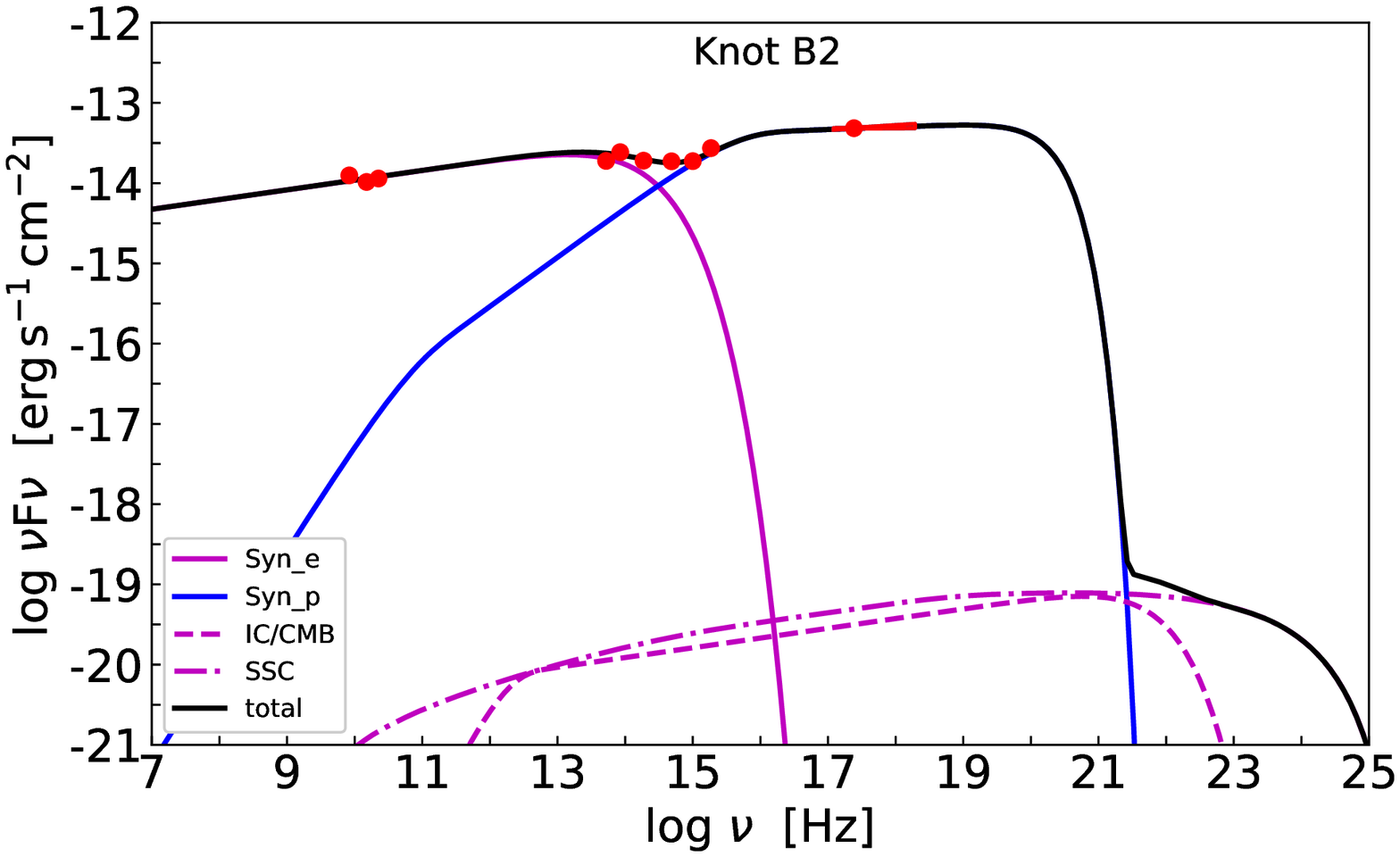}
\includegraphics[angle=0,scale=0.26]{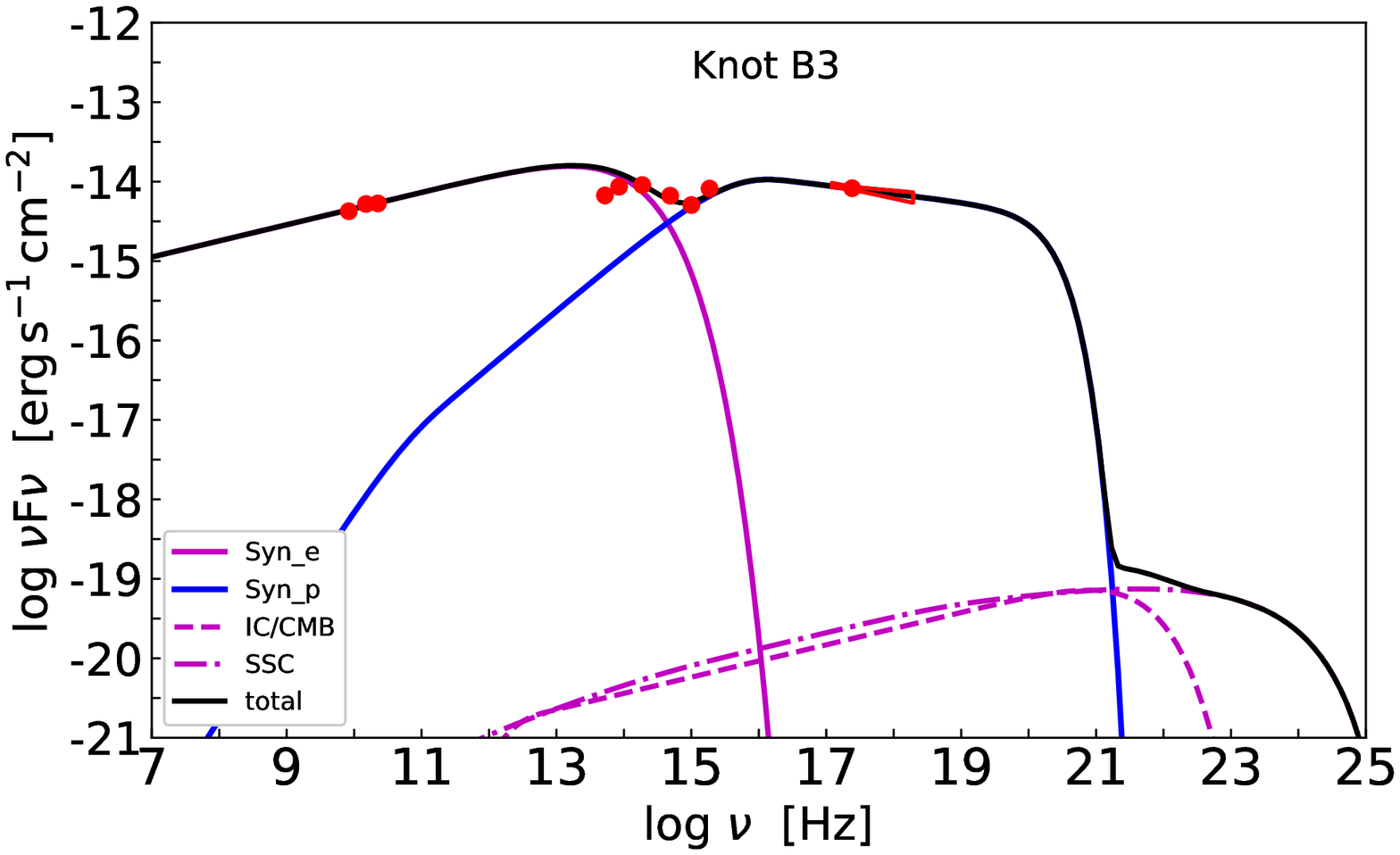}
\includegraphics[angle=0,scale=0.26]{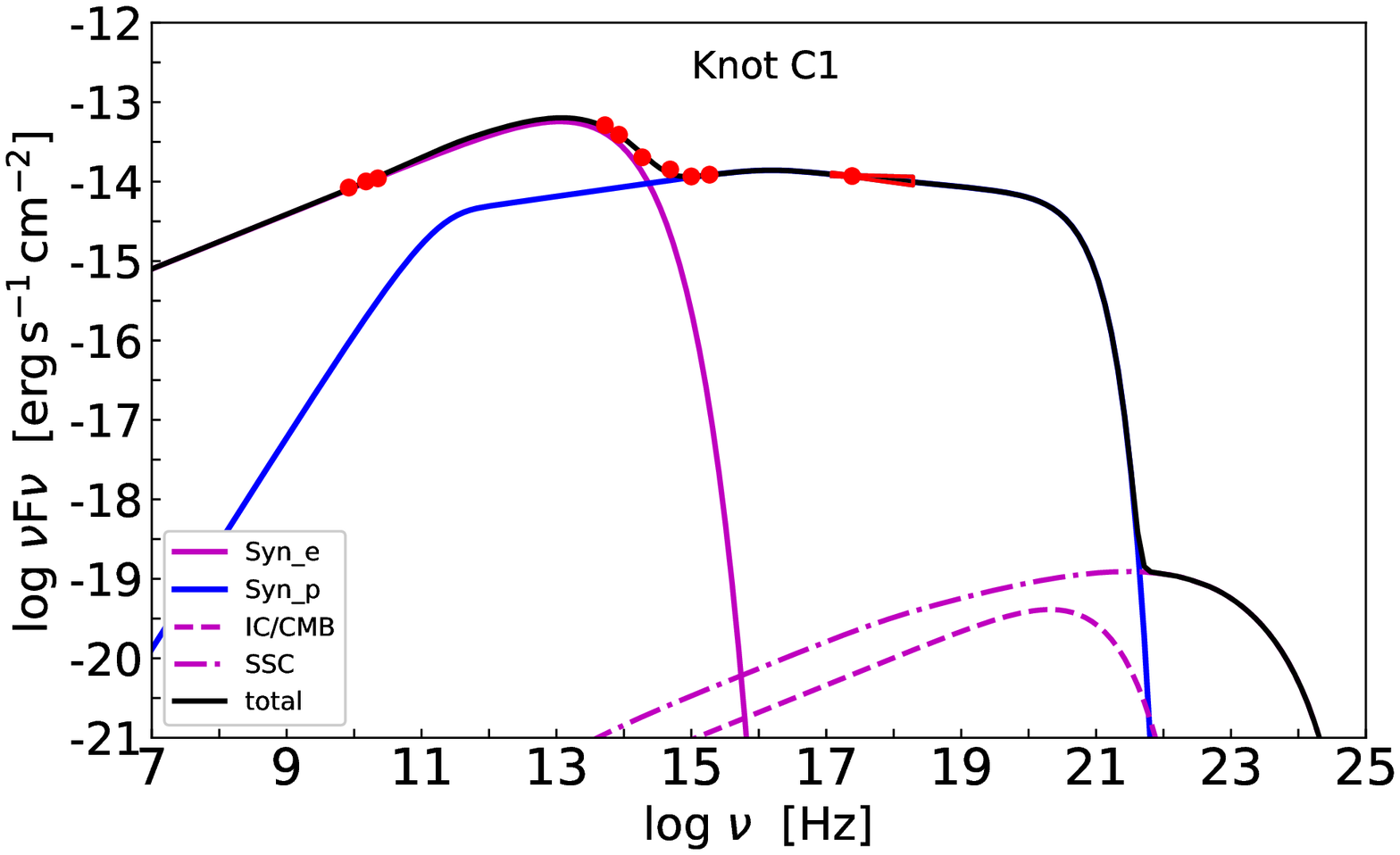}
\includegraphics[angle=0,scale=0.26]{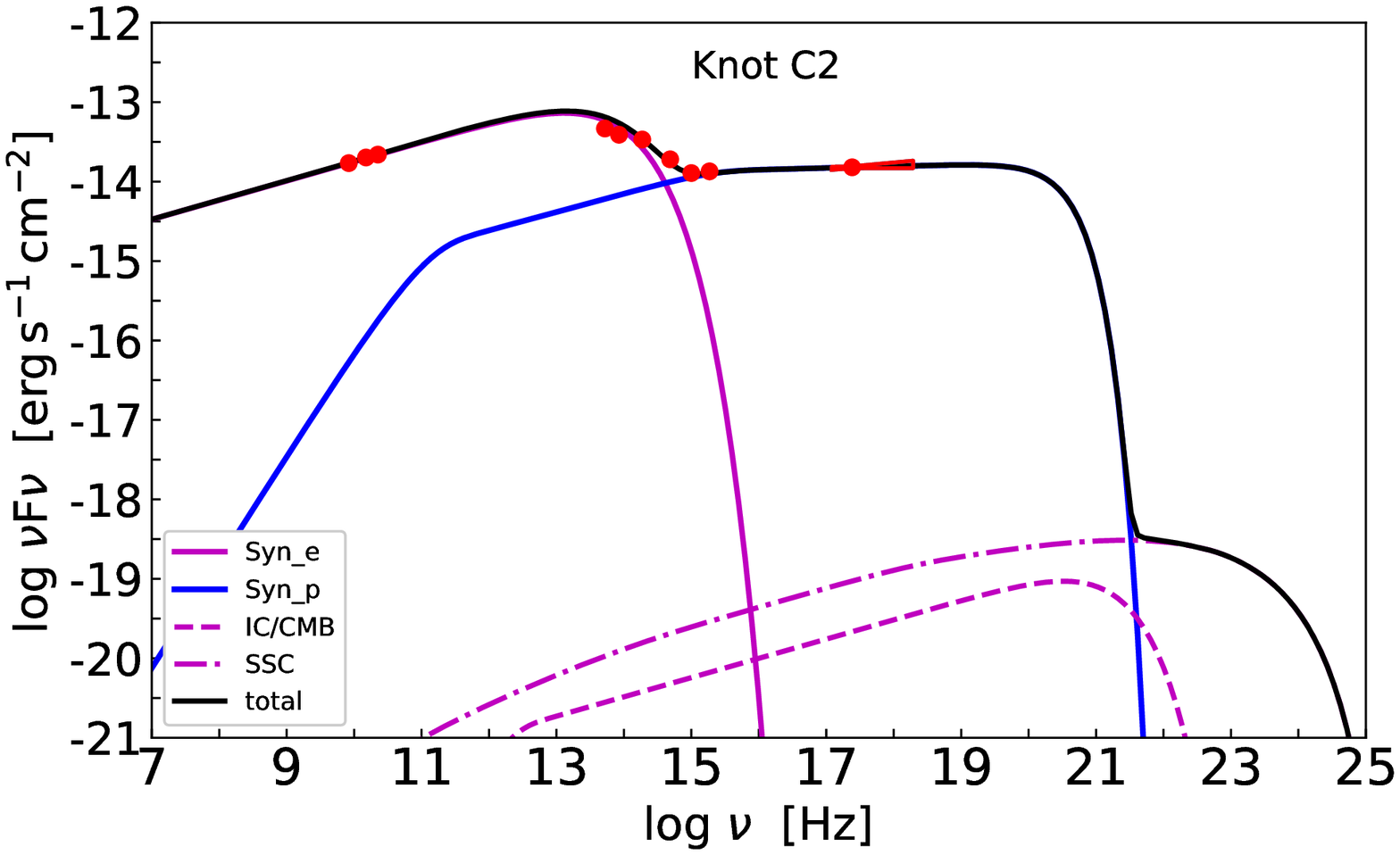}
\includegraphics[angle=0,scale=0.26]{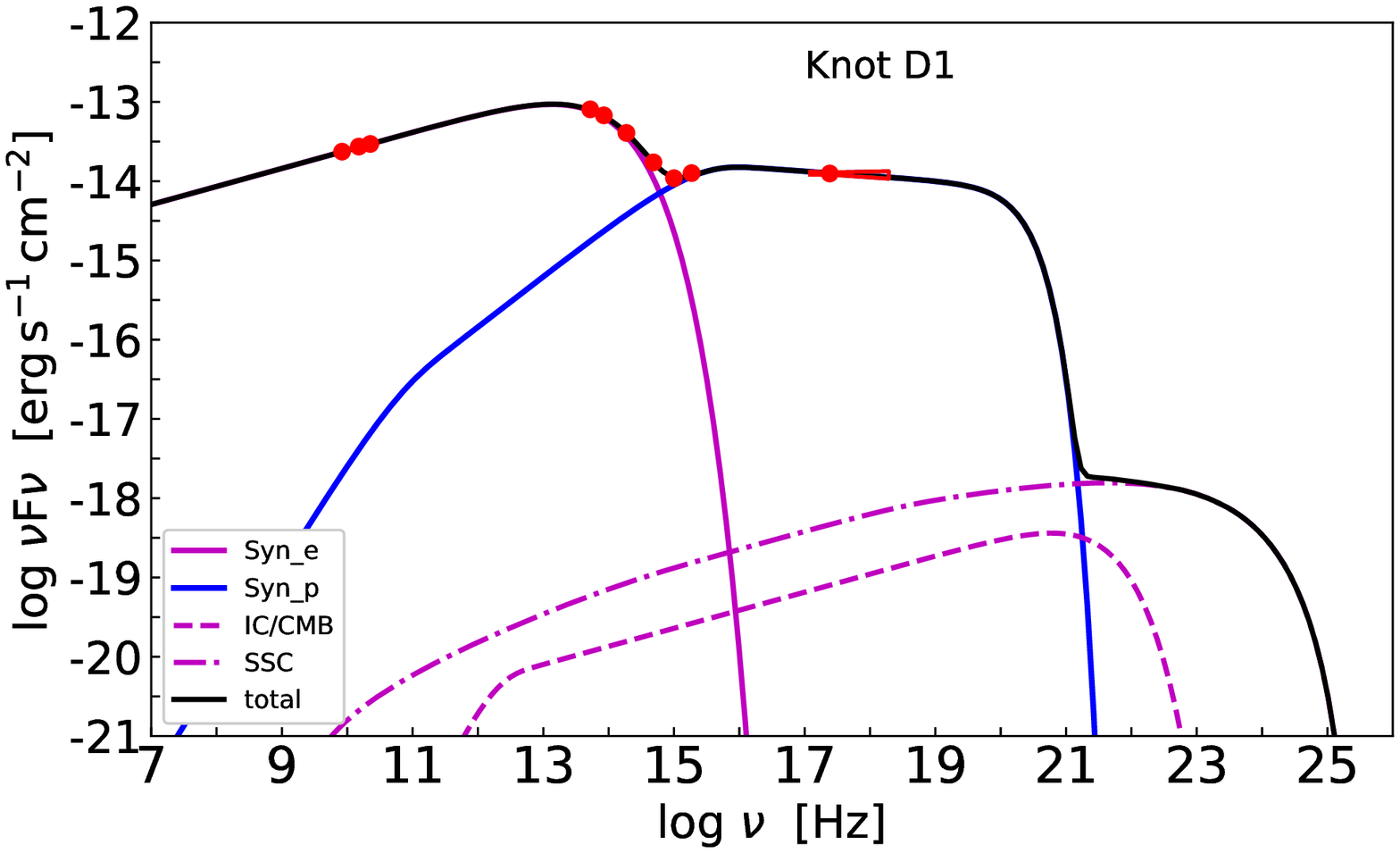}
\includegraphics[angle=0,scale=0.26]{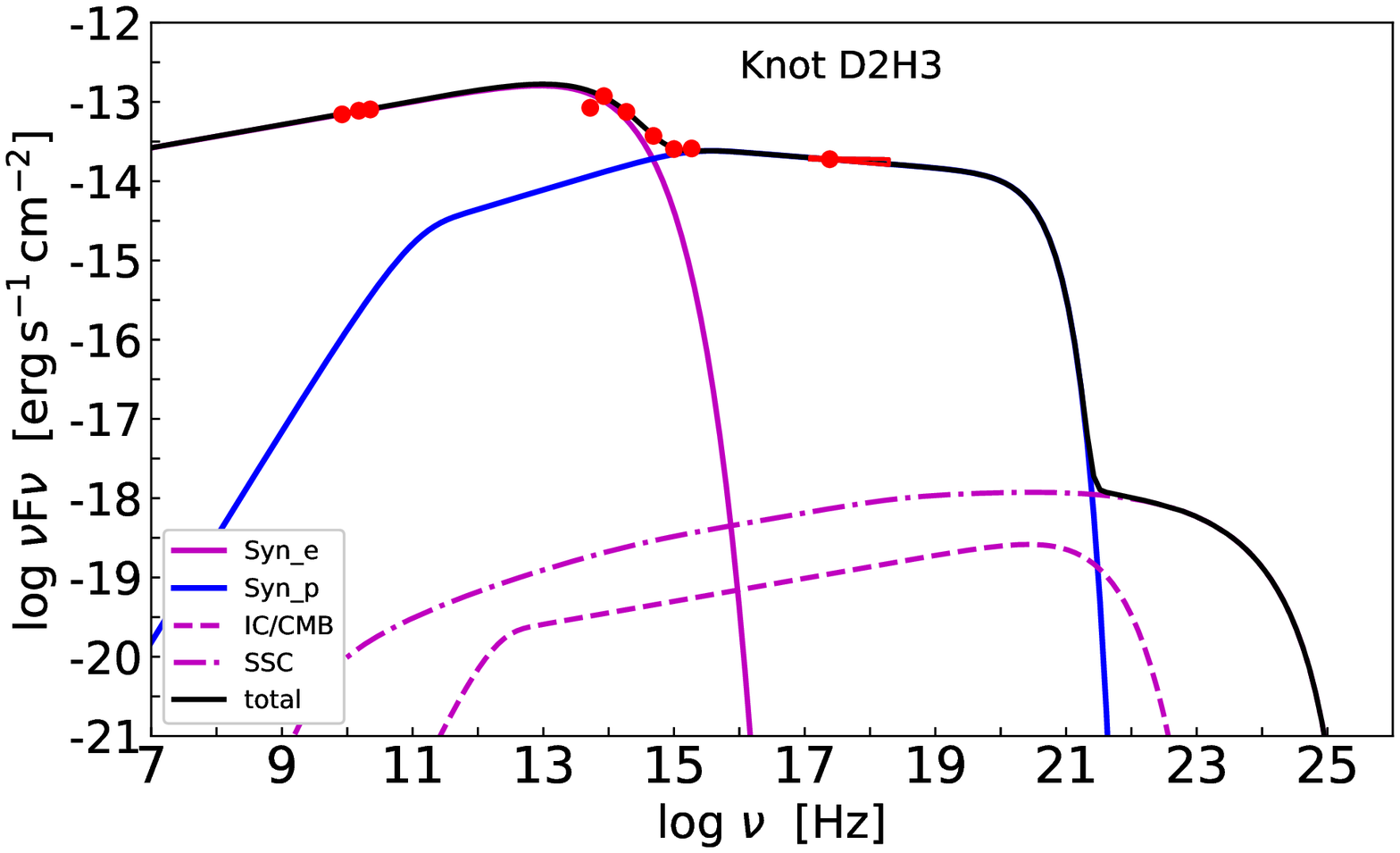}
\hfill
\includegraphics[angle=0,scale=0.26]{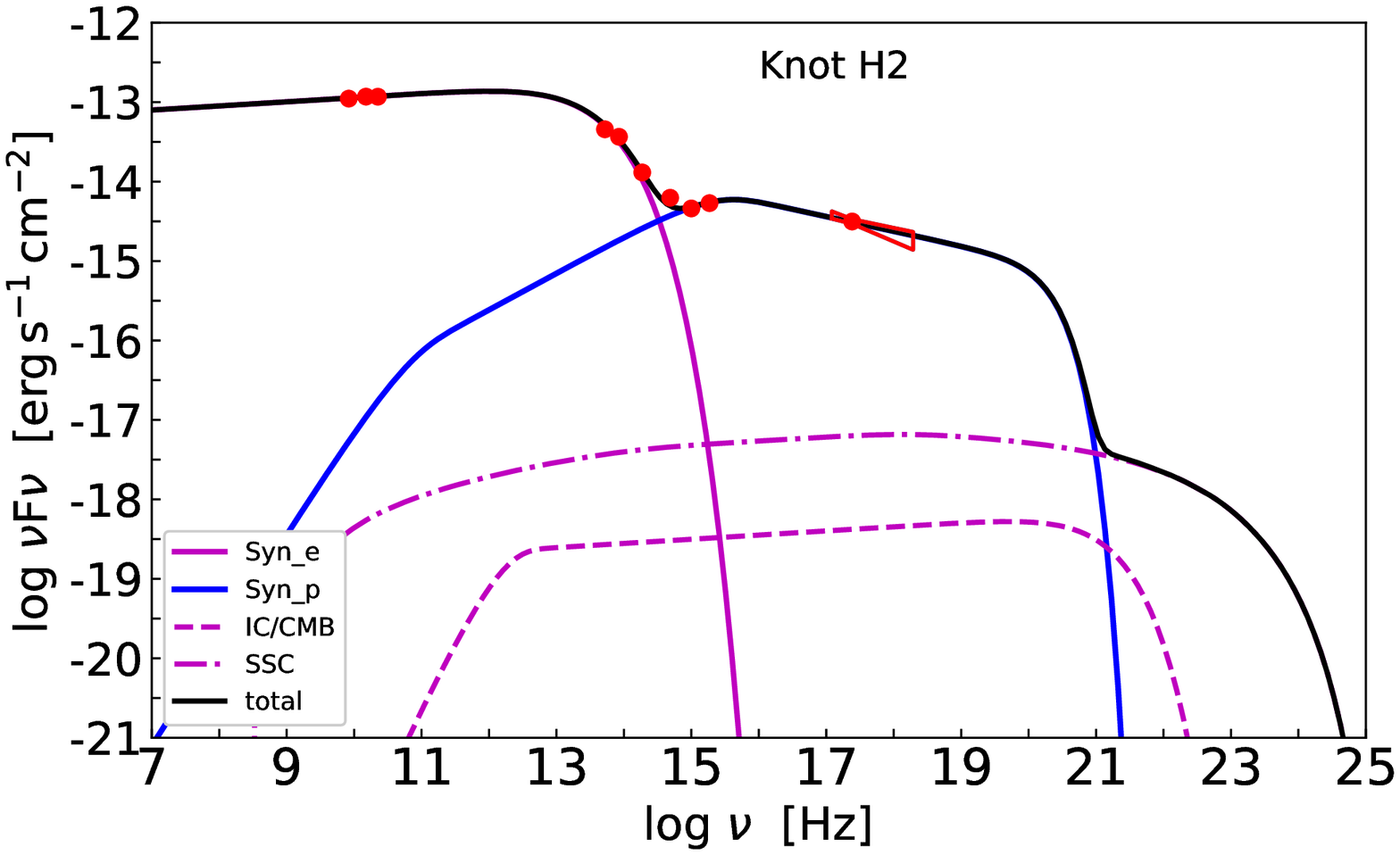}
\caption{The same as Figure \ref{single1}, but for the model fits of an electron population plus a proton population (scenario III) in case of $\delta=1$. The purple and blue lines display the radiation components from the electron population and the proton population, respectively. }\label{Two_P1}
\end{figure*}

It was suggested that the X-ray emission of large-scale jets in some AGNs may be from the synchrotron radiation of protons (e.g., Aharonian 2002; Kundu \& Gupta 2014). We thus try to explain the X-ray emission with the synchrotron radiation of the accelerated protons in these knots, i.e., the synchrotron radiations of an exponential cutoff power-law electron population (Equation (2)) and a broken power-law proton population (Equation (1)) are considered in this scenario. The minimum and maximum energies of electrons and protons are $E_{\rm e,min}$=1 MeV, $E_{\rm e,max}$=200 TeV, $E_{\rm p,min}$=100 TeV, and $E_{\rm p,max}$=5 EeV, respectively, where the minimum energy of protons roughly makes the equal numbers between electrons and protons in the knots. The equipartition magnetic field strength is calculated with $U_{B}=U_{\rm e}+U_{\rm p}$, where $U_{\rm p}$ is the energy density of non-thermal protons. In case of $\delta=1$, the free parameters of the SED modeling are $A_{0}$, $p_0$, $E_{\rm c}$, $A_1$, $p_1$, $p_2$, $E_{\rm b}$. The SED fitting results are shown in Figure \ref{Two_P1} and the corresponding SSC and IC/CMB contributions of the electron population are also presented in Figure \ref{Two_P1}. The fitting parameters are listed in Table 3.

\begin{figure*}
\includegraphics[angle=0,scale=0.26]{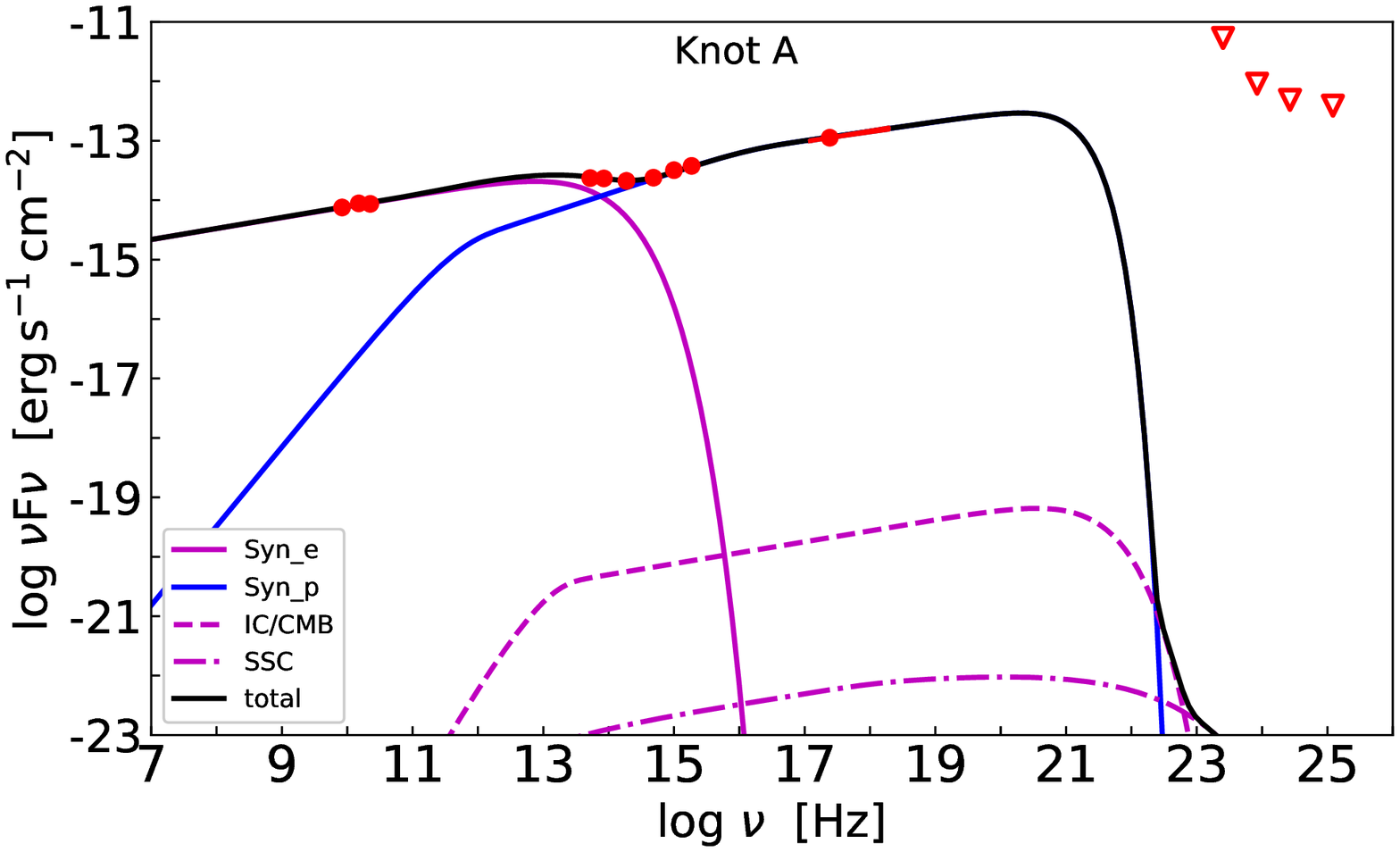}
\includegraphics[angle=0,scale=0.26]{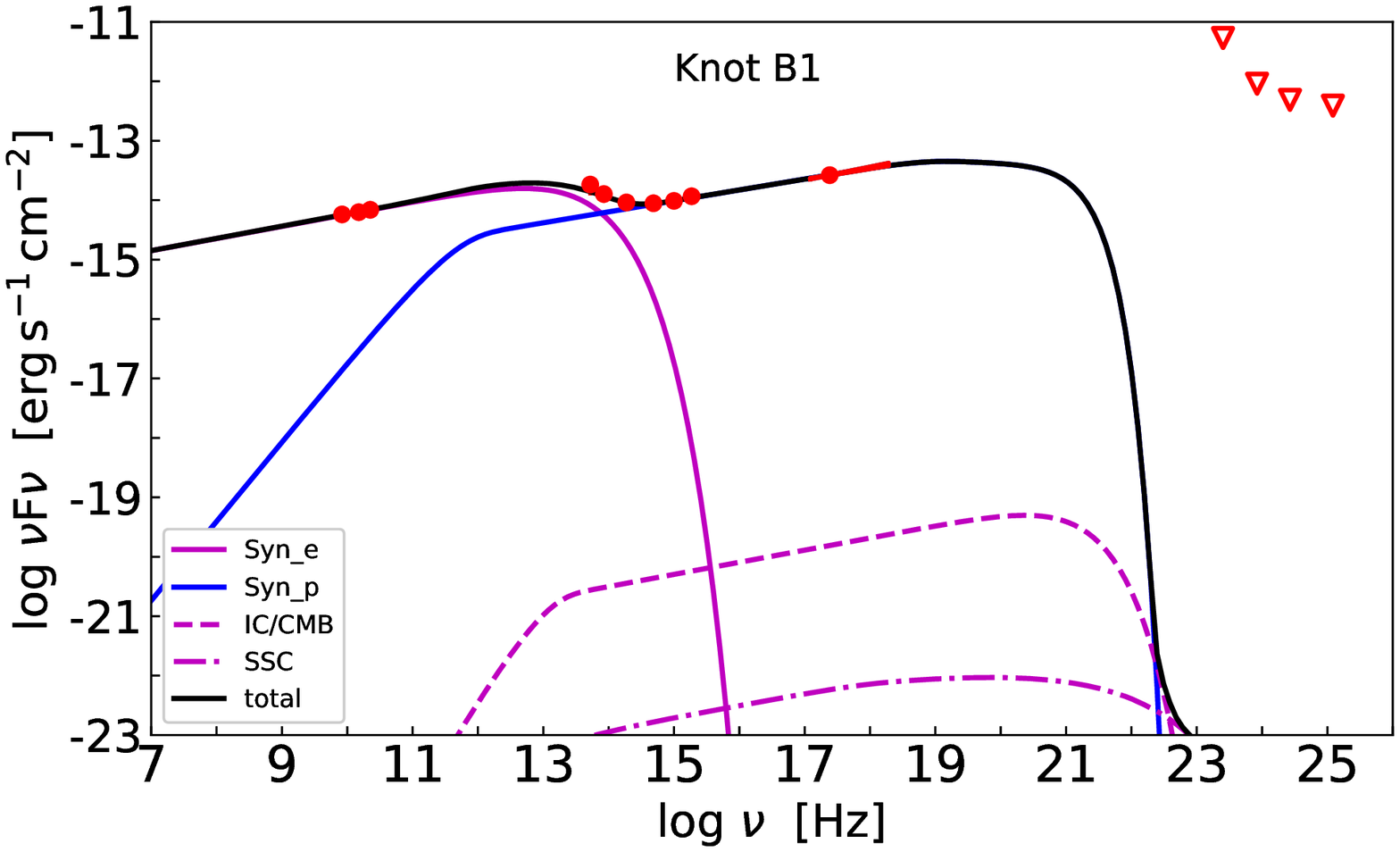}
\includegraphics[angle=0,scale=0.26]{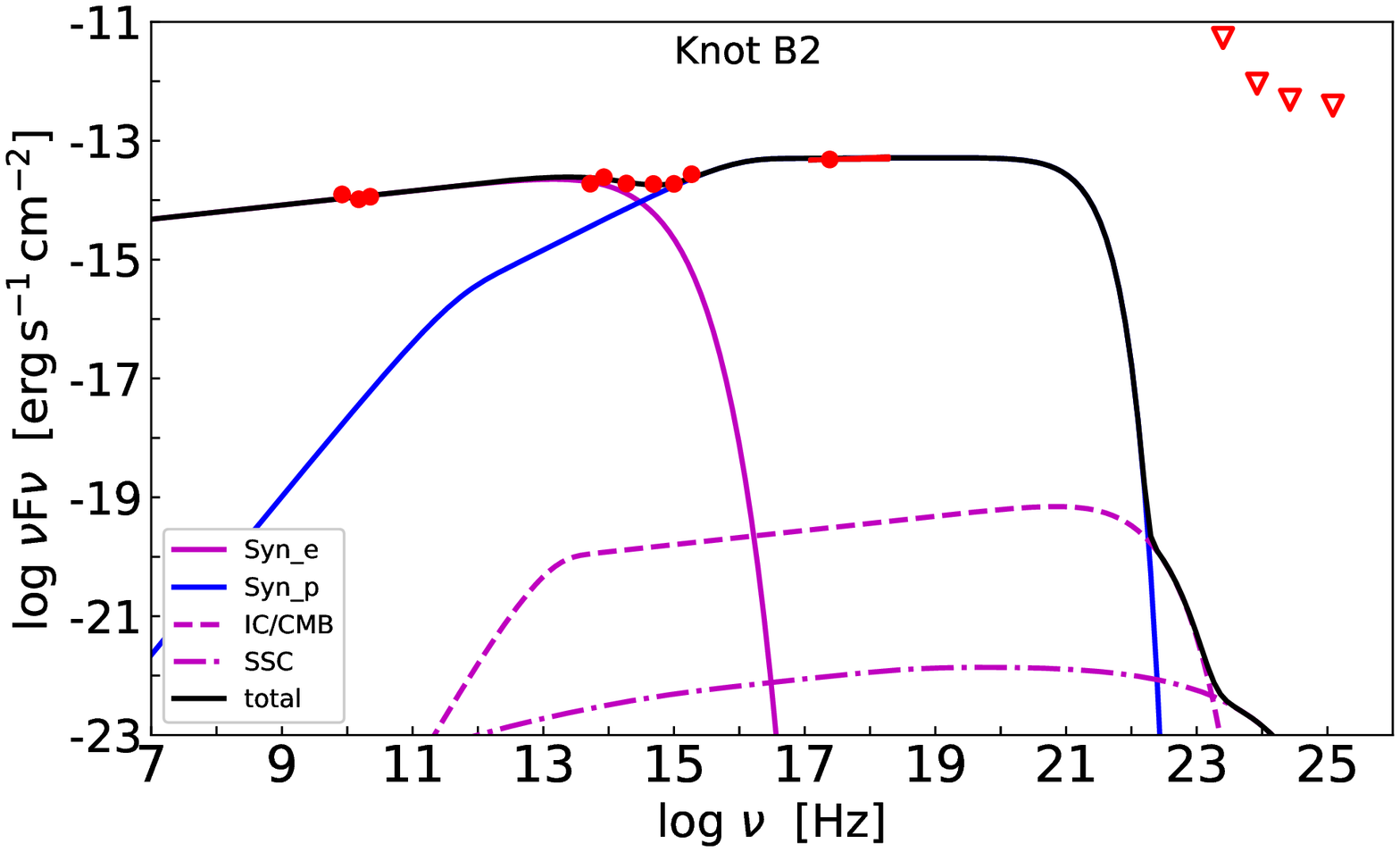}
\includegraphics[angle=0,scale=0.26]{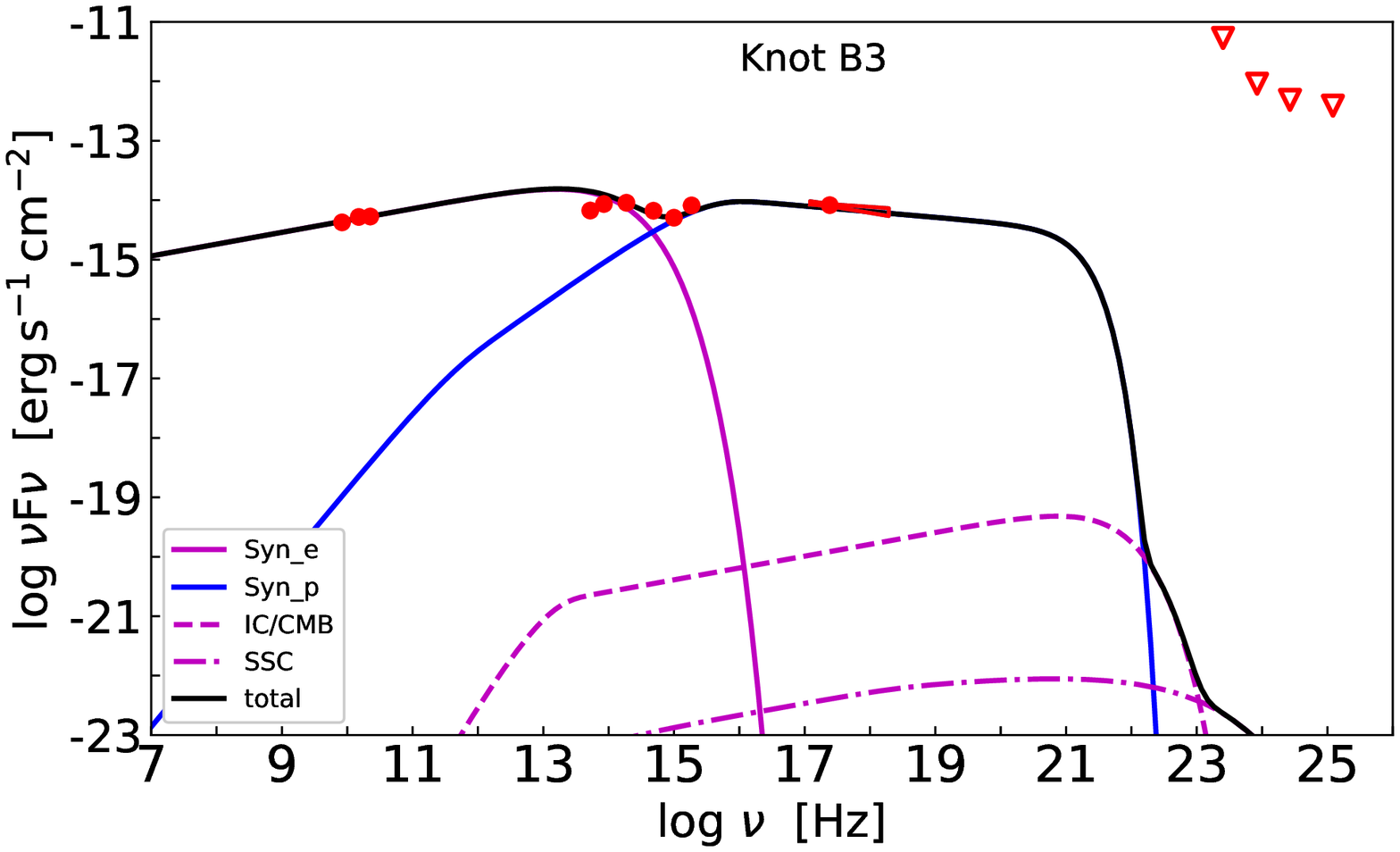}
\includegraphics[angle=0,scale=0.26]{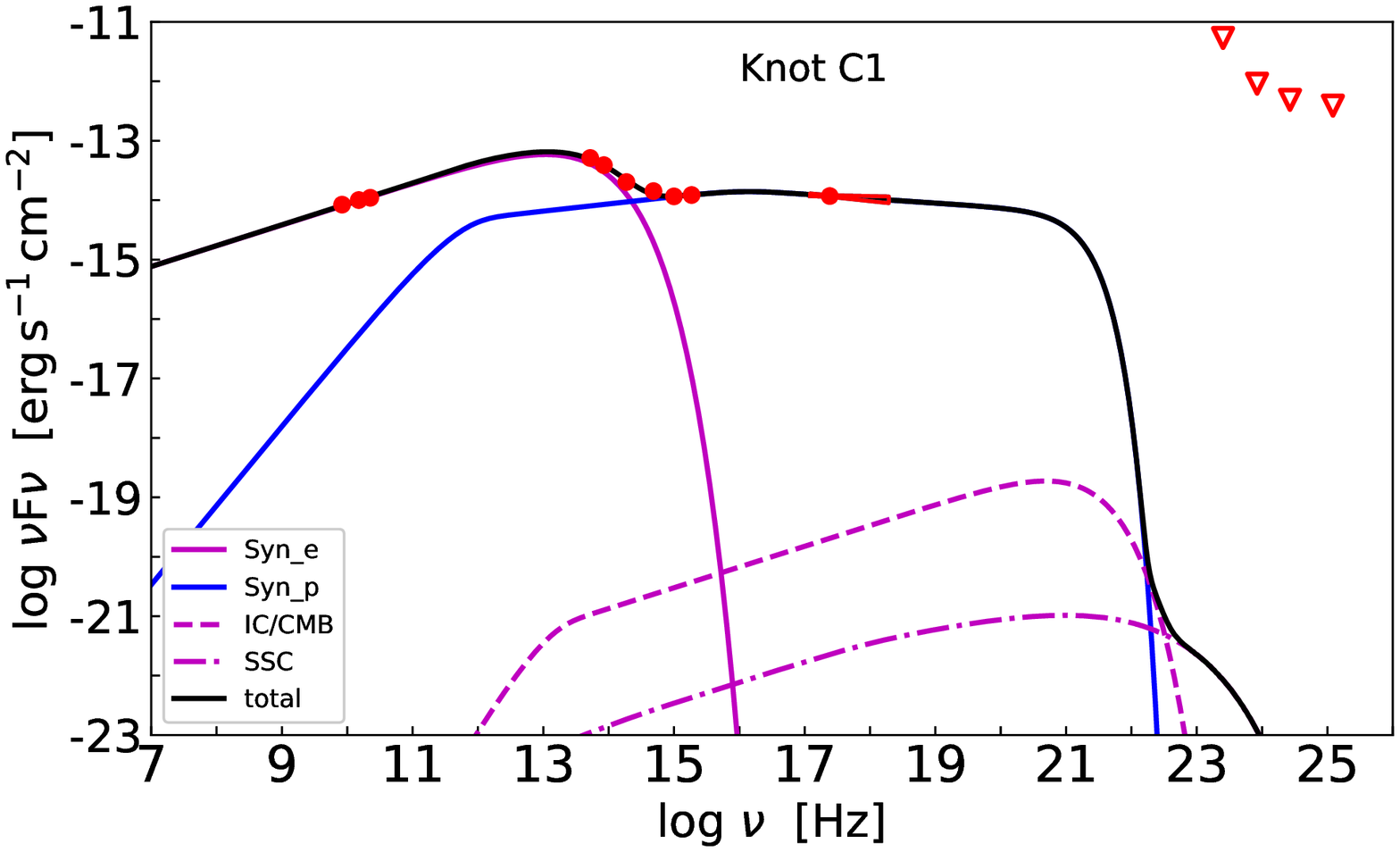}
\includegraphics[angle=0,scale=0.26]{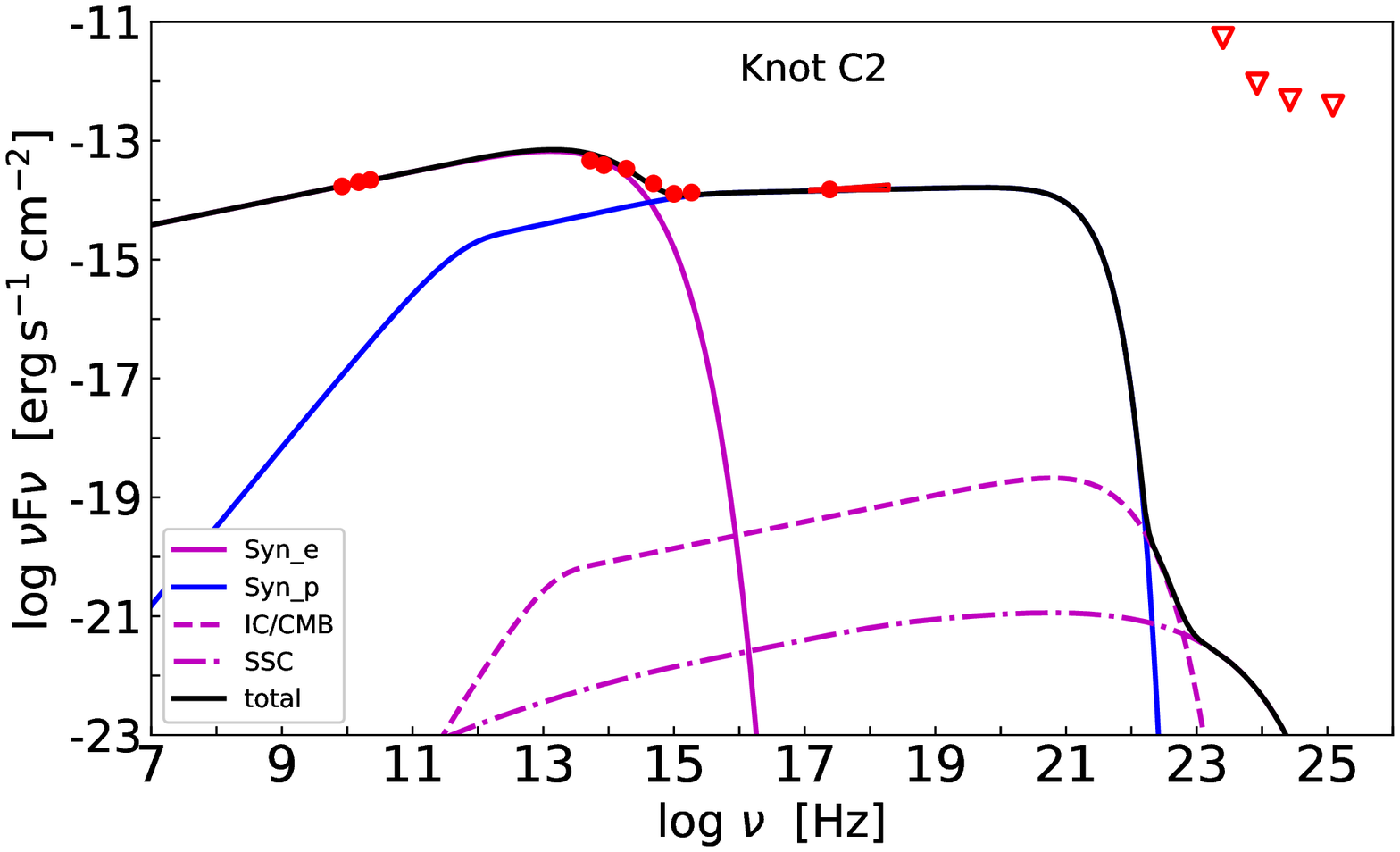}
\includegraphics[angle=0,scale=0.26]{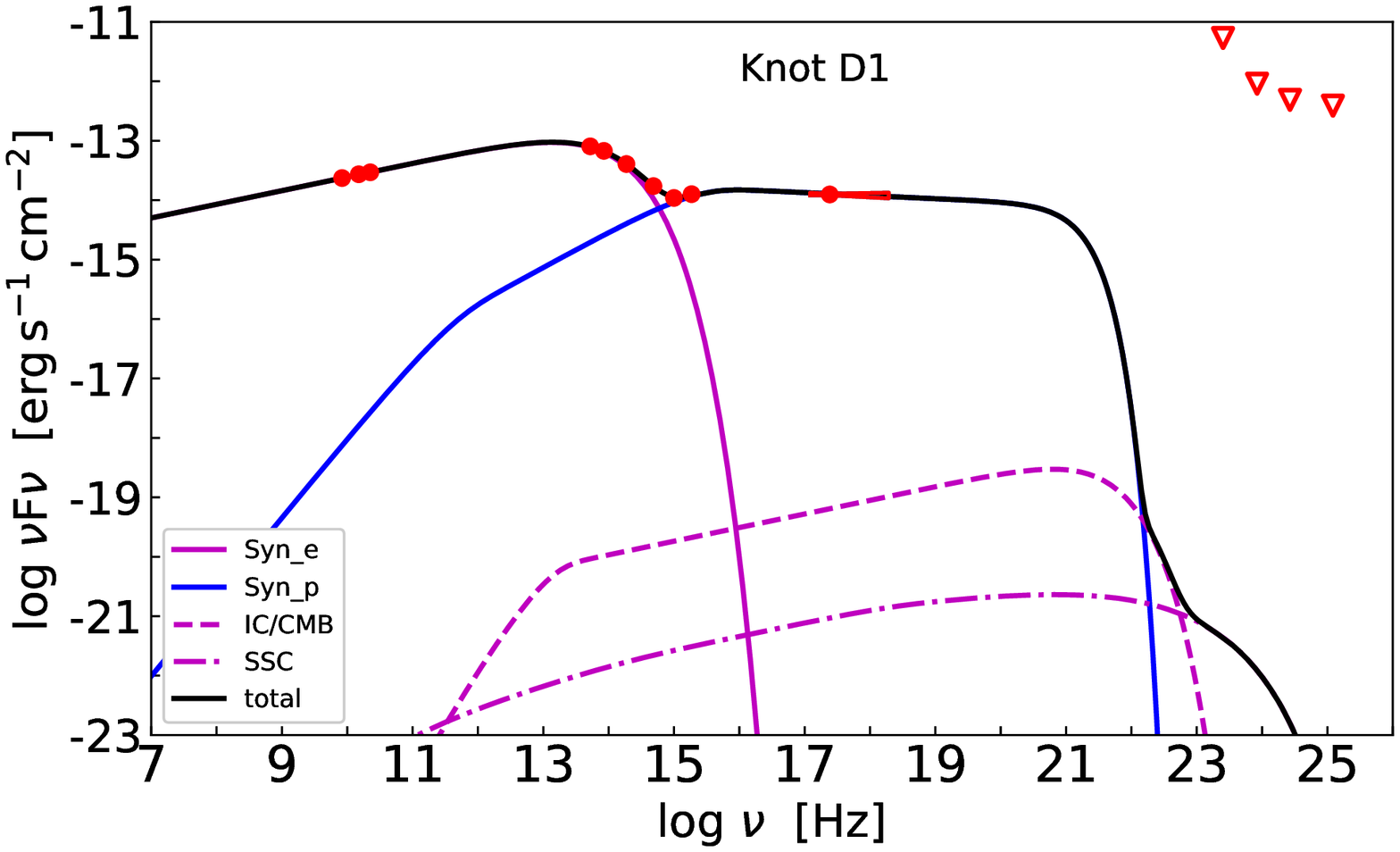}
\includegraphics[angle=0,scale=0.26]{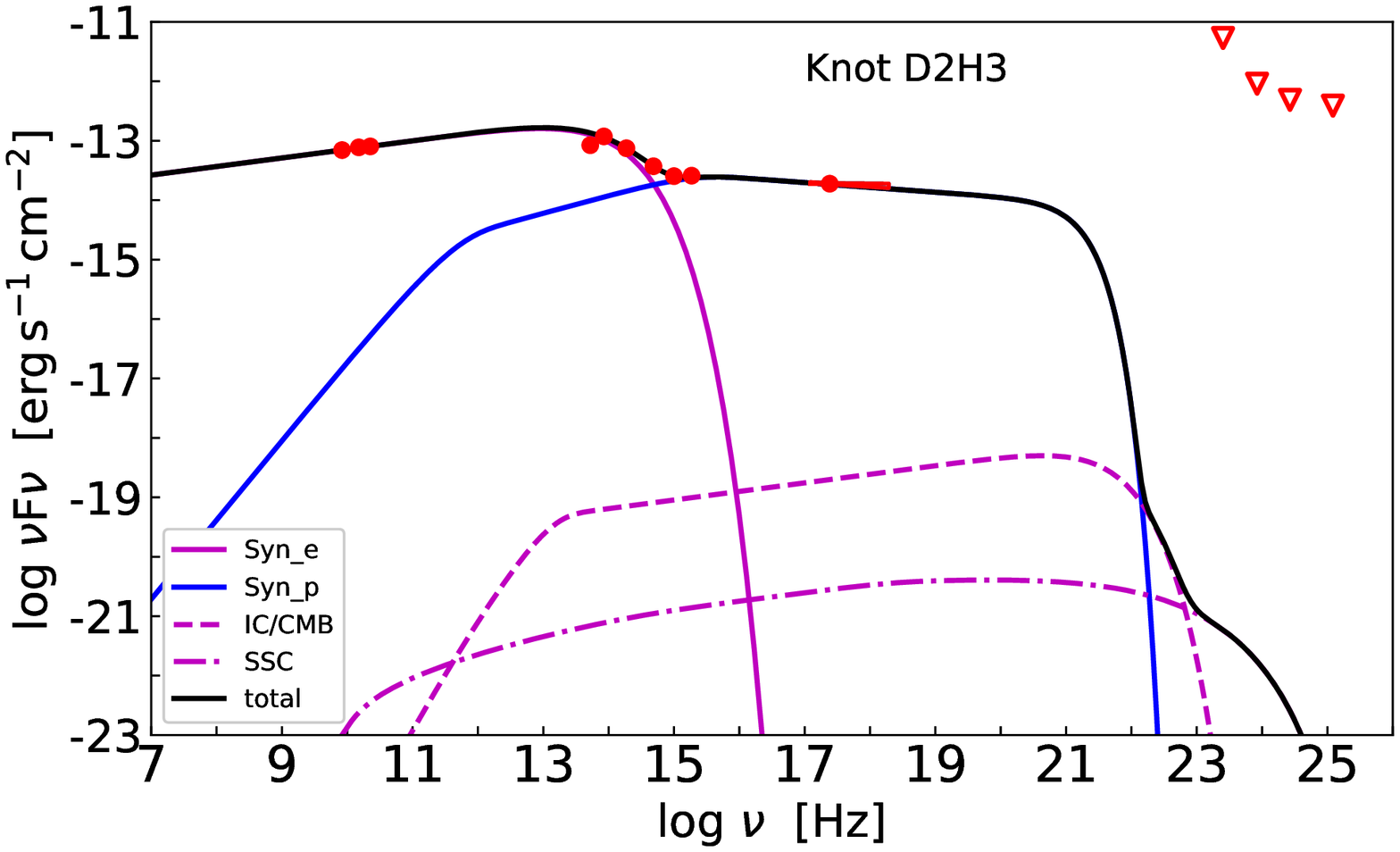}
\hfill
\includegraphics[angle=0,scale=0.26]{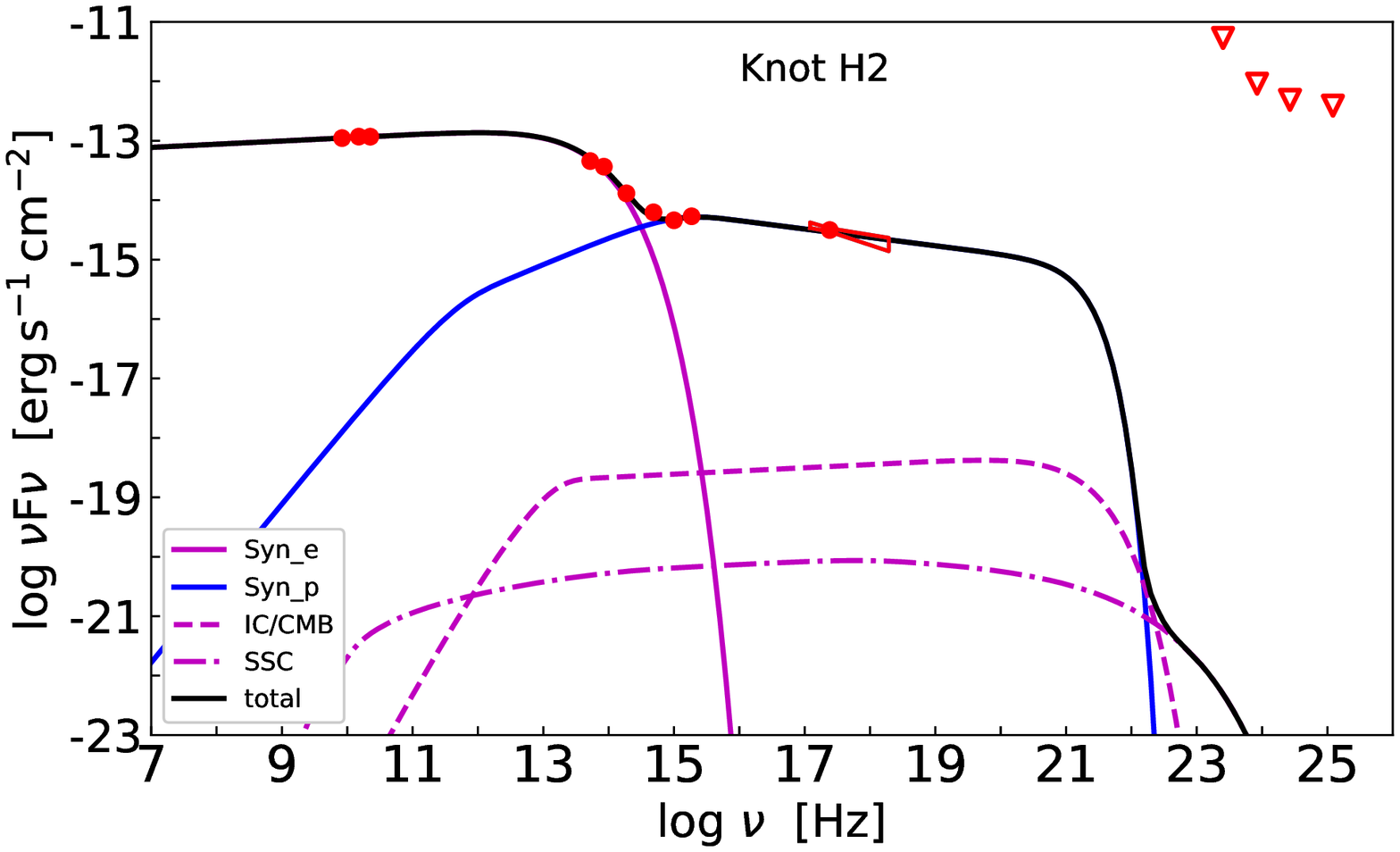}
\caption{The same as Figure \ref{single1}, but for the model fits of an electron population plus a proton population (scenario III) in case of $\delta>1$. The purple and blue lines display the radiation components from the electron population and the proton population, respectively. }\label{Two_P2}
\end{figure*}

In case of $\delta>1$, same as in the scenario II, we also adopt $\delta=3.7$, $\theta=3.3^{\circ}$, and $\Gamma=2$ in the SED modeling. Since the derived equipartition magnetic field strengths (as listed in Table 3) of knots are smaller than 1 mG, which conflicts with the condition that the energy loss of protons is dominated by the synchrotron cooling (more details to see Section 3), we take $B=5$ mG (see also Aharonian 2002) for all knots. In this case, the free parameters of the SED modeling are $A_{0}$, $p_0$, $E_{\rm c}$, $A_1$, $p_1$, $p_2$, $E_{\rm b}$. The SED fitting results are shown in Figure \ref{Two_P2} and the corresponding SSC and IC/CMB contributions of the electron population are also presented in Figure \ref{Two_P2}. The fitting parameters are given in Table 3. One can observe that the SEDs of knots can well be represented by the model either in case of $\delta=1$ or $\delta>1$.

\section{Discussion}

The above analysis shows that although the X-rays together with the UV emission of knot-A and knot-B1 can be explained with a single electron population via the IC processes (Scenario I) in both cases of $\delta=1$ and $\delta>1$, the broadband SEDs of the other knots cannot be represented in this scenario. In this scenario, the predicted $\gamma$-ray flux is normally much higher than the upper-limits set by {\em Fermi}/LAT observations, such as that seen in knot-A. This is also the reason to rule out the IC/CMB model for large-scale jet of 3C 273 (Meyer et al. 2015). The SED of knot-B1 apparently can be represented in this scenario, but the derived $B$ value is 1.83 $\mu$G in case of $\delta=1$, which is much lower than the derived equipartition magnetic field strength of $B_{\rm eq}=10.5$ mG. Zhang et al. (2018) suggested that using the SSC process to explain the X-ray emission of large-scale jet substructures would result in an extremely high jet power. We estimate the powers of the non-thermal electrons using $P_{\rm e}=\pi R^2\Gamma^2cU_{\rm e}$ (see also Zargaryan et al. 2017 for the large-scale jet knots), and find that $P_{\rm e}$ ranges from $1.9\times10^{47}$ erg s$^{-1}$ to $2.3\times10^{49}$ erg s$^{-1}$ in case of $\delta=1$ for the knots (as shown in Table 1). The black hole mass of 3C 273 is $10^{9.693}M_{\odot}$ (Gu et al. 2001) and the corresponding Eddington luminosity is $6.2\times10^{47}$ erg s$^{-1}$. $P_{\rm e}$ is $\sim10^{49}$ erg s$^{-1}$ for knot-B1, which is far beyond the Eddington luminosity of the source. In case of $\delta>1$, a large $\delta$ value ($\delta=16.6$) is required to model the SED of knot-B1. This value is comparable with the derived variability Doppler factor of $\delta=17$ for pc-scale jet in Hovatta et al. (2009), but is much larger than that reported by Liodakis et al. (2017, $\delta=3.7$), and even larger than the core-jet value of $\delta=7.4\pm0.9$ derived by SED fitting (Zhang et al. 2014, 2015). Therefore, Scenario I could not present a reasonable explanation for the SEDs of knots.

\begin{figure*}
\includegraphics[angle=0,scale=0.27]{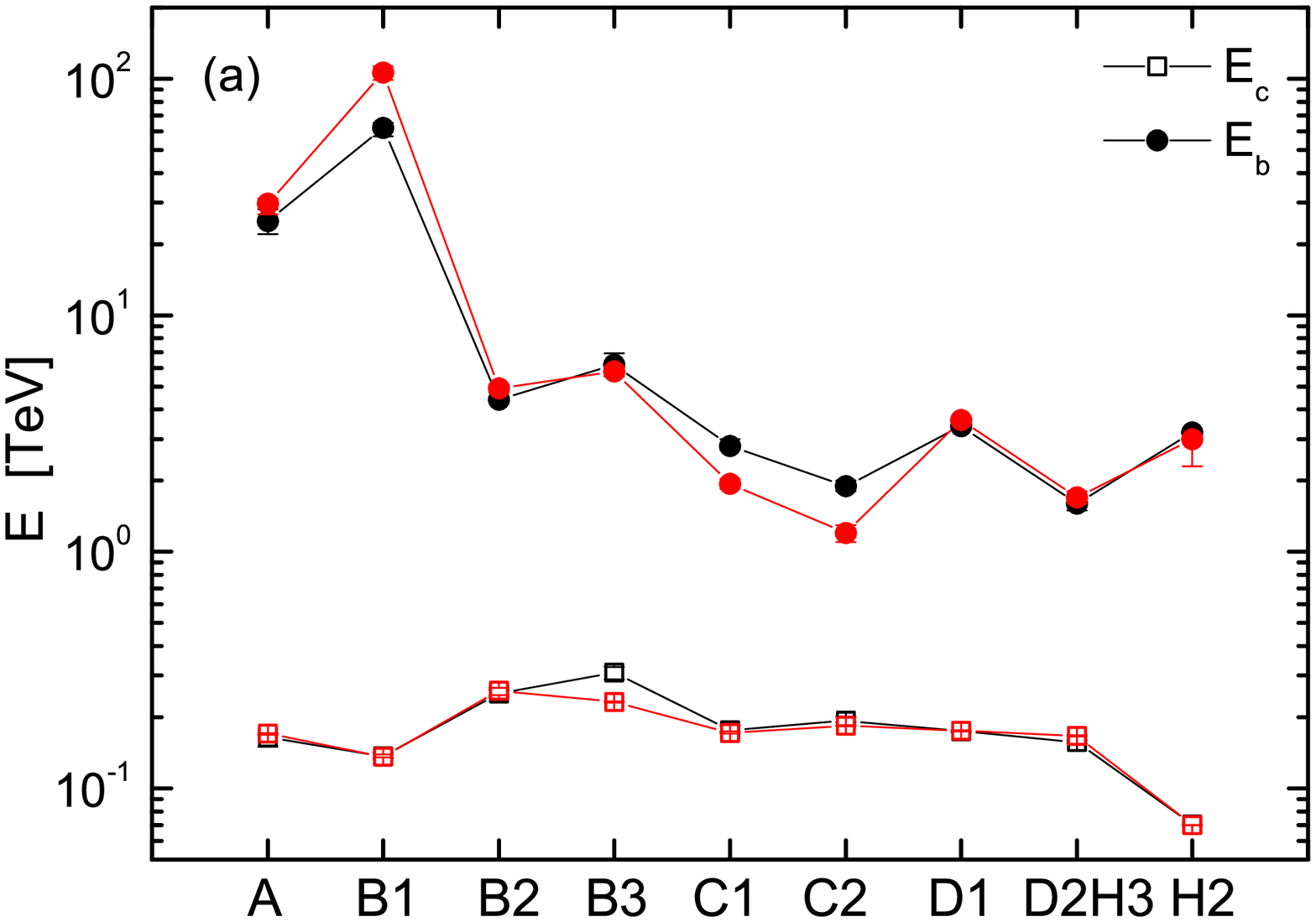}
\includegraphics[angle=0,scale=0.27]{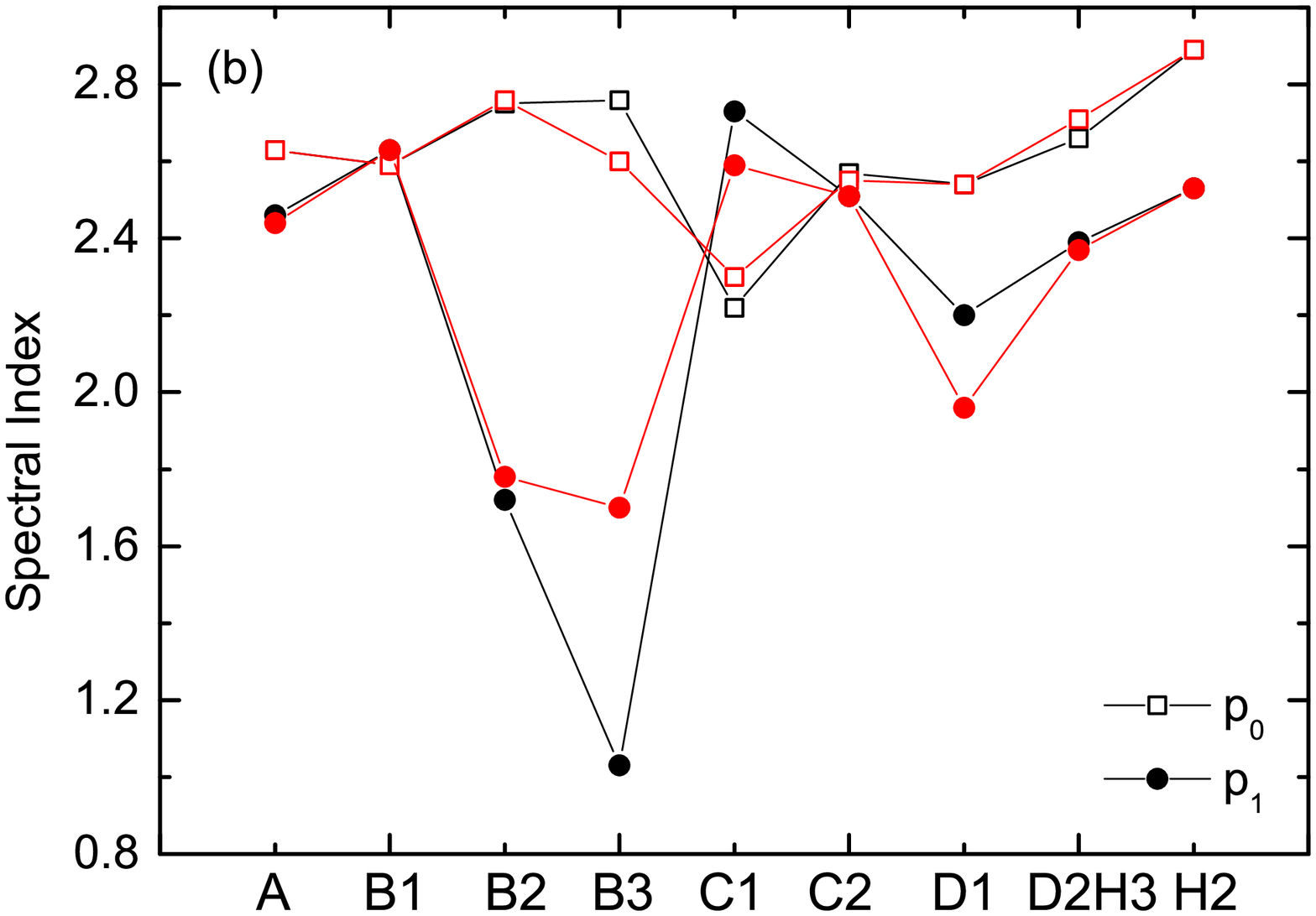}
\includegraphics[angle=0,scale=0.27]{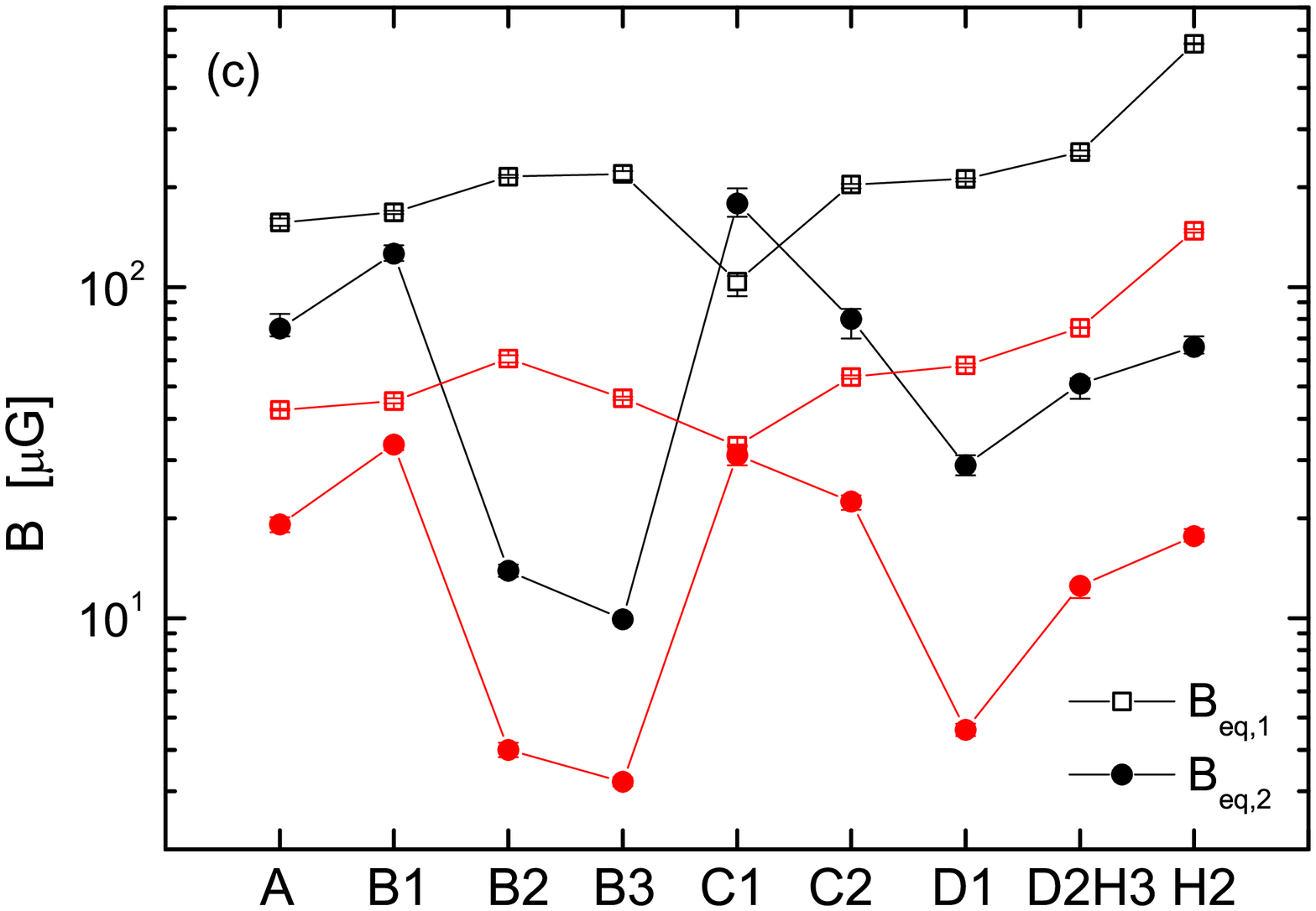}
\caption{Parameter distributions along the jet derived by the model fits of two electron populations (scenario II) in case of $\delta=1$ (black symbols) and $\delta>1$ (red symbols), including the cutoff energy ($E_{\rm c}$) of the exponential cutoff power-law electron spectrum and the break energy ($E_{\rm b}$) of the broken power-law electron spectrum, the spectral indices ($p_0$ and $p_1$) of the two electron population distributions, the equipartition magnetic field strengths ($B_{\rm eq,1}$ and $B_{\rm eq,2}$) of the two zones for each knot. }\label{Dis_E}
\end{figure*}

The scenario of two electron populations (Scenario II) can well represent the SEDs either in case of $\delta=1$ or $\delta>1$. Figure \ref{Dis_E} illustrates the distributions of the derived parameters along the jet. We do not find any evolution feature for the parameters. The medians of $E_{\rm c}$ and $E_{\rm b}$ are $\sim$170 GeV and 4 TeV, respectively, indicating that the electrons are effectively accelerated. Note that the synchrotron cooling time of electrons is $t_{\rm cool}=\frac{6\pi m_{\rm e}^2c^4}{\sigma_{\rm T}cE_{\rm e}B^2}$, and the electron travel distance can be estimated from $ct_{\rm cool}\sim3850E^{-1}_{\rm e,TeV}B^{-2}_{\rm \mu G}$ kpc, where the electron energy ($E_{\rm e,TeV}$) is in units of TeV and magnetic field strength ($B_{\rm \mu G}$) is in units of $\mu$G. The high-energy electrons cannot travel more than 1 kpc before exhausting their energies owing to synchrotron cooling, and thus the electrons should be accelerated in situ. Except for knot-B2 and knot-B3, $p_0$ and $p_1$ of other knots are roughly consistent with the particle acceleration and cooling in shocks, hence the acceleration mechanisms of the two electron populations may be the same. One can find that $B_{\rm eq,1}$ of knot-H2 is higher than
other knots. The substructure H2 in fact is historically called as a hot-spot and has the higher flux ratio of radio to X-ray than other knots. As reported by Zhang et al (2010, 2018b), the observed luminosity ratio of radio to X-ray can be an indicator to distinguish between hot-spots and knots. In this scenario, $P_{\rm e}$ is from $4.2\times10^{45}$ erg s$^{-1}$ to $3.9\times10^{46}$ erg s$^{-1}$ in case of $\delta=1$ and from $7.8\times10^{44}$ erg s$^{-1}$ to $1.2\times10^{46}$ erg s$^{-1}$ in case of $\delta>1$ among the knots. Although it has been suggested that the X-ray emission could be possibly from the synchrotron radiation of an additional shock accelerated electron population (Hardcastle 2006; Jester et al. 2006; Uchiyama et al. 2006; Zargaryan et al. 2017), it is not known what physical process can produce the second electron population in a single radiation region. Although there are some observational evidence of the complex morphologies for the knots in 3C 273 (e.g., Jester et al. 2005), no compact subcomponent, similar to in the hot-spot of Pictor A (Tingay et al. 2008; Zhang et al. 2009), is observationally resolved in the knots to support the two radiation regions. Due to the severe synchrotron cooling, the high-energy electrons that contribute the X-ray emission could not propagate far from their birth/acceleration places. In consequence, the X-ray morphologies of these knots should be like point-sources, unless there are multi compact regions inside the knots to accelerate electrons, just as the expectation of the multi-zone leptonic model.

\begin{figure*}
\includegraphics[angle=0,scale=0.27]{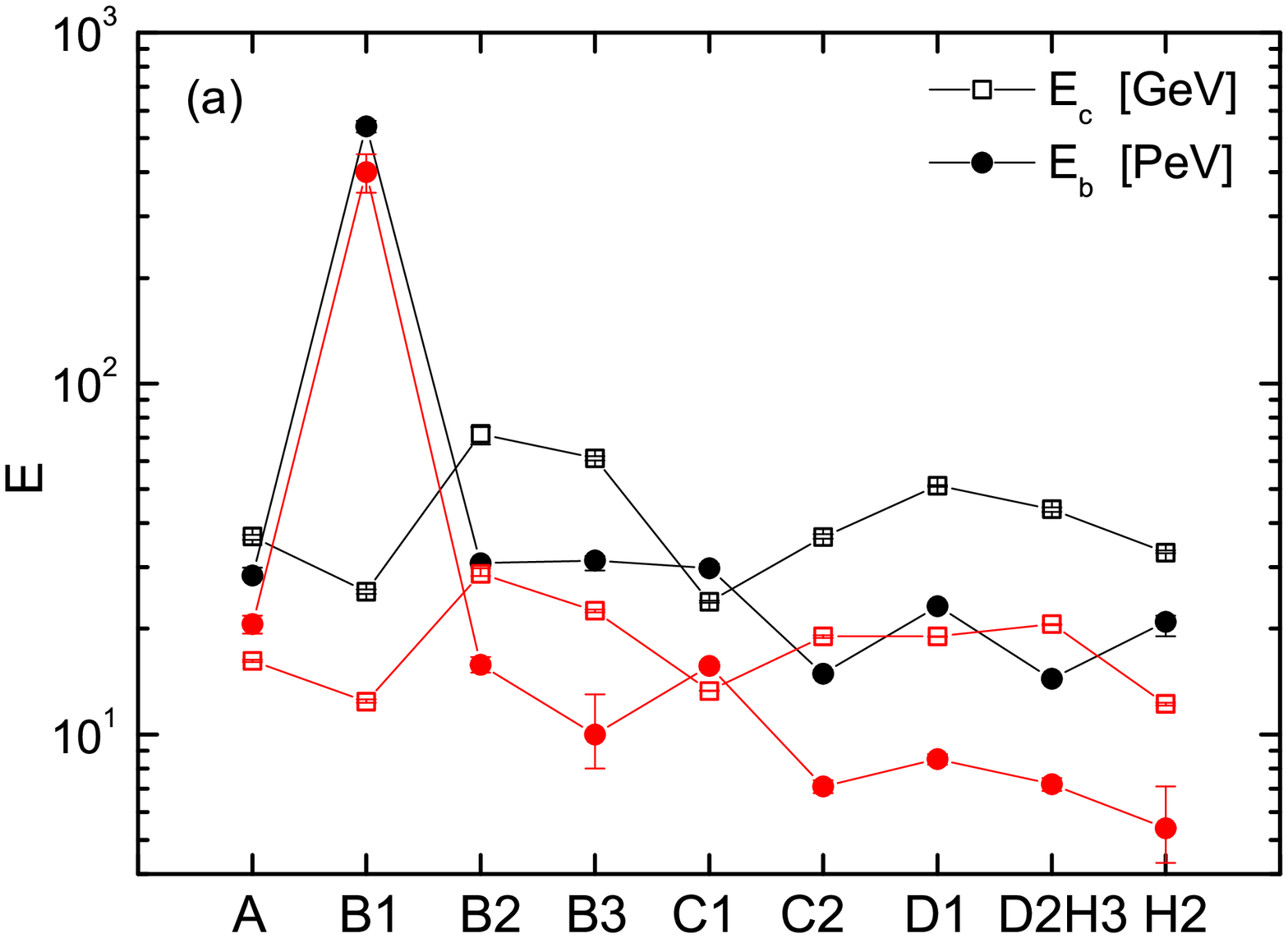}
\includegraphics[angle=0,scale=0.27]{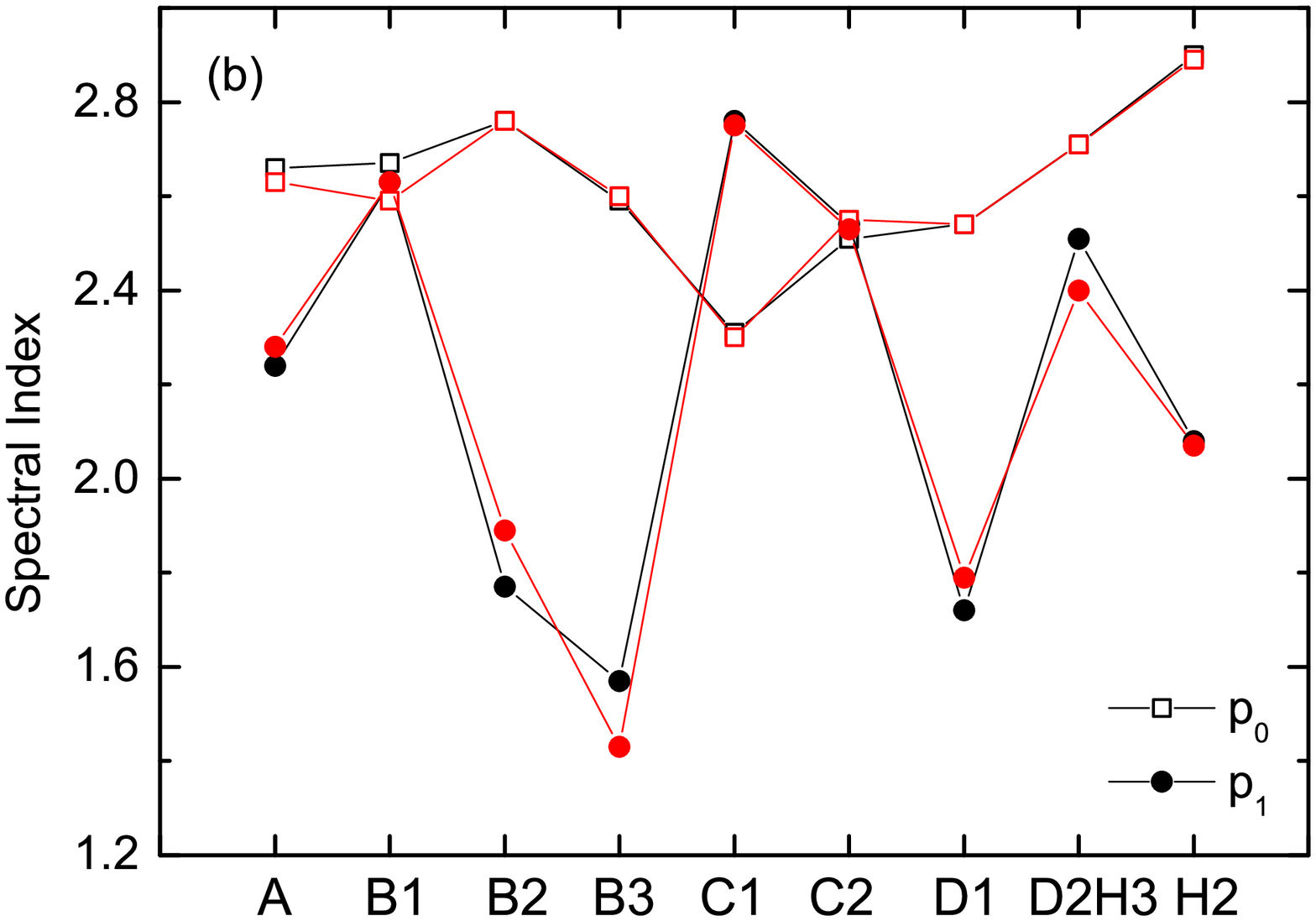}
\includegraphics[angle=0,scale=0.27]{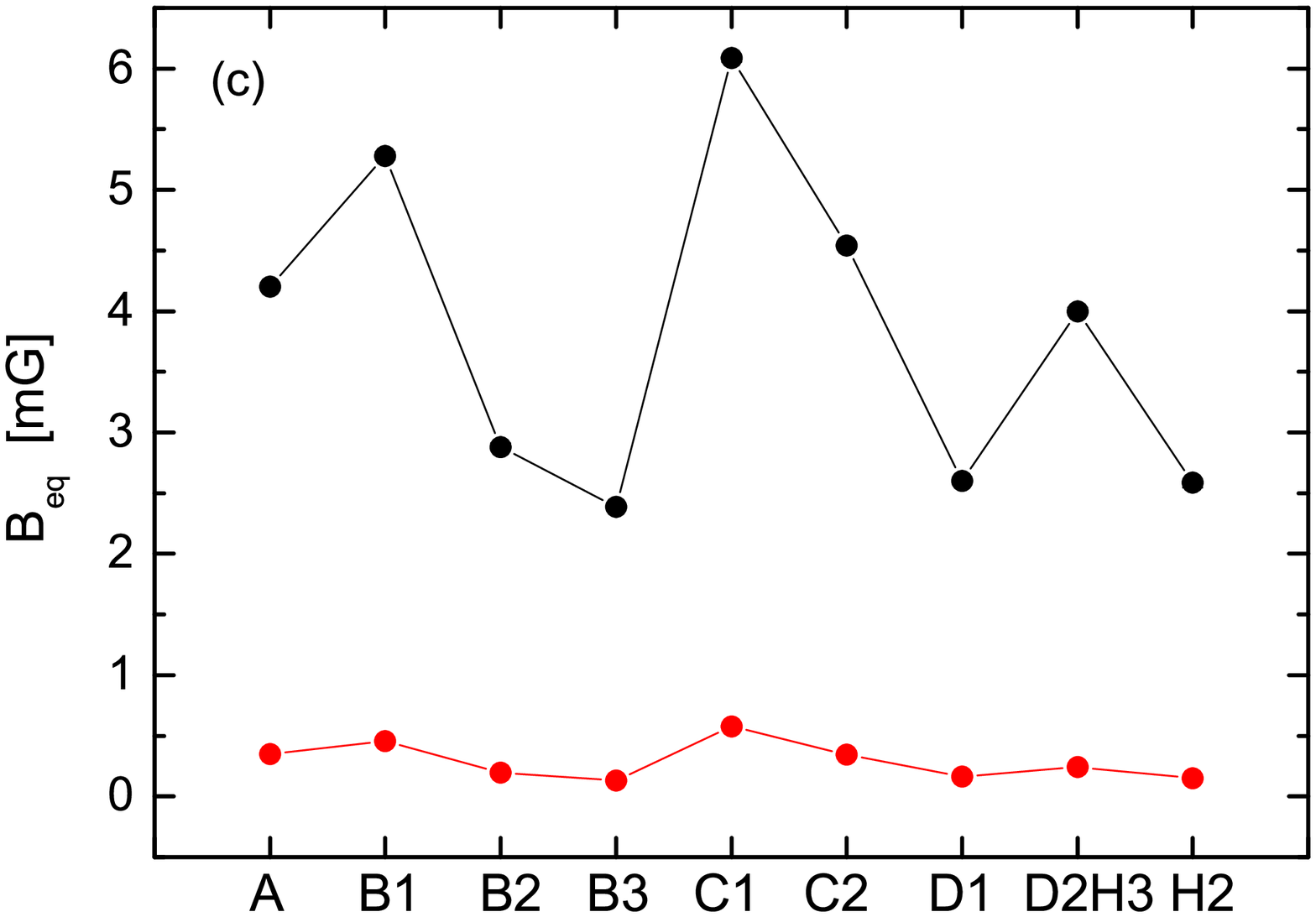}
\caption{Parameter distributions along the jet derived by the model fits of an electron population plus a proton population (scenario III) in case of $\delta=1$ (black symbols) and $\delta>1$ (red symbols), including the cutoff energy ($E_{\rm c}$) of the exponential cutoff power-law electron spectrum and the break energy ($E_{\rm b}$) of the broken power-law proton spectrum, the spectral indices of electron population ($p_0$) and proton population ($p_1$), the equipartition magnetic field strengths ($B_{\rm eq}$) for each knot.}\label{Dis_P}
\end{figure*}

The scenario of an electron population plus a proton population (Scenario III) also reproduces the SEDs of knots well. In case of $\delta=1$, $E_{\rm c}$ of the electron spectra narrowly clusters at 24--72 GeV while $E_{\rm b}$ of the proton spectra is in the range of 14--32 PeV, except for knot-B1 with $E_{\rm b}\sim$ 540 PeV, i.e., $\gamma_{\rm p}\sim10^7$, where $\gamma_{\rm p}$ is the Lorenz factors of protons. The X-ray spectrum of knot-B1 perfectly agrees with the extrapolation of its optical--UV spectrum as displayed in Figure \ref{Two_P1}, and thus a large peak energy is presented in the broadband SED and results in a large $E_{\rm b}$. In case of $\delta>1$, the derived equipartition magnetic fields of knots are smaller than 1 mG (as listed in Table 3) and $B=5$ mG is taken in our SED fitting. Hence the knots would deviate from equipartition condition if the X-ray emission is dominated by the synchrotron radiation of protons. We also illustrate the distributions of the derived parameters along the jet in Figure \ref{Dis_P} and do not find any evolution feature for the parameters. In this scenario, the powers of radiation particles are dominated by the powers of non-thermal protons ($P_{\rm p}$), which is estimated using $P_{\rm p}=\pi R^2\Gamma^2cU_{\rm p}$. $P_{\rm p}$ ranges from $5.5\times10^{47}$ erg s$^{-1}$ to $4.3\times10^{48}$ erg s$^{-1}$ in case of $\delta=1$ and ranges from $6.5\times10^{45}$ erg s$^{-1}$ to $1.3\times10^{47}$ erg s$^{-1}$ in case of $\delta>1$. It seems like that the derived jet powers in large-scale using the hadronic model parameters are super-Eddington in case of $\delta=1$. However, one can see that the total energies of radiation particles for these knots are significantly reduced in the presence of a relativistic bulk flow (see also Aharonian 2002).

For the Scenario III, the maximum proton energy of 5 EeV is required to explain the X-ray emission of the knots. Rachen \& Bierman (1993) reported that protons can be even accelerated up to 100 EeV in hot-spots with magnetic field of $\sim0.5$ mG and size of $\sim1$ kpc by the mildly relativistic jet terminal shocks. The observations with the very long baseline interferometry (VLBI) technique have demonstrated that the blazar jets are accelerated to relativistic velocities (e.g., Lister et al. 2009, 2019), and thus they could produce the strong shocks in the ambient medium. These shocks have been confirmed to propagate to kpc-scale in powerful (Nulsen et al. 2005; Simionescu et al. 2009; Gitti et al. 2010; Croston et al. 2011) and even less powerful jets (Kraft et al. 2007; Perucho et al. 2014, and references therein). Protons can also be accelerated to extremely high energy at the jet shear boundary layer (Ostrowski 1998). Hence the protons in large-scale jet knots can be accelerated up to several EeV.

Note that Ultra-High Energy Cosmic Rays (UHECRs) above 1 EeV have a predominant extragalactic origin component (e.g., Abreu et al. 2013; Aab et al. 2018), in which the energy range of $10^{18}-10^{18.5}$ eV is dominated by protons (Abbasi et al. 2017; Schr{\"o}der et al. 2019), and AGNs, especially blazars, may be the candidates of UHECR origin. The escape time and the synchrotron cooling time of the high energy protons (Aharonian 2002) respectively are
\begin{equation}
t_{\rm esc}\simeq4.2\times10^5\eta^{-1}B_{\rm mG}R^2_{\rm kpc}E^{-1}_{19}~~yr,
\end{equation}
\begin{equation}
t_{\rm syn}\simeq1.4\times10^7B^{-2}_{\rm mG}E^{-1}_{19}~~yr,
\end{equation}
where $\eta=1$ is the gyrofactor in the Bohm limit, $B_{\rm mG}$ is the magnetic field strength in units of mG, $R_{\rm kpc}$ is the size of knots in units of kpc, $E_{19}$ is the proton energy in units of $10^{19}$ eV. If the protons are cooled down by the synchrotron radiation before escape, then $B^{3}_{\rm mG}R^2_{\rm kpc}\eta^{-1}\geq34$. Taking the knot size of $\sim1$ kpc, in the Bohm regime $\eta=1$, it requires $B>3$ mG. As listed in Table 3, the derived equipartition magnetic field strengths cluster at 2.4--6.1 mG in case of $\delta=1$, and the knot sizes are larger than 1 kpc, so the high-energy protons that contribute the X-ray emission would be cooled down by the synchrotron radiation before escape. However, the derived equipartition magnetic field strengths of knots would be smaller than 1 mG in case of $\delta>1$, hence the knots would deviate from equipartition condition if the X-ray emission is dominated by the synchrotron radiation of protons. Note that the cooling times of the pp and p$\gamma$ interactions are much longer than the escape time and the proton synchrotron cooling time, and the plasma density in knots should be at least 0.1--1 cm$^{-3}$ for the effective pp interaction (Aharonian 2002), hence the synchrotron radiation of secondary electrons is not considered in this paper.

The X-ray emission of knot-A were previously explained with the proton synchrotron radiations (Aharonian 2002; Kundu \& Gupta 2014). Aharonian (2002) assumed the continuous injection of relativistic protons with a constant rate during the jet age ($3\times10^7$ yr; see also Kundu \& Gupta 2014). They derived a total energy of protons of $\sim10^{60}\sim 10^{62}$ erg and the corresponding proton acceleration rate is $L_{\rm p}=10^{45}\sim 10^{47}$. Kundu \& Gupta (2014) proposed that the X-ray fluxes of knot-A originate from the proton synchrotron radiation ($\delta=1$) and obtained $L_{\rm p}\sim10^{43}-10^{44}$ erg s$^{-1}$ by assuming the jet age of $1.4\times10^7$ yr. Recently, Kusunose \& Takahara (2018) used a photo-hadronic model to explain the X-ray emission of knot-A with the proton energy of $10^{61}-10^{62}$ erg. They estimated the proton power with $L_{\rm p}\sim E_{\rm p, total}/(3R/c)\sim 10^{50}$ erg s$^{-1}$, where $3R/c$ is the escape time of protons, and reported that the proton power can be lowed down to nearly Eddington power if the core photons are more beamed toward the X-ray knots than toward to the line of sight. Chen (2018) also reported that the jet powers of some blazars may be super-Eddington. These results are consistent with ours. However, the super-Eddington issue could be relaxed by cutting down the low-energy protons, which are incapable to produce the observable photon flux, and might be avoided in case of $\delta>1$ (see also Aharonian 2002).

As illustrated in Figure \ref{Two_E1}, the predicted fluxes in the GeV--TeV band by the SSC and IC/CMB processes of the two electron populations are low, even in case of $\delta>1$ (Figure \ref{Two_E2}), the predicted fluxes in the $\gamma$-ray band still cannot be detected by the available $\gamma$-ray detectors. As reported by Aharonian (2002), if the X-ray emission of large-scale jet is really dominated by the proton synchrotron, the most energetic protons with energy larger than $10^{19}$ eV may eventually escape the jet, which would result in different $\gamma$-ray emission characters depending on the magnetic field strength of the environment. Hence, the detection of the $\gamma$-ray emission around the knots would help to exclude and constrain the radiation mechanism of the X-ray emission in large-scale jet.

\section{Summary}

Based on a comprehensively theoretical analysis on the SED modeling, we suggested that the SEDs from radio to X-ray bands of the large-scale jet knots in 3C 273 cannot be represented well with the synchrotron, SSC, and IC/CMB radiations of a single electron population when the Doppler boosting effect is either considered or not. We then considered two synchrotron radiation components to explain the broadband SEDs of knots, from two independent electron populations (the leptonic model) or from an electron population plus a proton population (the hadronic model). Both models can represent the broadband SEDs of knots well. However, there is no observational evidence for different zones to accelerate two distinct electron populations. Especially the electrons with high-energy that contribute the X-ray emission have very short cooling time, and thus it is hard to explain the observed X-ray morphologies of knots. In this respect, protons lose energy very slowly, and thus their acceleration sites do not have to be within the emission sites. The proton synchrotron radiation model would result in the high jet powers, the so-called ``super-Eddington" jet powers. However, this issue might be avoided if these knots have relativistic motion towards the observer (see also Aharonian 2002). We also note that the knots may deviate the equipartition condition if the X-ray emission is dominated by the synchrotron radiation of protons in case of $\delta>1$. The predicted $\gamma$-ray fluxes by the leptonic model through the IC process are very low, so the detection of the $\gamma$-ray emission from the knots in the future would be help to exam models. Multiwavelength polarimetry observations with high resolution together with the constrains on the magnetic field strength by Faraday rotation measurements, may shed light on the issue of particle acceleration and the emitting volume of particles. Further observations are required to confirm or rule out different models.

\acknowledgments
We thank the anonymous referee for the valuable suggestions. This work is supported by the National Natural Science Foundation of China (grants 11973050, 11573034, 11533003, 11851304, and U1731239). En-Wei Liang acknowledges support from the special funding from the Guangxi Science Foundation for Guangxi distinguished professors (grant 2017AD22006 for Bagui Yingcai \& Bagui Xuezhe).

\clearpage

\clearpage

\clearpage

\begin{deluxetable}{lccccccccccccccc}
\tabletypesize{\scriptsize }\rotate\tablecolumns{15}\tablewidth{40pc}
\setlength{\tabcolsep}{1.mm}
\tablecaption{Parameters in the Scenario I: a Single Electron Population}
\tablenum{1}\tablehead{\colhead{Knot}&\colhead{{$A_{1}$}}&\colhead{{$E_{\rm b}$}}&\colhead{$p_{1}$}&\colhead{$p_{2}$}&\colhead{$E_{\rm total}$}&\colhead{{$\Gamma$}}&\colhead{{$\delta$}}
&\colhead{$B$\tablenotemark{a}}&\colhead{$B_{\rm eq}$}&\colhead{{$\log P_{\rm e}$}}
&\colhead{$R$}\\
&\colhead{\tiny{$[\rm 1/eV]$}}&\colhead{\tiny{[GeV]}}&\colhead{}&\colhead{}&\colhead{$[\rm erg]$}
&\colhead{}&\colhead{}&\colhead{\tiny{[$\mu$G]}}&\colhead{\tiny{[$\mu$G]}}&\colhead{\tiny{[erg s$^{-1}$]}}&\colhead{[arcsec]}}
\startdata
\scriptsize{\textbf{$\delta=1$}}\\
\hline
A &3.1$^{+0.5}_{-0.4}$E45&1230$^{+70}_{-170}$&2.47$^{+0.02}_{-0.02}$&5.0$^{+0.4}_{-0.4}$&6.9$^{+0.6}_{-0.6}$E60&1&1&0.65$^{+0.03}_{-0.03}$&11600$^{+500}_{-500}$&49.36&0.8\\
B1&2.5$^{+0.3}_{-0.3}$E44&900$^{+90}_{-90}$&2.59$^{+0.01}_{-0.01}$&5.6$^{+0.5}_{-0.5}$&2.4$^{+0.1}_{-0.2}$E60&1&1&1.8$^{+0.1}_{-0.1}$&10500$^{+200}_{-400}$&49.02&0.6\\
B2&3.5$^{+1.3}_{-0.6}$E45&490$^{+60}_{-40}$&2.28$^{+0.03}_{-0.04}$&3.76$^{+0.09}_{-0.06}$&9.9$^{+1.1}_{-1.1}$E59&1&1&1.0$^{+0.1}_{-0.1}$&4400$^{+200}_{-200}$&48.52&0.8\\
B3&2.3$^{+0.9}_{-0.4}$E45&510$^{+200}_{-50}$&2.10$^{+0.03}_{-0.05}$&3.93$^{+0.40}_{-0.11}$&1.1$^{+0.1}_{-0.1}$E59&1&1&1.1$^{+0.1}_{-0.1}$&2290$^{+60}_{-80}$&47.70&0.6\\
C1&5.8$^{+1.1}_{-1.1}$E45&575$^{+70}_{-15}$&1.90$^{+0.02}_{-0.02}$&4.51$^{+0.09}_{-0.03}$&6.6$^{+1.1}_{-0.3}$E58&1&1&1.5$^{+0.1}_{-0.2}$&1740$^{+150}_{-40}$&47.47&0.6\\
C2&7.2$^{+0.7}_{-1.2}$E45&380$^{+100}_{-40}$&1.89$^{+0.02}_{-0.02}$&4.25$^{+0.30}_{-0.11}$&7.4$^{+1.5}_{-0.7}$E58&1&1&2.1$^{+0.2}_{-0.2}$&1470$^{+140}_{-70}$&47.45&0.7\\
D1&9.9$^{+1.5}_{-1.0}$E45&374$^{+15}_{-15}$&1.76$^{+0.02}_{-0.02}$&4.73$^{+0.06}_{-0.06}$&5.5$^{+0.3}_{-0.2}$E58&1&1&2.8$^{+0.1}_{-0.1}$&1260$^{+40}_{-30}$&47.32&0.7\\
D2H3&1.7$^{+0.3}_{-0.3}$E46&217$^{+30}_{-9}$&1.72$^{+0.03}_{-0.03}$&4.49$^{+0.19}_{-0.06}$&7.0$^{+0.7}_{-0.3}$E58&1&1&4.4$^{+0.2}_{-0.3}$&834$^{+40}_{-17}$&47.27&1.0\\
H2&4.0$^{+0.8}_{-0.5}$E43&25$^{+8}_{-8}$&2.25$^{+0.03}_{-0.03}$&4.2$^{+0.2}_{-0.2}$&7.7$^{+1.1}_{-1.1}$E57&1&1&72.4$^{+1.1}_{-1.1}$&470$^{+30}_{-30}$&46.47&0.7\\
\hline
\scriptsize{\textbf{$\delta>1$}}\\
\hline
A&6.8$^{+1.2}_{-0.8}$E38&112$^{+19}_{-10}$&2.54$^{+0.01}_{-0.02}$&4.05$^{+0.40}_{-0.17}$&3.5$^{+0.2}_{-0.2}$E54&22.4$^{+1.9}_{-1.9}$&17$^{+0.3}_{-0.4}$&8.3$^{+0.2}_{-0.2}$&8.3$^{+0.2}_{-0.2}$&45.77&0.8\\
B1&4.5$^{+0.4}_{-0.4}$E38&72$^{+7}_{-7}$&2.53$^{+0.01}_{-0.01}$&3.99$^{+0.11}_{-0.11}$&2.1$^{+0.1}_{-0.1}$E54&12.6$^{+0.3}_{-0.2}$&16.6$^{+0.1}_{-0.1}$&9.8$^{+0.2}_{-0.2}$&9.8$^{+0.2}_{-0.2}$&45.16&0.6\\
B2&7.8$^{+1.2}_{-0.8}$E38&81$^{+9}_{-9}$&2.55$^{+0.02}_{-0.02}$&3.94$^{+0.13}_{-0.13}$&4.6$^{+0.2}_{-0.2}$E54&12.3$^{+0.1}_{-0.1}$&16.5$^{+0.1}_{-0.1}$&9.5$^{+0.3}_{-0.3}$&9.5$^{+0.3}_{-0.3}$&45.36&0.8\\
B3&1.3$^{+0.3}_{-0.1}$E40&59$^{+6}_{-6}$&2.08$^{+0.01}_{-0.04}$&3.91$^{+0.12}_{-0.12}$&4.6$^{+0.2}_{-0.5}$E53&13.6$^{+1.3}_{-0.5}$&16.9$^{+0.3}_{-0.2}$&4.6$^{+0.1}_{-0.3}$&4.6$^{+0.1}_{-0.3}$&44.58&0.6\\
C1&4.2$^{+0.3}_{-0.2}$E39&104$^{+1}_{-1}$&2.31$^{+0.01}_{-0.01}$&4.57$^{+0.04}_{-0.01}$&1.6$^{+0.1}_{-0.1}$E54&10.2$^{+0.2}_{-0.1}$&15.2$^{+0.2}_{-0.1}$&8.5$^{+0.1}_{-0.3}$&8.5$^{+0.1}_{-0.3}$&44.85&0.6\\
C2&3.2$^{+0.4}_{-0.3}$E40&67$^{+2}_{-5}$&2.12$^{+0.01}_{-0.02}$&4.24$^{+0.04}_{-0.09}$&1.6$^{+0.1}_{-0.1}$E54&10.6$^{+0.4}_{-0.2}$&15.5$^{+0.3}_{-0.1}$&6.9$^{+0.2}_{-0.3}$&6.9$^{+0.2}_{-0.3}$&44.84&0.7\\
D1&2.2$^{+0.3}_{-0.2}$E40&86$^{+2}_{-1}$&2.23$^{+0.01}_{-0.02}$&4.76$^{+0.07}_{-0.03}$&3.3$^{+0.1}_{-0.2}$E54&8.3$^{+0.2}_{-0.1}$&13.5$^{+0.2}_{-0.1}$&9.8$^{+0.2}_{-0.3}$&9.8$^{+0.2}_{-0.3}$&44.94&0.7\\
D2H3&2.1$^{+0.2}_{-0.2}$E41&53$^{+3}_{-6}$&2.09$^{+0.02}_{-0.02}$&4.34$^{+0.06}_{-0.14}$&8.2$^{+0.5}_{-0.7}$E54&7.2$^{+0.2}_{-0.1}$&12.3$^{+0.3}_{-0.2}$&9.0$^{+0.3}_{-0.4}$&9.0$^{+0.3}_{-0.4}$&45.05&1.0\\
H2&2.5$^{+0.4}_{-0.3}$E41&16$^{+7}_{-5}$&2.19$^{+0.02}_{-0.03}$&4.3$^{+0.3}_{-0.2}$&2.5$^{+0.2}_{-0.4}$E55&3.7$^{+0.2}_{-0.1}$&7.0$^{+0.3}_{-0.2}$&26.9$^{+1.0}_{-2.0}$&26.9$^{+1.0}_{-2.0}$&45.12&0.7\\
\enddata
\tablenotetext{a}{$B<B_{\rm eq}$ in case of $\delta=1$ and $B=B_{\rm eq}$ in case of $\delta>1$, more details to see the text. }
\end{deluxetable}

\begin{deluxetable}{lccccccccccccccc}
\tabletypesize{\scriptsize }\rotate\tablecolumns{20}\tablewidth{40pc}
\setlength{\tabcolsep}{1.mm}
\tablecaption{Parameters in the Scenario II: Two Electron Populations}
\tablenum{2}\tablehead{\colhead{Knot}&\colhead{{$A_{0}$}}&\colhead{{$E_{\rm c}$}}&\colhead{{$p_{0}$\tablenotemark{a}}}&\colhead{{$A_{1}$}}&\colhead{{$E_{\rm b}$}}&\colhead{{$p_{1}$}}&\colhead{{$p_{2}$\tablenotemark{a}}}&\colhead{{$E_{\rm total}$}}&\colhead{{$B_{\rm eq, 1}$}}&\colhead{{$B_{\rm eq, 2}$}}&\colhead{{$\log P_{\rm e}$}}\\
\colhead{}&\colhead{\tiny{$[\rm 1/eV]$}}&\colhead{\tiny{[GeV]}}&\colhead{}&\colhead{\tiny{$[\rm 1/eV]$}}&\colhead{\tiny{[TeV]}}&\colhead{}&\colhead{}&\colhead{\tiny{$[\rm erg]$}}&\colhead{[$\mu$G]}&\colhead{[$\mu$G]}&\colhead{\tiny{[erg s$^{-1}$]}}}
\startdata
\scriptsize{\textbf{$\delta=1$}}\\
\hline
A&8.2$^{+0.7}_{-1.2}$E40&164$^{+5}_{-14}$&2.63&1.5$^{+0.1}_{-0.3}$E41&25$^{+3}_{-3}$&2.46$^{+0.03}_{-0.02}$&2.91$^{+0.19}_{-0.15}$&1.6$^{+0.1}_{-0.1}$E57&157$^{+4}_{-4}$&75$^{+8}_{-4}$&45.71\\
B1&6.5$^{+0.4}_{-0.4}$E40&137$^{+5}_{-3}$&2.59&2.2$^{+0.2}_{-0.2}$E40&62$^{+3}_{-5}$&2.63$^{+0.02}_{-0.01}$&3.15&9.6$^{+0.4}_{-0.4}$E56&168$^{+2}_{-2}$&126$^{+8}_{-6}$&45.63\\
B2&3.7$^{+0.2}_{-0.2}$E40&252$^{+9}_{-9}$&2.75&8.5$^{+0.3}_{-0.3}$E41&4.4$^{+0.1}_{-0.1}$&1.72$^{+0.06}_{-0.1}$&2.83$^{+0.04}_{-0.02}$&2.4$^{+0.1}_{-0.1}$E57&216$^{+2}_{-2}$&14$^{+1}_{-1}$&45.90\\
B3&1.4$^{+0.3}_{-0.2}$E40&308$^{+19}_{-19}$&2.76&1.3$^{+0.1}_{-0.1}$E41&6.2$^{+0.7}_{-0.7}$&1.03$^{+0.06}_{-0.06}$&3.26$^{+0.11}_{-0.08}$&1.1$^{+0.1}_{-0.1}$E57&220$^{+4}_{-9}$&9.9$^{+0.1}_{-0.1}$&45.67\\
C1&1.7$^{+0.1}_{-0.2}$E42&176$^{+8}_{-4}$&2.22&1.3$^{+0.1}_{-0.1}$E40&2.8$^{+0.2}_{-0.1}$&2.73$^{+0.02}_{-0.02}$&3.26$^{+0.10}_{-0.10}$&9.2$^{+1.5}_{-1.1}$E56&104$^{+4}_{-10}$&179$^{+20}_{-16}$&45.62\\
C2&1.9$^{+0.1}_{-0.1}$E41&193$^{+8}_{-3}$&2.57&6.2$^{+0.1}_{-0.3}$E40&1.9$^{+0.1}_{-0.1}$&2.51$^{+0.02}_{-0.02}$&3.04$^{+0.04}_{-0.08}$&1.6$^{+0.1}_{-0.1}$E57&204$^{+1}_{-5}$&80$^{+6}_{-10}$&45.80\\
D1&3.1$^{+0.1}_{-0.1}$E41&174$^{+1}_{-2}$&2.54&2.4$^{+0.1}_{-0.1}$E41&3.4$^{+0.1}_{-0.1}$&2.20$^{+0.02}_{-0.01}$&3.14$^{+0.08}_{-0.03}$&1.6$^{+0.1}_{-0.1}$E57&212$^{+1}_{-4}$&29$^{+2}_{-2}$&45.78\\
D2H3&3.0$^{+0.2}_{-0.3}$E41&157$^{+3}_{-2}$&2.66&2.8$^{+0.1}_{-0.1}$E41&1.6$^{+0.1}_{-0.1}$&2.39$^{+0.01}_{-0.02}$&3.19$^{+0.08}_{-0.06}$&6.8$^{+0.2}_{-0.4}$E57&256$^{+3}_{-7}$&51$^{+2}_{-5}$&46.26\\
H2&2.7$^{+0.1}_{-0.1}$E40&70.2$^{+0.3}_{-0.2}$&2.89&3.4$^{+0.1}_{-0.4}$E40&3.2$^{+0.1}_{-0.2}$&2.53$^{+0.02}_{-0.01}$&3.56$^{+0.05}_{-0.05}$&1.0$^{+0.1}_{-0.1}$E58&544$^{+2}_{-2}$&66$^{+5}_{-3}$&46.59\\
\hline
\scriptsize{\textbf{$\delta=3.7$}}\\
\hline
A&6.1$^{+0.1}_{-0.1}$E39&170$^{+1}_{-1}$&2.63&1.2$^{+0.1}_{-0.1}$E40&30$^{+2}_{-3.0}$&2.44$^{+0.02}_{-0.02}$&3.18$^{+0.2}_{-0.14}$&1.1$^{+0.1}_{-0.1}$E56&42.6$^{+0.2}_{-0.2}$&19.2$^{+1.0}_{-1.0}$&45.17\\
B1&4.7$^{+0.2}_{-0.1}$E39&136$^{+3}_{-1}$&2.59&1.6$^{+0.1}_{-0.1}$E39&106$^{+7}_{-7}$&2.63$^{+0.01}_{-0.01}$&3.15&6.8$^{+0.3}_{-0.3}$E55&45.2$^{+0.9}_{-0.6}$&33.4$^{+0.8}_{-1.2}$&45.08\\
B2&2.5$^{+0.1}_{-0.1}$E39&259$^{+7}_{-11}$&2.76&6.4$^{+0.2}_{-0.2}$E40&4.9$^{+0.2}_{-0.2}$&1.78$^{+0.05}_{-0.05}$&2.88$^{+0.04}_{-0.04}$&1.9$^{+0.1}_{-0.1}$E56&60.7$^{+1.5}_{-1.5}$&4.0$^{+0.2}_{-0.2}$&45.40\\
B3&4.3$^{+0.1}_{-0.1}$E39&232$^{+1}_{-1}$&2.6&2.0$^{+0.1}_{-0.1}$E40&5.8$^{+0.3}_{-0.2}$&1.70$^{+0.04}_{-0.06}$&3.18$^{+0.02}_{-0.02}$&4.6$^{+0.1}_{-0.1}$E55&46.2$^{+0.5}_{-0.5}$&3.2$^{+0.1}_{-0.1}$&44.92\\
C1&7.3$^{+0.1}_{-0.1}$E40&171$^{+3}_{-1}$&2.3&2.1$^{+0.4}_{-0.3}$E39&1.9$^{+0.1}_{-0.1}$&2.59$^{+0.03}_{-0.03}$&3.14$^{+0.02}_{-0.02}$&4.4$^{+0.3}_{-0.3}$E55&33.2$^{+0.2}_{-0.3}$&31$^{+2}_{-2}$&44.89\\
C2&1.7$^{+0.1}_{-0.1}$E40&184$^{+1}_{-1}$&2.55&4.7$^{+0.1}_{-0.1}$E39&1.2$^{+0.1}_{-0.1}$&2.51$^{+0.01}_{-0.01}$&3.00$^{+0.01}_{-0.02}$&1.2$^{+0.1}_{-0.1}$E56&53.5$^{+0.7}_{-0.7}$&22.5$^{+1.0}_{-1.3}$&45.25\\
D1&2.3$^{+0.1}_{-0.1}$E40&175$^{+1}_{-1}$&2.54&3.6$^{+0.1}_{-0.1}$E40&3.6$^{+0.1}_{-0.1}$&1.96$^{+0.02}_{-0.03}$&3.11$^{+0.01}_{-0.01}$&1.2$^{+0.1}_{-0.1}$E56&58.1$^{+0.7}_{-1.0}$&4.6$^{+0.2}_{-0.2}$&45.25\\
D2H3&1.4$^{+0.1}_{-0.1}$E40&169$^{+1}_{-1}$&2.71&2.1$^{+0.1}_{-0.1}$E40&1.7$^{+0.1}_{-0.1}$&2.37$^{+0.01}_{-0.02}$&3.17$^{+0.06}_{-0.02}$&5.8$^{+0.1}_{-0.1}$E56&75.4$^{+0.1}_{-0.1}$&12.5$^{+0.3}_{-1.0}$&45.79\\
H2&1.9$^{+0.1}_{-0.1}$E39&70$^{+0.1}_{-0.2}$&2.89&2.4$^{+0.1}_{-0.2}$E39&3.0$^{+0.1}_{-0.7}$&2.53$^{+0.02}_{-0.01}$&3.55$^{+0.08}_{-0.08}$&7.6$^{+0.2}_{-0.2}$E56&148$^{+2}_{-2}$&17.7$^{+0.9}_{-0.7}$&46.07\\
\enddata
\tablenotetext{a}{$p_0$ is not fixed during SED fits, but the derived errors are $\sim$0.01 and thus are not shown in the Table. $p_2$ of knot-B1 is fixed as the average value of other knots. }
\end{deluxetable}
\clearpage

\begin{deluxetable}{lccccccccccccccc}
\tabletypesize{\scriptsize }\rotate\tablecolumns{20}\tablewidth{46pc}
\setlength{\tabcolsep}{1.mm}\tablecaption{Parameters in the Scenario III: an Electron Population plus a Proton Population}
\tablenum{3}\tablehead{\colhead{Knot}&\colhead{{$A_{0}$}}&\colhead{{$E_{\rm c}$}}&\colhead{{$p_{0}$\tablenotemark{a}}}&\colhead{{$A_{1}$}}&\colhead{{$E_{\rm b}$}}&\colhead{{$p_{1}$}}&\colhead{{$p_{2}$\tablenotemark{a}}}&\colhead{{$E_{\rm e, total}$}}&\colhead{{$E_{\rm p, total}$}}&\colhead{$B$\tablenotemark{b}}&\colhead{{$B_{\rm eq}$}}&\colhead{{$\log P_{\rm e}$}}&\colhead{{$\log P_{\rm p}$}}\\
\colhead{}&\colhead{\tiny{$[\rm 1/eV]$}}&\colhead{\tiny{[GeV]}}&\colhead{}&\colhead{\tiny{$[\rm 1/eV]$}}&\colhead{\tiny{[PeV]}}&\colhead{}&\colhead{}&\colhead{\tiny{$[\rm erg]$}}&\colhead{\tiny{$[\rm erg]$}}
&\colhead{[$\mu$G]}&\colhead{[$\mu$G]}&\colhead{\tiny{[erg s$^{-1}$]}}&\colhead{\tiny{[erg s$^{-1}$]}}}
\startdata
\scriptsize{\textbf{$\delta=1$}}\\
\hline
A&2.0$^{+0.1}_{-0.1}$E38&36.6$^{+0.5}_{-0.7}$&2.66&4.9$^{+0.1}_{-0.2}$E47&28.4$^{+1.5}_{-0.7}$&2.24$^{+0.01}_{-0.01}$&2.62$^{+0.03}_{-0.02}$&9.5$^{+0.1}_{-0.1}$E53&9.1$^{+0.1}_{-0.1}$E59&4204$^{+10}_{-15}$&4204$^{+10}_{-15}$&42.50&48.48\\
B1&9.1$^{+0.7}_{-0.5}$E37&25.5$^{+0.4}_{-0.4}$&2.67&4.4$^{+0.2}_{-0.2}$E48&540$^{+20}_{-20}$&2.63$^{+0.01}_{-0.01}$&3.07&4.8$^{+0.1}_{-0.1}$E53&6.0$^{+0.1}_{-0.1}$E59&5280$^{+30}_{-30}$&5280$^{+30}_{-30}$&42.33&48.43\\
B2&2.7$^{+0.1}_{-0.1}$E38&72$^{+3}_{-5}$&2.76&5.7$^{+0.2}_{-0.2}$E45&30.7$^{+0.6}_{-0.4}$&1.77$^{+0.01}_{-0.01}$&2.93$^{+0.03}_{-0.02}$&1.9$^{+0.1}_{-0.1}$E51&4.3$^{+0.1}_{-0.1}$E59&2880$^{+30}_{-20}$&2880$^{+30}_{-20}$&39.80&48.15\\
B3&5.4$^{+0.2}_{-0.2}$E38&61.2$^{+0.8}_{-0.8}$&2.59&3.1$^{+0.1}_{-0.1}$E44&31.3$^{+1.0}_{-2.0}$&1.57$^{+0.01}_{-0.01}$&3.21$^{+0.04}_{-0.05}$&7.2$^{+0.1}_{-0.1}$E50&1.2$^{+0.1}_{-0.1}$E59&2388$^{+8}_{-11}$&2388$^{+8}_{-11}$&39.50&47.74\\
C1&1.6$^{+0.1}_{-0.1}$E39&23.9$^{+0.2}_{-0.2}$&2.31&1.3$^{+0.1}_{-0.1}$E49&29.7$^{+0.9}_{-0.4}$&2.76$^{+0.01}_{-0.01}$&3.16$^{+0.01}_{-0.01}$&2.7$^{+0.1}_{-0.1}$E53&8.0$^{+0.1}_{-0.1}$E59&6086$^{+16}_{-30}$&6086$^{+16}_{-30}$&42.08&48.55\\
C2&1.2$^{+0.1}_{-0.1}$E39&36.5$^{+0.7}_{-0.4}$&2.51&2.9$^{+0.1}_{-0.1}$E48&14.9$^{+0.3}_{-0.3}$&2.54$^{+0.01}_{-0.01}$&2.96$^{+0.01}_{-0.02}$&1.4$^{+0.1}_{-0.1}$E54&7.1$^{+0.1}_{-0.1}$E59&4541$^{+15}_{-15}$&4541$^{+15}_{-15}$&42.72&48.43\\
D1&3.5$^{+0.1}_{-0.1}$E39&51.1$^{+0.3}_{-0.3}$&2.54&2.4$^{+0.1}_{-0.1}$E45&23.2$^{+0.3}_{-0.3}$&1.72$^{+0.01}_{-0.01}$&3.12$^{+0.01}_{-0.01}$&2.9$^{+0.1}_{-0.1}$E51&2.3$^{+0.1}_{-0.1}$E59&2601$^{+16}_{-6}$&2601$^{+16}_{-6}$&40.05&47.95\\
D2H3&1.4$^{+0.1}_{-0.1}$E39&43.8$^{+0.5}_{-0.7}$&2.7&5.5$^{+0.1}_{-0.1}$E48&14.4$^{+0.2}_{-0.2}$&2.51$^{+0.01}_{-0.01}$&3.13$^{+0.02}_{-0.02}$&1.1$^{+0.1}_{-0.1}$E55&1.6$^{+0.1}_{-0.1}$E60&4000$^{+30}_{-20}$&4000$^{+30}_{-20}$&43.46&48.63\\
H2&1.2$^{+0.1}_{-0.1}$E39&33.0$^{+0.4}_{-0.2}$&2.9&4.2$^{+0.1}_{-0.1}$E46&20.9$^{+0.9}_{-1.9}$&2.08$^{+0.01}_{-0.01}$&3.37$^{+0.04}_{-0.04}$&6.5$^{+0.2}_{-0.1}$E55&2.3$^{+0.1}_{-0.1}E59$&2587$^{+16}_{-40}$&2587$^{+16}_{-40}$&44.40&47.94\\
\hline
\scriptsize{\textbf{$\delta=3.7$}}\\
\hline
A&1.2$^{+0.1}_{-0.1}$E36&16.2$^{+0.1}_{-0.1}$&2.63&4.5$^{+0.2}_{-0.3}$E45&20.6$^{+1.2}_{-1.2}$&2.28$^{+0.01}_{-0.01}$&2.68$^{+0.01}_{-0.01}$&1.9$^{+0.1}_{-0.1}$E52&6.2$^{+0.1}_{-0.1}$E57&5000&348$^{+2}_{-3}$&41.40&46.92\\
B1&1.3$^{+0.1}_{-0.1}$E36&12.4$^{+0.2}_{-0.1}$&2.59&3.2$^{+0.2}_{-0.1}$E46&400$^{+50}_{-50}$&2.63$^{+0.01}_{-0.01}$&3.07&1.2$^{+0.1}_{-0.1}$E52&4.5$^{+0.1}_{-0.1}$E57&5000&456$^{+4}_{-4}$&41.33&46.90\\
B2&5.8$^{+0.1}_{-0.1}$E35&28.7$^{+1.1}_{-0.4}$&2.76&8.8$^{+0.6}_{-0.4}$E43&15.8$^{+0.8}_{-0.8}$&1.89$^{+0.01}_{-0.01}$&2.99$^{+0.02}_{-0.02}$&2.4$^{+0.1}_{-0.1}$E49&2.0$^{+0.1}_{-0.1}$E57&5000&195$^{+3}_{-5}$&38.51&46.42\\
B3&9.3$^{+0.1}_{-0.1}$E35&22.5$^{+0.2}_{-0.2}$&2.6&4.6$^{+0.4}_{-0.4}$E41&10$^{+3}_{-2}$&1.43$^{+0.01}_{-0.02}$&3.19$^{+0.12}_{-0.12}$&5.4$^{+0.1}_{-0.1}$E48&3.7$^{+0.1}_{-0.1}$E56&5000&130$^{+1}_{-1}$&37.98&45.81\\
C1&2.1$^{+0.1}_{-0.1}$E37&13.3$^{+0.1}_{-0.1}$&2.3&1.1$^{+0.1}_{-0.1}$E47&15.7$^{+0.4}_{-0.4}$&2.75$^{+0.01}_{-0.01}$&3.15$^{+0.01}_{-0.01}$&6.6$^{+0.1}_{-0.1}$E51&7.2$^{+0.1}_{-0.2}$E57&5000&579$^{+5}_{-9}$&41.07&47.11\\
C2&5.6$^{+0.1}_{-0.1}$E36&19.0$^{+0.2}_{-0.2}$&2.55&1.6$^{+0.1}_{-0.1}$E46&7.1$^{+0.3}_{-0.3}$&2.53$^{+0.01}_{-0.01}$&2.95$^{+0.01}_{-0.01}$&3.2$^{+0.1}_{-0.1}$E52&4.1$^{+0.1}_{-0.1}$E57&5000&347$^{+3}_{-3}$&41.69&46.80\\
D1&8.2$^{+0.1}_{-0.1}$E36&19.0$^{+0.1}_{-0.1}$&2.54&2.3$^{+0.1}_{-0.2}$E43&8.5$^{+0.3}_{-0.3}$&1.79$^{+0.01}_{-0.01}$&3.10$^{+0.01}_{-0.01}$&2.3$^{+0.1}_{-0.1}$E49&9.1$^{+0.1}_{-0.1}$E56&5000&163$^{+1}_{-1}$&38.54&46.14\\
D2H3&5.8$^{+0.1}_{-0.1}$E36&20.6$^{+0.1}_{-0.1}$&2.71&1.1$^{+0.1}_{-0.1}$E46&7.2$^{+0.3}_{-0.3}$&2.40$^{+0.01}_{-0.01}$&3.16$^{+0.04}_{-0.03}$&2.4$^{+0.1}_{-0.1}$E53&6.0$^{+0.1}_{-0.1}$E57&5000&245$^{+2}_{-2}$&42.41&46.80\\
H2&2.0$^{+0.1}_{-0.1}$E36&12.2$^{+0.1}_{-0.1}$&2.89&1.7$^{+0.2}_{-0.2}$E44&5.4$^{+1.7}_{-1.1}$&2.07$^{+0.01}_{-0.01}$&3.28$^{+0.05}_{-0.05}$&8.0$^{+0.1}_{-0.1}$E53&7.8$^{+0.7}_{-0.7}$E56&5000&151$^{+7}_{-7}$&43.08&46.07\\
\enddata
\tablenotetext{a}{$p_0$ is not fixed during SED fits, but the derived errors are $\sim$0.01 and thus are not shown in the Table. $p_2$ of knot-B1 is fixed as the average value of other knots. }
\tablenotetext{b}{$B=B_{\rm eq}$ in case of $\delta=1$, but $B=5$ mG in case of $\delta>1$, more details to see the text. }
\end{deluxetable}

\end{document}